\documentclass[]{aastex631}

\usepackage{tablefootnote}
\usepackage{comment}

\def\s{\phantom}

\def\amin{\ifmmode^{\prime}\else$^{\prime}$\fi}
\def\asec{\ifmmode^{\prime\prime}\else$^{\prime\prime}$\fi}

\def\simgt{\lower.5ex\hbox{$\; \buildrel > \over \sim \;$}}
\def\simlt{\lower.5ex\hbox{$\; \buildrel < \over \sim \;$}}

\newcommand{\chandra}{\textit{Chandra}}

\newcommand{\kms}{\ensuremath{\mathrm{km}\,\mathrm{s}^{-1}}}

\begin{document}

\title{Modeling isolated magnetar spin-down evolution and implications for long-period radio transients}

\author[0009-0008-0402-1243]{Jon Kwong}
\affiliation{Columbia Astrophysics Laboratory, Columbia University, New York, NY 10027, USA}

\author[0000-0002-9709-5389]{Kaya Mori}
\affiliation{Columbia Astrophysics Laboratory, Columbia University, New York, NY 10027, USA}



\begin{abstract}
Long-period radio transients (LPTs) are a new class of radio sources characterized by long spin periods ($P_{\rm spin} > 10^3$ s) and highly variable radio emission. While known magnetars are relatively young ($\tau < 10^5$ yrs) with spin periods clustered between 1–10 sec, it has been proposed that LPTs may be linked to a missing population of older magnetars. In this paper, we present an extensive parametric analysis of isolated magnetar spin evolution using various propeller spin-down models. In general, at higher initial magnetar B-fields ($B_0 \simgt 10^{15}$ G) and larger ambient densities ($n_0 \simgt 10^2$~cm$^{-3}$), magnetars will transition to the propeller phase earlier, and they start accreting gas from the ISM or molecular clouds after $\tau \sim 10^8$ yrs. We found that a transition from the pulsar to the propeller phase is required to reach the observed LPT period range of $P > 10^3$ s. More specifically, our population synthesis study based on Monte-Carlo simulations shows that two propeller models can account for most of the observed LPT periods ($P \sim 1\rm{-}400$ [min]) and their period derivative constraints ($\dot{P} < 10^{-9}$ s s$^{-1}$). Our spin-down models predict that (1) nearby radio-quiet neutron stars with the estimated dipole B-field range of $B\sim(1\rm{-}5)\times10^{13}$ G will transition to the propeller phase eventually after $\tau \simgt 10^{7}$ yrs; (2) thermal X-ray emission from accretion-phase magnetars becomes too faint for detection after traveling ($d\simgt 10$~kpc) from their birth places; (3) sporadic radio outbursts observed from LPTs may not be explained by regular radio pulsar and magnetar emission mechanisms that operate during the propeller phase. 
\end{abstract}

\keywords{}


\section{Introduction} \label{sec:intro}

The formation of neutron stars (NSs) from the core collapse of massive stars concentrates angular momentum into a fraction of the original volume, making rapid rotation a defining characteristic of young NSs. Although these NSs lose a significant portion of their rotational energy through a variety of processes over their lifetimes, nearly all known pulsars and magnetars exhibit pulsation periods shorter than $P\sim 10$ s. In recent years, however, the advent of wide-sky radio interferometers through the SKA precursors and pathfinders has unveiled a new class of radio sources characterized by periods ($P > 10^3$ s) and highly variable radio emission with irregular outbursts \citep[e.g.,][]{18min_hurley-walker_2022}. To date, 12 long-period transients (LPTs) have been found, primarily at low Galactic latitudes (Table \ref{tab:LPT list}). Their transient nature suggests a large population of LPTs exists but has remained undetected due to selection biases in previous searches. The number of LPT detections is expected to increase rapidly to $\sim15$--30 over the next few years, owing to extensive ongoing searches in the radio band \citep{rodriguez_2025}. 

The radio emission from LPTs is coherent and highly linearly polarized, which, along with second-timescale structures, suggests likely compact object progenitors with highly ordered magnetic fields. Yet, the mechanisms powering LPT emissions remain unclear, as the radio luminosities are $>3$ orders of magnitude greater than the upper limits for rotation powered spin-down in isolated NSs. The stark distinction between the long periods observed in LPTs and the much shorter periods of well-established groups of radio sources, such as pulsars and magnetars, coupled with the mystery surrounding their emission mechanisms, has earned LPTs a fair amount of notoriety over the past few years. Stranger still are the distinct emission properties amongst LPTs. For example, most LPTs, such as GLEAM-X\,J1627$-$52, showed no evidence of an X-ray counterpart despite follow-up observations by {\it SWIFT} \citep{18min_hurley-walker_2022} and {\it Chandra} \citep{18min_Rea_2022}, setting stringent upper limit on its X-ray luminosity of ($L_X \sim 10^{29}\mathrm{-}10^{30}$ erg\,s$^{-1}$), suggesting that LPTs are generally X-ray faint. However, another recently discovered LPT ASKAP J1832$-$0911 was serendipitously discovered in the X-ray band by a \chandra\ observation targeted at a nearby supernova remnant, G22.7$-$0.2 \citep{44min_wang_rea_2024}. Due to these transient radio and X-ray properties, it is generally accepted that LPTs represent a new, unique class of compact objects or binaries whose emissions are neither rotation powered, nor accretion powered.

\begin{deluxetable*}{l|cccc}
\tablewidth{0pt} 
\tablecaption{List of long-period radio transients in ascending order of period. Only four have had their identities confirmed as either NS pulsars or WD+M dwarf binaries, and only a further three have been strongly favored as WD systems, as indicated below. This paper will focus on the remaining seven who have not yet been ruled out as NSs. 
}\label{tab:LPT list}
\tablehead{
\colhead{Object} & \colhead{$P$ (s [min])} & \colhead{$\dot{P}$ (s s$^{-1}$)} & \colhead{Counterpart} & \colhead{Citation}
}
\startdata 
{CHIME J0630+25}& 421 [7.0] & $\leq2.5\times10^{-12}$ & none & \cite{7min_dong_2024}\\
{ILT/CHIME J1634+45}$^{\text{c}}$& 841 [14.0]& $-9.0\times 10^{-12}$ & UV & \cite{14min_bloot_2025}\\ 
{GLEAM--X J1627--52}& 1091 [18.1] & $\leq1.2\times10^{-9}$ & none  & \cite{18min_hurley-walker_2022}\\
{GPM J1839--10}& 1318 [22.0] & $\leq3.6\times10^{-13}$ & none & \cite{21min_hurley-walker_2023}\\
{ASKAP J1832--09}& 2655 [44.3] & $\leq9.8\times10^{-10}$ & X-ray & \cite{44min_wang_rea_2024}\\
{ASKAP J1935+21}& 3225 [53.8] & $\leq1.2\times10^{-10}$ & none & \cite{54min_caleb_2024}\\
{ASKAP J1755--25}$^{\text{c}}$& 4186 [69.8] & $-1.0\times10^{-11}$ & none & \cite{70min_mcsweeney_2025}\\
{GCRT J1745--30}& 4620 [77.0] & -- & none  & \cite{77min_hyman_2005}\\
{ASKAP J1448--68}$^{\text{c}}$& 5631 [93.9] & $>-2.2\times10^{-8}$ & X-ray, optical & \cite{94min_anumarlapudi_2025}\\
{ILT J1101+55}$^{\text{b}}$ & 7531 [125.5] & $\leq3.0\times10^{-11}$ &  optical & \cite{125min_deRuiter_2024}\\
{GLEAM--X J0704--37}$^{\text{b}}$ & 10497 [174.95] & $\leq1.3\times10^{-11}$ & optical  & \cite{2.9hr_hurley-walker_2024}\\
{ASKAP J1839--08}& 23222 [387.0] & $\leq1.6\times10^{-8}$ &none & \cite{6.5hr_lee_2025}
\enddata
\footnote{Confirmed as NS pulsar}
\footnote{Confirmed as WD+M dwarf binary}
\footnote{Favored as WD system}
\end{deluxetable*}

Among the various proposed models for LPTs, two scenarios, one associated with magnetars \citep{Beniamini2023,  Cooper2024} and the other with white dwarfs (WDs) \citep{Katz2022, Qu2025}, seem to be the most promising as they account for some, but not all, of the radio and multi-wavelength properties of known LPTs. To date, a few of the recently discovered LPTs have been identified or favored as WD binary systems via the confirmation of optical/infrared counterparts, such as GLEAM-X~J0704–37 \citep{rodriguez_2025,2.9hr_hurley-walker_2024} and ILT J1101+5521 \citep{125min_deRuiter_2024}. 
Yet, multi-wavelength observations of the remaining LPTs have failed to confirm the presence of a companion \citep[e.g.,][]{Pelisoli_2025}, casting doubt on a similar WD origin. However, the magnetar interpretations similarly face obstacles, since the low quiescent X-ray luminosity would imply an older system ($\tau \gtrsim$~0.5\,Myr) with a dipolar B-field of $\lesssim10^{13}\,$G \citep{Rea2024}. In such cases, producing bright radio emission is challenging due to insufficient rotational or magnetic energy. Furthermore, a spin-down mechanism that can bridge the currently known population of young, $P<10$ s magnetars to this population of LPTs has yet to be established. Most proposed models suggest more dramatic spin-down mechanisms invoking short-lived accretion disks in binary systems \citep{Cary_2026} or the formation of a fallback disk by supernova ejecta \citep{Alpar_2001,Caliskan_2013,Gencali_2022,Fan_2024}. However, there is no clear observational evidence of fallback disks around long-period pulsars or magnetars, such as the magnetar 4U 0142+61, where this possibility was investigated using observations by JWST \citep{Hare_2024}.  

In this paper, we will focus on the magnetar interpretation and specifically aim to determine whether isolated NSs can be spun down to LPT periods of $P\gtrsim 10^3$ s in the absence of fall-back accretion. The spin evolution of magnetars has been widely studied and, like pulsars, can be segmented into three distinct phases -- (1) pulsar, (2) propeller, and (3) accretion. pulsar phase spin-down of magnetars has been rigorously studied in \cite{rea_2024}, which demonstrated that even highly-magnetized NSs with $B\sim10^{15}$ G are unlikely to achieve $P\gtrsim 10^3$ s through rotation-powered dipole radiation alone. During the propeller phase, magnetars lose angular momentum through interactions with the ISM or molecular clouds. Various models of propeller mechanisms have been proposed over the last five decades and have been summarized in detail in \cite{mori_2003}. Finally, isolated NS reaches the end of its spin evolution in the accretion phase, wherein surrounding material is no longer spun away but instead falls onto the NS surface, producing weak but potentially detectable X-rays described by \cite{rutledge_2001}. This paper takes a broader view by combining these processes into a complete picture of isolated magnetar spin-down to better understand the upper limits on magnetar periods across a wide parameter space. We will then use our results to assess whether the magnetar hypothesis is consistent with the observed timing properties of LPTs and discuss whether old magnetars are detectable in the radio and X-ray bands. In \S\ref{sec:spin-down description}, we present our generic NS spin-down model by considering the propeller effect and magnetic field decay. In \S\ref{sec:qual_model_analysis} and \S\ref{sec:population_synthesis}, we will fully explore the parameter space effects on the spin-down efficiency of the various propeller models. In \S\ref{sec:discussion}, we discuss implications from our magnetar spin-down models and Monte-Carlo simulations, including a special case for ASKAP J1832--09. Throughout the paper, we assumed the canonical NS mass ($M$) and radius ($R$) of $1.4 M_{\odot}$ and 10~km, respectively. 

\section{Description of the Propeller spin-down Model} \label{sec:spin-down description}
The spin of an isolated NS evolves over time according to a torque equation, which essentially encodes the rate of rotational energy loss. A generic spin-down evolution often depends strongly on the angular velocity ($\Omega$) and is described in the form of  
\begin{equation}
    I\dot{\Omega} \propto -\Omega^n, 
\end{equation}
where $I$ is the moment of inertia of the NS, and $n$ is known as a braking index, which is determined by the specific spin-down mechanism generating the torque. The canonical spin-down mechanism for isolated NSs is attributed to magnetic dipole  radiation, which converts the rotational energy of the NS into electromagnetic radiation via pair production, and has an associated braking index of $n=3$. However, other spin-down mechanisms can also contribute to the evolution of NS spin, each characterized by a distinct braking index. Typical examples, such as multipole magnetic radiation, gravitational waves, and pulsar winds, can make minor contributions to pulsar spin-down, and the observed braking indices of known pulsars rarely match the canonical value. Another such alternative spin-down mechanism is the propeller mechanism, often observed in NS binaries, in which the NS loses angular momentum by propelling the incoming particles at the magnetospheric outer boundary. Even for isolated systems, this mechanism can play a significant role in the later stages of the NS's life through its interactions with surrounding material. The spin evolution of isolated NSs could be classified into three stages with distinct spin-down mechanisms: 1) pulsar phase, 2) propeller phase, and 3) accretion phase. The transitions between these phases are governed by the relation between three characteristic radii: the light cylinder radius ($R_{\rm LC}=c/\Omega$), the magnetospheric radius ($R_m$), and the Keplerian circular orbital radius ($R_{K}=(GM/\Omega^2)^{1/3}$). Below, we follow the description of \cite{mori_2003} to outline our generic spin-down models through these evolution stages.

\subsection{Overview of Isolated Neutron Star Spin-down Evolution}
Young, rapidly rotating NSs typically have an extremely small light cylinder such that $R_{LC}<R_m$, and incoming material from the ISM, which is unable to penetrate $R_m$ does not interact with the co-rotating magnetic field lines found within $R_{LC}$. In such a case, we can associate the losses in rotational energy to the Poynting flux and the spin-down follows the standard magneto-dipole formula
\begin{equation}
    I\dot{\Omega} = -\frac{2\mu^2}{3c^3}\Omega^3\sin^2\chi \approx-\frac{2\mu^2}{3c^3}\Omega^3 \propto \mu^2 \Omega^3,
\end{equation}
where $\mu$ its magnetic dipole moment and $\chi$ the angle between the rotational and magnetic axes. For simplicity, we approximate the expression by setting $\chi=\pi/2$. 
To better make the connection with the observed quantities of period ($P$) and period derivative ($\dot{P}$), we rewrite the dipole spin-down formula as
\begin{equation}\label{eq:Pdot_dipole}
    \dot{P} = -\frac{8\pi^2\mu^2}{3Ic^3}P^{-1} \propto \mu^2 P^{-1}.
\end{equation}

Once the NS slows down enough so that $R_{LC}$ exceeds $R_m$, it enters the propeller phase, in which the surrounding medium can penetrate the light cylinder, quenching the dipole radiation. As a further consequence, charged particles in the surrounding medium are now free to interact with the closed magnetic field lines, inducing significant spin-up and ejecting them with this newfound angular momentum. Naturally, the spin-up of the surrounding medium generates a corresponding spin-down of the NS, which is the so-called propeller spin-down mechanism. Since the commencement of propeller spin-down is accompanied by the cessation of dipole spin-down, the condition $R_{LC}=R_m$ can be said to mark the transition from the pulsar phase into the propeller phase. 

For this propeller-driven picture of spin-down, the torque induced onto the NS would depend strongly on the rate of incoming material ($\dot{M}$) on top of $\Omega$. $\dot{M}$ in turn depends on the velocity of the surrounding medium at the magnetospheric radius relative to the motion of the NS ($v_m$), the density of the surrounding medium at the magnetospheric radius ($\rho_m$) and $\mu$ (which controls $R_m$). Assuming that there are no other physical quantities relevant to the propeller mechanism, a general propeller spin-down equation takes the functional form \citep{mori_2003}
\begin{equation} \label{eq:torque}
    I\dot{\Omega} = f(\mu,\rho_m,v_m,\Omega) = -\kappa \mu^{(3+n)/3}\rho_m^{(3-n)/6}v_m^{(3-4n)/3}\Omega^n,
\end{equation}
where $\kappa$ is a dimensionless constant. The exact behavior of the propeller torque is characterized by the choice of braking index $n$ (to be discussed in \S\ref{subsec:propeller models}), and the remaining exponents of the different quantities are determined in terms of $n$ via dimensional analysis. Since it is more common for the dipole moment to be expressed in terms of magnetic field strength ($B = \mu_m/R^3$ [G]) and density to be expressed in terms of a hydrogen number density ($n_m = N_A\rho_m$ [cm$^{-3}$]), we will use those quantities interchangeably in later sections.

Finally, when the NS slows down enough such that $R_{K}$ exceeds $R_m$, we expect that a majority of the incoming material will not gain enough angular momentum to be ejected. Instead, they are gravitationally captured and eventually accrete onto the NS surface. Since there is no longer a significant loss of angular momentum from the NS system, we can assume that the propeller spin-down ceases and mass accretion occurs when $R_{K}=R_m$. During the accretion phase, we assume that further spin-up or spin-down effects are negligible (unlike NS binaries with accretion disks), and it marks the endpoint for the spin-evolution of isolated NSs.

\subsection{Parameterization of Various Propeller Model} \label{subsec:propeller models}
In order to parameterize the propeller model, we begin with two simpler equations that describe how the model works. The first equation describes propeller torque as proportional to $\dot{M}$, which is something we would physically expect for such a mechanism
\begin{equation} \label{eq:gamma}
    I\dot{\Omega} = -\dot{M}R_mv_m\mathcal{M}^\gamma.
\end{equation}
The second equation describes the radial location of the magnetosphere by balancing the magnetic and ram pressures
\begin{equation} \label{eq:delta}
    \frac{\mu^2}{8\pi R_m^6} = \rho_m v_m^2 \mathcal{M}^\delta.
\end{equation}
In either equation, we have added a dimensionless ratio of rotational and translational velocities $\mathcal{M}\equiv R_m\Omega/v_m$ and introduced two exponents ($\gamma$ and $\delta$). Note that if we adopt $\gamma=1$, equation (\ref{eq:gamma}) reflects the torque on the NS if all the incoming gas at $R_m$ is spun up to the angular velocity $\Omega$ of the NS. Meanwhile, if we adopt $\delta=0$, equation (\ref{eq:delta}) reflects the ram pressure of the gas balances for spherical accretion. Therefore, the two exponents most significantly determine the propeller spin-down and overall NS evolution in the $P$--$\dot{P}$ diagram. In general, $\gamma$ controls the efficiency of the spin-down mechanism, whereas $\delta$ determines the location of $R_m$ and transitions between the pulsar, propeller, and accretion phases. \citet{mori_2003} detailed six possible propeller spin-down models (Table \ref{tab:propmodels}), each characterized by particular values of $\gamma$ and $\delta$. 
\begin{deluxetable*}{l|cc|cccc|c|c|c}
\label{tab:propmodels}
\tablewidth{0pt} 
\tablecaption{A list of various propeller models as detailed in \cite{mori_2003}. Their qualitative descriptions are based on our results in \S3-4.} 
\tablehead{
\colhead{Model} & \colhead{$\gamma$} & \colhead{$\delta$} & \colhead{$n$} & \colhead{$n_\mu$} & \colhead{$n_v$} & \colhead{$n_\rho$} & 
\colhead{Spin-down efficiency} & \colhead{Propeller transition} & \colhead{Citation}
}
\startdata 
{Model A}& -1 & 0 & -1 & 2/3 & 1/3 & 7/6 & Low & Late & \citet{modelA_illarionov_1975}\\
{Model B}& 0 & 0 & 0 & 1 & 1 & 1 & Low & Late & \citet{modelB_davidson_1973}\\
{Model C}& 1 & 0 & 1 & 4/3 & 5/3 & 5/6 & High & Late & \citet{modelC_menou_1999}\\
{Model D}& 2 & 0 & 2 & 5/3 & 7/3 & 2/3 & Very High & Late & \citet{modelD_davies_1979}\\
{Model E}& 1 & 1 & 3/7 & 8/7 & 3/7 & 6/7 & High & Early & \citet{mori_2003}\\
{Model F}& 2 & 2 & 3/4 & 5/4 & 0 & 3/4 & Very High & Early & \citet{modelF_romanova_2003}\\
\enddata
\end{deluxetable*}

Using equation (\ref{eq:delta}), we can obtain an explicit expression for $R_m$
\begin{equation}\label{eq:R_m}
    R_m = \Biggl(\frac{1}{8\pi }\mu^2v_m^{\delta-2}\rho_m^{-1}\Omega^{-\delta}\Biggr)^{\frac{1}{\delta+6}}.
\end{equation}
Finally, by adopting $\dot{M}=\rho_m(\pi R_m^2)v_m$ in equation \ref{eq:gamma}, we can obtain an analytical expression for the spin-down torque\footnote{There is a missing factor of 1/8 in \cite{mori_2003} equation 8 which has been added here.} expressed in terms of $\mu$, $R_m$ and $\mathcal{M}$
\begin{equation} \label{eq:torque2}
    I\dot{\Omega} = -\frac{1}{8}\mu^2 R_m^{-3}\mathcal{M}^{\gamma-\delta} = -\frac{1}{8}\mu^2 R_m^{\gamma-\delta-3}v_m^{\delta-\gamma}\Omega^{\gamma-\delta}\propto \Omega^{\frac{3(2\gamma-\delta)}{\delta+6}}\mu^{\frac{2\gamma+6}{\delta+6}}v_m^{\frac{2(\gamma-\delta)(\delta+2) - 3(\delta-2)}{\delta+6}}\rho_m^{\frac{\delta-\gamma+6}{\delta+6}}. 
\end{equation}
Alongside the braking index $n$, we define corresponding indices for the exponents for the magnetic moment, velocity, and density terms, as $I\dot\Omega \propto \Omega^n \mu^{n_\mu} v_m^{n_v} \rho_m^{n_\rho}$, whose numerical values are presented in Table \ref{tab:propmodels} for each propeller model. We can convert equation (\ref{eq:torque2}) into an expression of in terms of $P$, $B$, and $n_m$, in order to obtain the power law relation of $\dot{P}$ to the rest of the quantities:
\begin{equation}\label{eq:Pdot}
    \dot{P}\propto P^{\frac{5\delta-6\gamma+12}{\delta+6}}B^{\frac{2\gamma+6}{\delta+6}}v_m^{\frac{2(\gamma-\delta)(\delta+2) - 3(\delta-2)}{\delta+6}}n_m^{\frac{\delta-\gamma+6}{\delta+6}}, 
\end{equation}
where universally, the exponent of $P$ will be $-n+2$. We should note that in the propeller model, the magnetic moment ($\mu$) becomes time-dependent if magnetic decay is taken into consideration. Consequently, $R_m$ is non-constant and thus, $v_m$ and $\rho_m$ could also vary across the lifespan of the NS. Nevertheless, the simplest approach to prescribe these values is to fix $\mu$ to its initial value for the magnetic dipole moment ($\mu_0$), $v_m$ to the translational velocity of the NS ($v_0$), and $\rho_m$ to the density of the surrounding material in the absence of the NS ($\rho_0$). Within such a context, equation (\ref{eq:torque2}) simplifies to an analytically solvable differential equation for $\Omega$. We will address more intricate descriptions for $\mu$, $v_m$, and $\rho_m$ in the following two sections. 

\subsection{Gravitational Correction}\label{subsec:grav_correction}
The generic description at the end of \S\ref{subsec:propeller models} assumes that $v_m$ is entirely dominated by the translational motion of the surrounding material relative to the NS, neglecting gravitational effects. For example, if $v_m=v_0 \approx 0$, the ram pressure term in equation (\ref{eq:delta}) vanishes, although physically, it should match the ram pressure from spherical accretion. Thus, the propeller model described in \S\ref{subsec:propeller models} is inadequate for very slow-moving NSs at $v_m\simlt 100$ \kms. We incorporated the gravitational correction in the model by altering $v_m$ and $\rho_m$ with their appropriate dependencies on $R_m$. 

We begin by introducing a critical radius $R_0= 2GM/v_0^2$, where the Keplerian velocity becomes comparable to $v_0$.
Next, we consider gravitational effects negligible for $R_m>R_0$, but when $R_m<R_0$, $v_m$ and $\rho_m$ scale with $R_m$ as they would in a conventional gravity-dominating system. The scaling for $v_m$ is given by energy conservation in a gravitational potential 
\begin{equation}
    v_m\sim R_m^{-1/2} \Rightarrow v_m = v_0\sqrt{\frac{R_0}{R_m}} = \sqrt{\frac{2GM}{R_m}}. 
\end{equation}
For the gas density $\rho_m$, this scaling is instead determined by the continuity equation for spherical accretion
\begin{equation}
    \rho_m\sim R_m^{-3/2} \Rightarrow \rho_m = \rho_0\Biggl(\frac{R_0}{R_m}\Biggr)^{3/2} = \frac{\rho_0}{v_0^3}\biggl(\frac{2GM}{R_m} \biggr)^{3/2}. 
\end{equation}
It is important to note that these revised definitions for $v_m$ and $\rho_m$ are not merely corrections -- they are fully consistent with the expressions obtained for a spherically accreting NS, where the accreting gas has density $\rho_0$ at a radial distance $R_m$. In this context, $R_m<R_0$ marks the transition from a velocity-dominated $\dot{M}$ to a gravitationally-dominated $\dot{M}$. By substituting these new definitions for $v_m$ and $\rho_m$ into equations (\ref{eq:R_m}) and (\ref{eq:torque2}), we find 
\begin{equation} \label{eq:R_m_grav}
    R_m = (2GM)^{\frac{\delta-5}{3\delta+7}}\Biggl(\frac{1}{8\pi}\mu^2v_0^{3}\rho_0^{-1}\Omega^{-\delta}\Biggr)^{\frac{2}{3\delta +7}},
\end{equation}
\begin{equation}\label{eq:torque2_grav}
    I\dot{\Omega} = -\frac{1}{8}\mu^2 R_m^{-3}\mathcal{M}^{\gamma-\delta} = -\frac{1}{8}(2GM)^{\frac{\delta-\gamma}{2}}\mu^2 R_m^{\frac{3(\gamma-\delta)-6}{2}}\Omega^{\gamma-\delta} \propto \Omega^{\frac{7\gamma - \delta}{3\delta + 7}}\mu^{\frac{6\gamma - 3\delta -5}{3\delta + 7}}v_0^{\frac{9(\gamma - \delta -2)}{3\delta + 7}}\rho_0^{\frac{3(\delta - \gamma +2)}{3\delta + 7}}. 
\end{equation}
Converting this into $P$ and $\dot{P}$, we again obtain the power law relation of $\dot{P}$ to the rest of the quantities
\begin{equation}\label{eq:Pdot_grav}
    \dot{P}\propto P^{\frac{7(\delta-\gamma+2)}{3\delta+7}}\mu^{\frac{6\gamma - 3\delta -5}{3\delta + 7}}v_0^{\frac{9(\gamma - \delta -2)}{3\delta + 7}}n_0^{\frac{3(\delta - \gamma +2)}{3\delta + 7}}.
\end{equation}
Crucially, the gravitationally corrected torque in equation (\ref{eq:torque2_grav}) introduced here is not a generalization of equation (\ref{eq:torque2}), and neither does it constitute a new phase in the spin-down evolution. The appropriate equation for the torque must be chosen according to the condition $R_m<R_0$, with a transition from gravitationally-dominated $\dot{M}$ transition to a velocity-dominated $\dot{M}$ during spin-down evolution being quite rare. Thus, the dominant parameter determining the appropriate torque equation is $v_0$, and the results in \S3.2 will allow us to better understand how $v_0$ can affect spin-down evolution.

\subsection{Magnetic Field Decay}
We also consider that the magnetic moment ($\mu$) varies with time throughout the NS spin evolution lifetime. In particular, strong magnetic fields in magnetars are believed to decay due to  Ohmic dissipation, ambipolar diffusion, and Hall drift  \citep{Goldreich1992}. While the magnetic field decay processes are complex and intrinsically tied to the magneto-thermal evolution in the NS crust and core, various models predict that the decay time scales are $\simgt 10^6$ yrs. In our simulation, we implemented a phenomenological and parametrized model for the magnetic field decay presented in \cite{aguilera_2008}, 
\begin{equation}
    B (t) = B_0\frac{\exp(-t/\tau_{\text{O}})}{1+\frac{\tau_{\text{O}}}{\tau_{\text{H}}} - \frac{\tau_{\text{O}}}{\tau_{\text{H}}}\exp(-t/\tau_{\text{O})}}, 
\end{equation}
where $B_0$ is the initial magnetic field strength. $\tau_O$ and $\tau_H$ are the characteristic timescales of the Ohmic dissipation and Hall effect decay, respectively. We adopted typical values $\tau_{\text{O}} = 10^6$~yrs and $\tau_{\text{H}} = 10^3\text{yrs}/(B_0/10^{15} \text{G})$ \citep{aguilera_2008, Pons_2008}. The $B(t)$ evolution depends on $B_0$, and it keeps decaying toward zero in the form of equation (14). Magneto-thermal simulations of magnetic decay (e.g., \cite{Vigano_2013}) indicate that NS magnetic fields begin to plateau to a minimum value during the later stages of its life, though such simulations typically cease at $\tau\sim 10^6$. Since we will be investigating the spin evolution over much longer timescales ($\tau\sim 10^8$ yrs), we approximate this behavior by setting a floor on the magnetic field as assumed in \cite{boldin_popov_2010}
\begin{equation}
    B_{\text{min}} = \text{min}(B_0/2,2\times10^{13}\text{G}). 
\end{equation}
The lower threshold of $B = 2\times10^{13}$~G is consistent with the observed dipole B-fields measured from nearby radio-quiet NSs. These NSs are considered older magnetars, with characteristic ages of $\tau\sim10^6$~yrs, after their magnetic fields have decayed considerably. 

\section{Magnetar spin-down models and parameters} \label{sec:qual_model_analysis}
We simulate the spin evolution of an isolated NS using the equations and transition conditions described in \S\ref{sec:spin-down description}, with the adaptive-step RK45 method from Python's {\tt scipy} library. The model includes three physical parameters ($\mu_0, v_0$, and $n_0$) and two model parameters ($\gamma$ and $\delta$). We set the initial spin period $P_0 = 0.01$ s. In each simulation, we track the evolution of $P(t)$ and $\dot{P}(t)$, along with three key radii -- $R_{\rm LC}, R_m$ and $R_K$ -- which determine the transitions between the pulsar, propeller and accretion phases. Most results in the following sections are presented in $P$--$\dot{P}$ diagrams where we can directly compare the spin properties of LPTs with the evolutionary paths of simulated NSs.

\subsection{Model Comparison and evolution} \label{subsec:model_comparison}

\begin{figure}
    \centering
    \includegraphics[width=0.92\textwidth]{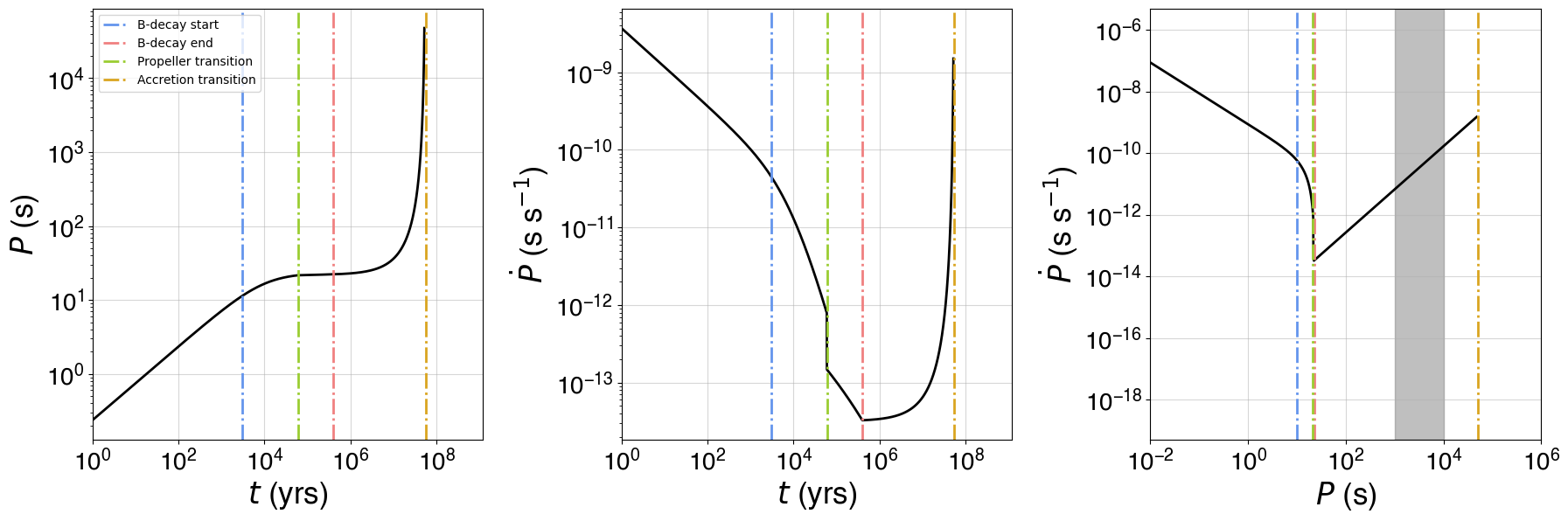}
    \caption{NS spin evolution for $B_0=10^{15}$ G, $v_0=100$ \kms, $n_0 = 1$ cm$^{-3}$ in case of the propeller model E. The vertical dashed lines indicate the four transition points as detailed in the text. In the rightmost panel, the typical LPT period range ($P = 10^3 - 10^4$~s) has been indicated in gray.}
    \label{fig:evolution_phases}
\end{figure} 

In general, the spin-down evolution of a simulated NS is characterized by four transition points: (i) onset of significant decay of the NS magnetic field, (ii) cessation of magnetic field decay, (iii) transition from the pulsar phase to propeller phase, and (iv) onset of accretion. These phases are indicated in blue, red, green, and yellow, respectively, in time evolution and the $P$--$\dot{P}$ diagram (Figures \ref{fig:evolution_phases}). One notes that the order of (ii) and (iii) is not fixed and would depend on both the choice of propeller models and initial parameters. Outside the duration of significant magnetic field decay (i.e., before (i) and after (ii)), the spin-down torque or $\dot{P}$ in the pulsar and propeller phases follows a power-law dependence on $P$, as expected from equations (\ref{eq:torque}) and (\ref{eq:torque2_grav}).

\begin{figure} 
    \centering
    \includegraphics[width=0.92\textwidth]{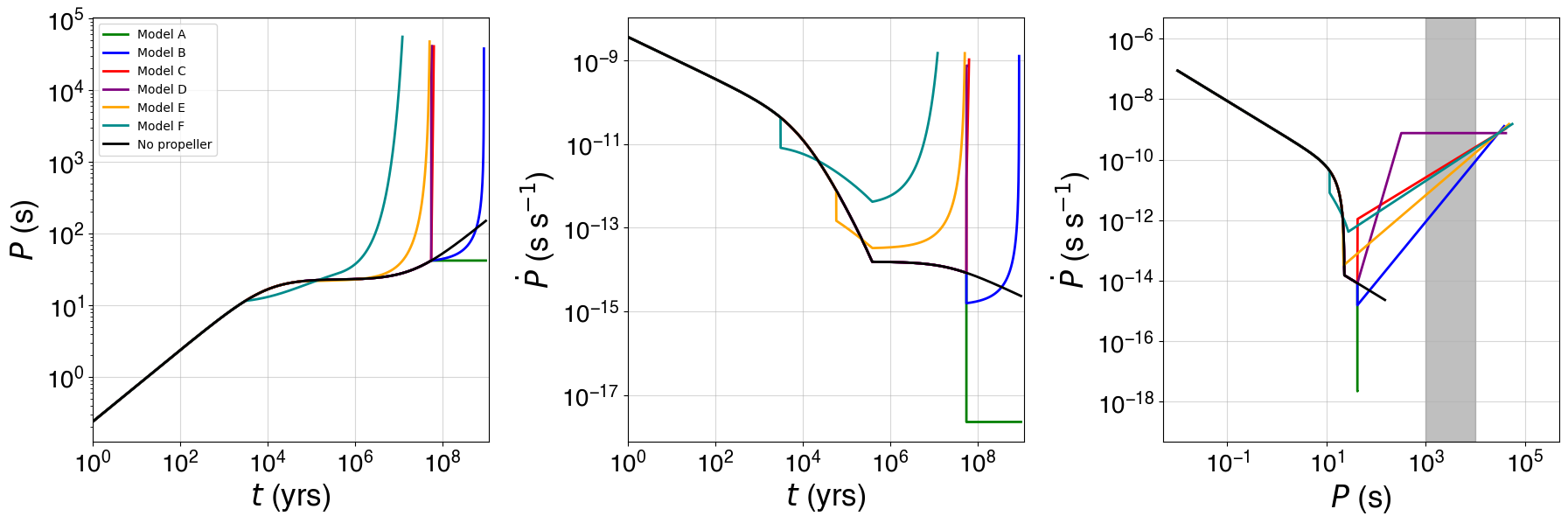}
    \caption{A comparison of the six propeller spin-down models for the same input parameters ($B_0=10^{15}$ G, $v_0=100$ \kms, and $n_0 = 1$ cm$^{-3}$), including the case where only dipole spin-down is considered. In the rightmost panel, the typical LPT period range ($P = 10^3 - 10^4$~s) has been indicated in gray.}
    \label{fig:model_comparison}
\end{figure} 

In all six propeller models, the early-stage evolution during the pulsar phase is identical when the same input parameters are adopted. The models begin to diverge in their trajectories on the $P$--$\dot{P}$ diagram only after the pulsar-propeller transitions. Note that this transition does not occur at a fixed time or spin period, but rather depends on the assumed propeller spin-down model. More specifically, equation (\ref{eq:R_m_grav}) dictates that the transition is determined most significantly by the value of  $\delta$, as evident in our simulations: models A, B, C, and D (all with $\delta=0$) transition to the propeller phase at the same time and spin period, whereas models E and F do much earlier. Furthermore, $\dot{P}$ is always discontinuous at the pulsar-propeller transition, due to an abrupt change in the spin-down torque. The magnitude of this discontinuity is model dependent: models with $\{\gamma=-1,\delta=0\}$ and $\{\gamma=2,\delta=0\}$ exhibit a $\dot{P}$ jump of a few orders of magnitude, whereas models with $\gamma = \delta$ produce only small jumps which are less than an order of magnitude. The relative sizes of these discontinuities had led \cite{mori_2003} to favor models with $\gamma = \delta$ (models B, E, F) as more physically plausible. 

\subsection{Parameter space study} \label{subsec:qualpar_study}
\begin{figure}
    \centering
    \includegraphics[width=0.92\textwidth]{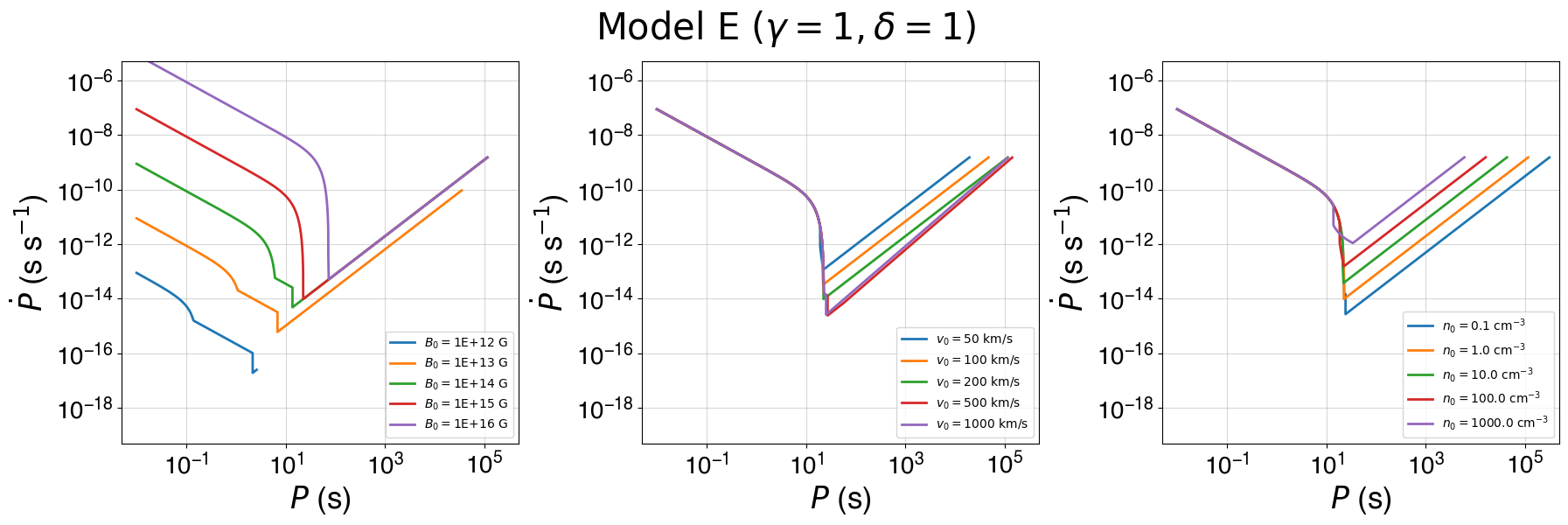}
    \caption{A qualitative study of the parameter space for the propeller model E, which traces the evolution of an NS in the $P-\dot{P}$ space. The leftmost plot varies the initial B-field $B_0$, the center plot varies the translational velocity of the NS $v_0$, and the right column varies the density of the surrounding material $\rho_0$. The remaining parameters, which are not being varied, are fixed to $B_0=10^{15}$ G, $v_m = 200$ \kms and $n_0$ = 1~cm$^{-3}$. Similar results for the remaining models can be found in appendix \ref{app:qual_par_study}.}
    \label{fig:qualpar_study}
\end{figure}

Aside from the two model parameters, the model includes three physical parameters: the initial magnetic field strength ($B_0$), the NS velocity ($v_0$), and the ambient gas density ($n_0$). From the general description of the propeller spin-down mechanism in \S\ref{sec:spin-down description}, it should be expected that increasing any of these parameters -- $B_0$, $v_0$, and $n_0$ -- would result in stronger spin-down torques. We will see that this is indeed the case except for low velocity values ($v_0\ll 100$ \kms). Our primary goal in this section is to investigate how varying these parameters affects the overall spin evolution of an NS and its trajectory in the $P$--$\dot{P}$ diagram. Hence, we conducted a single-parameter survey by running simulations for a wide range of each physical parameter, while keeping the other two parameters fixed. The parameters were fixed to $B_0 = 10^{15}$ G, $v_0 = 200$ \kms, and $n_0 = 1$ cm$^{-1}$. The results are presented in Figure~$\ref{fig:qualpar_study}$.

\begin{itemize}

\item[(1)]{Initial NS B-field strength ($B_0$):} Both the dipole radiation and propeller spin-down torques depend strongly on the magnetic moment $\mu$ and hence $B(t)$.  The obvious consequence of a higher $B_0$ value is that the spin-down efficiency of a NS should be generally greater over its lifetime. During the pulsar phase, this is manifested in the results of \ref{fig:qualpar_study}, where the trajectories for larger $B_0$ values start higher and progress further along the $\dot{P}$ axis before (time-dependent) B-field decay kicks in. Furthermore, although a lower B-field reduces the $P$ threshold required to transition to the propeller phase due to a smaller $R_m$, the lower B-field cases almost universally struggle to enter the propeller phase within the timescale of the simulation as a result of the lower spin-down rate during the pulsar phase. However, once in the propeller phase, the $B\geq 10^{14}$ G cases no longer follow this relationship between $B_0$ and $\dot{P}$. This is since NSs generally enter the propeller phase only much later in their evolution, when the magnetic field decay has ceased, and $\dot{P}$ becomes less insensitive to $B_0$. Nonetheless, a higher initial B-field enables NSs to reach the propeller phase more quickly and is generally crucial for achieving longer spin periods on shorter timescales. 

\item[(2)]{NS velocity ($v_0$):} The NS velocity plays no role during the pulsar phase, and trajectories in the $P$--$\dot{P}$ diagram for different $v_0$ values remain identical until the NS transitions into the propeller phase. By comparing the $v_0 = 10^3$ \kms and $v_0 = 10^4$ \kms cases in Figure \ref{fig:qualpar_study}, we can deduce that higher $v_0$ values enable NSs to enter the propeller phase slightly earlier and reach longer spin periods over shorter timescales, as expected from equation (\ref{eq:torque2}). However, it is also evident that at lower $v_0$ values, the behavior of the spin-evolution breaks from this expectation. In particular, in cases where $v_0 < 100$ \kms, NSs enter the propeller phase drastically earlier and consequently spin down far more efficiently as a result. This is a result of the gravitationally corrected propeller spin-down taking effect at $v_0\lesssim 100$ \kms, and per equation $(\ref{eq:R_m_grav})$, a smaller velocity leads to a smaller $R_m$ and thus an earlier transition to the propeller phase for all propeller models. This is the dominant effect for typical NS velocities of $v_0\sim 100$ \kms, and therefore a lower $v_0$ value actually leads to reaching longer spin periods more effectively on shorter timescales.

\item[(3)]{Ambient gas density ($n_0$):} Like $v_0$, the ambient gas density plays no role during the pulsar phase. As expected, higher $n_0$ values lead to earlier transitions to the propeller phase and higher spin-down rate (Figure \ref{fig:qualpar_study}). However, this effect is minute because of the weak dependence of the propeller spin-down torque on $n_0$ as seen in equation (\ref{eq:torque2}) and (\ref{eq:torque2_grav}). Thus, higher values of $n_0$ lead to a more effective spin-down, but it is typically less impactful than $B_0$ or $v_0$. 

\end{itemize}

\section{Population Synthesis} \label{sec:population_synthesis}
It is evident from \S\ref{sec:qual_model_analysis} that the six propeller models yield diverse NS spin evolution paths. In addition, the initial choice of $\mu_0,v_0$ and $\rho_0$ significantly affects the timing and duration of the propeller phase. Hence, a full exploration of the parameter space for these physical parameters for each propeller model is required to gain a complete picture of the long-term magnetar spin evolution, rather than tracking spin evolution across several initial conditions. Following \cite{rea_2024}, we employed a population synthesis method based on Monte-Carlo simulations by randomizing $\mu_0,v_0$ and $\rho_0$ values for a given propeller model.

\begin{deluxetable*}{l|c|c|c}
\tablewidth{0pt} 
\tablecaption{Random distributions for the initial parameters of the simulated population of NSs} \label{tab:initial_population_distributions}
\tablehead{
\colhead{Parameter} & \colhead{Units} & \colhead{Distribution type} & \colhead{Distribution values}
}
\startdata 
{NS age ($\tau$)} & yrs & Uniform & $\tau\sim U(0,10^9)$\\
{Initial B-field ($B_0$)}& G  & Log-normal & $\log B_0\sim \mathcal{N}(13.25,0.75^2)$\\
{NS velocity ($v_0$)}& \kms  & Maxwell-Boltzmann & $v_0\sim MB(265)$ \\
{Number density ($n_0$)}& cm$^{-3}$  & Fixed & $n_0 = 1$\\
{Initial period ($P_0$)}& s  & Log-normal & $\log P_0\sim \mathcal{N}(-0.6,0.3^2)$ \\
{NS Mass ($M$)}& $M_{\odot}$ & Fixed & $M = 1.4$ \\
{NS Radius ($R_{\text{NS}}$)}& km & Fixed & $R_{\text{NS}}=10$\\
\enddata
\end{deluxetable*}

\subsection{Simulation setup: initial parameter distributions} \label{subsec:initial_population}

Instead of tracking the trajectories of a small set of simulations, our population synthesis method begins with a large population of randomly generated NS parameters and analyzes the final states of these simulated NS. Following \cite{rea_2024}, to simulate the observable population of NS at present, we assign each NS an age ($\tau_{\text{NS}}$), which determines when its simulation ends. $\tau_{\text{NS}}$ is randomly drawn from a uniform distribution between 0 and $10^9$ years, reflecting the constant birthrate of NSs during the history of our galaxy. We also assign initial B-fields ($B_0$) and initial spin periods ($P_0$) using log-normal distributions as adopted in \citet{rea_2024}. For NS velocities ($v_0$), we assume that they follow a Maxwell-Boltzmann distribution centered at 265 \kms\ \citep{sartore_2010}. For the ambient gas density ($n_0$) is fixed to $n_0=1$ cm$^{-3}$, assuming that all NSs travel entirely within the ISM.  The details for the initial distributions of all physical parameters are summarized in Table \ref{tab:initial_population_distributions}. After randomly generating an initial population of $10^5$ NSs based on these statistical distributions, we simulated their spin evolutions for each of the six models described in Table \ref{tab:propmodels}.

\begin{figure} 
    \centering
    \includegraphics[width=0.32\textwidth]{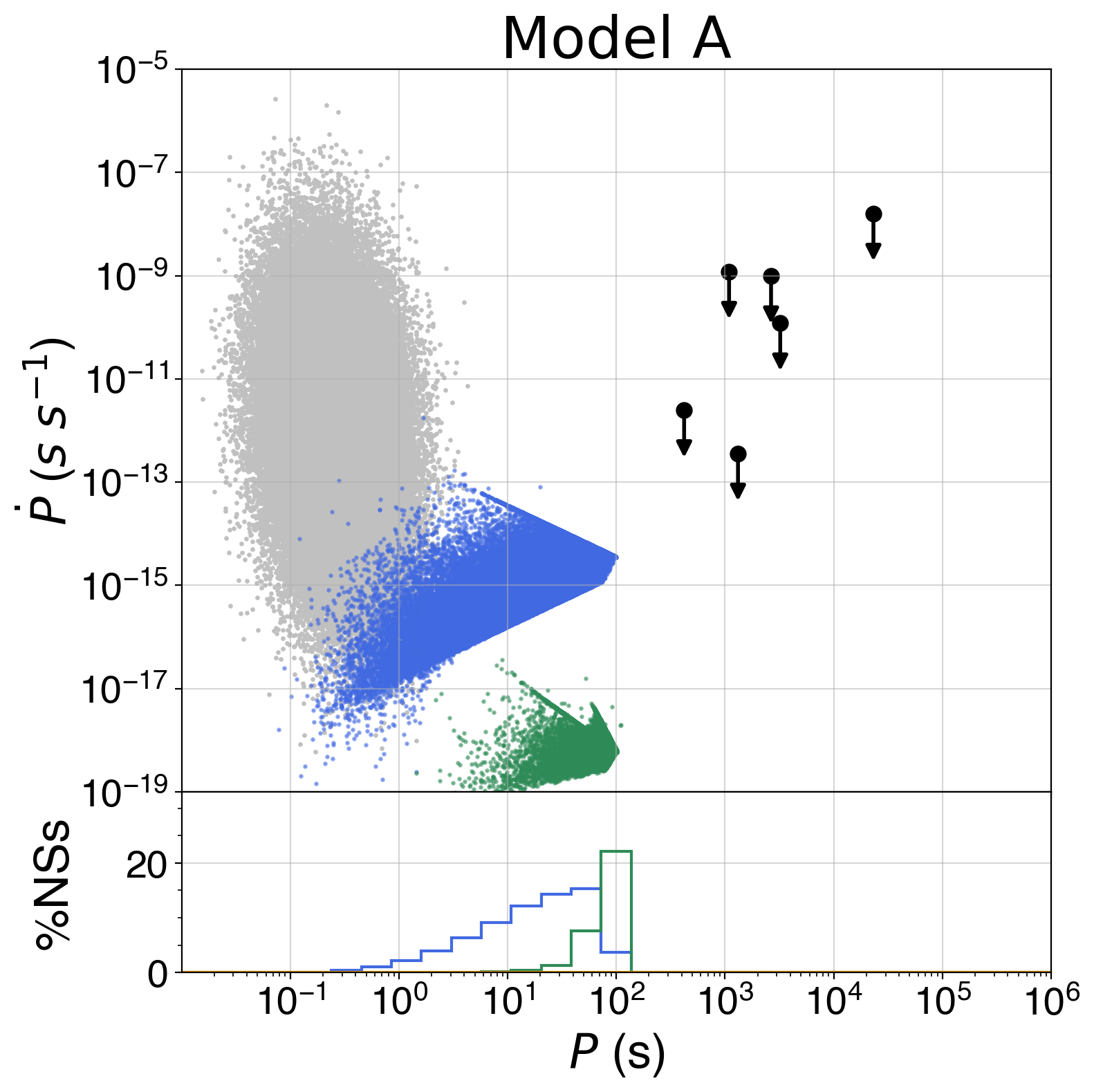}
    \includegraphics[width=0.32\textwidth]{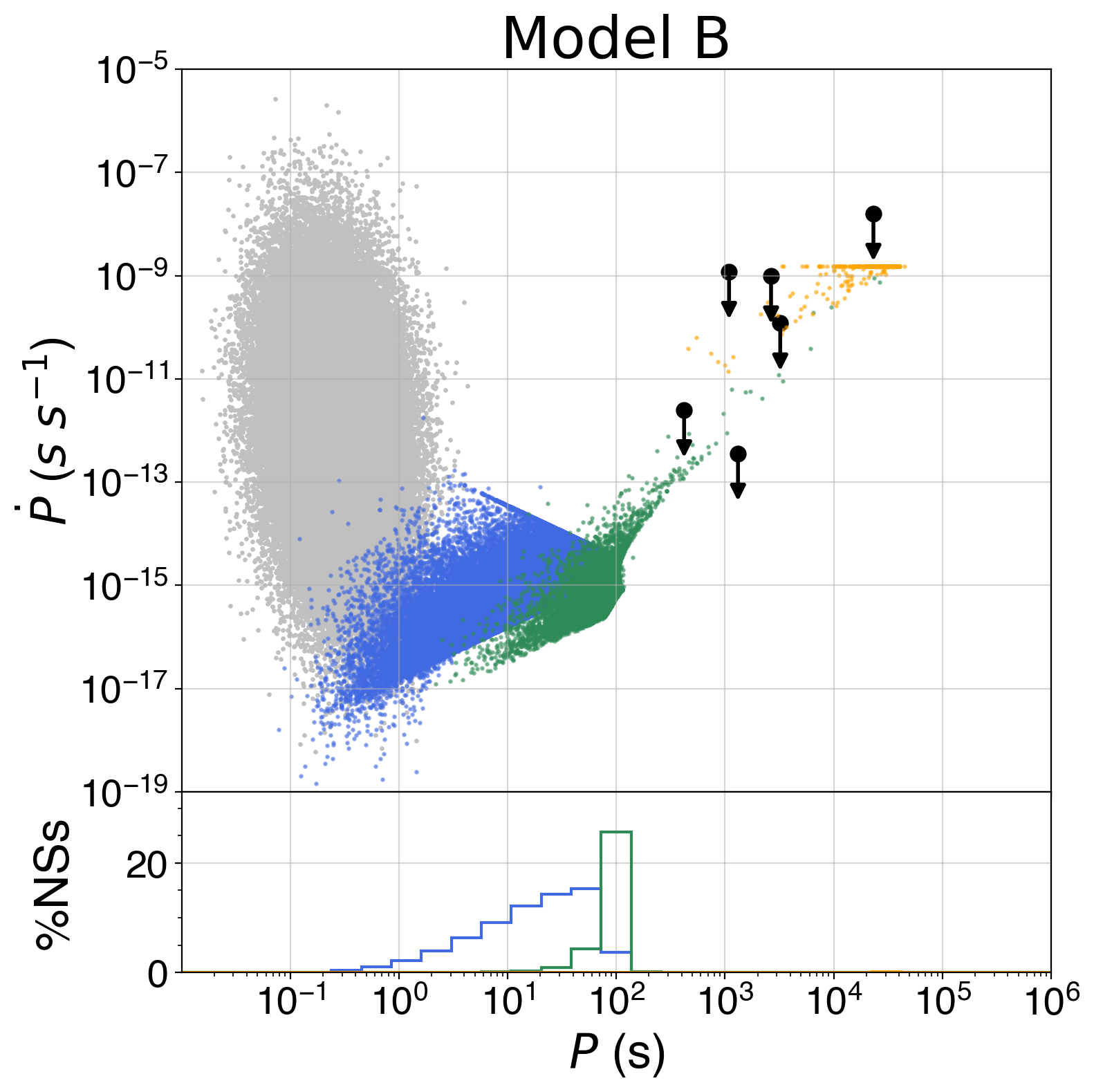}
    \includegraphics[width=0.32\textwidth]{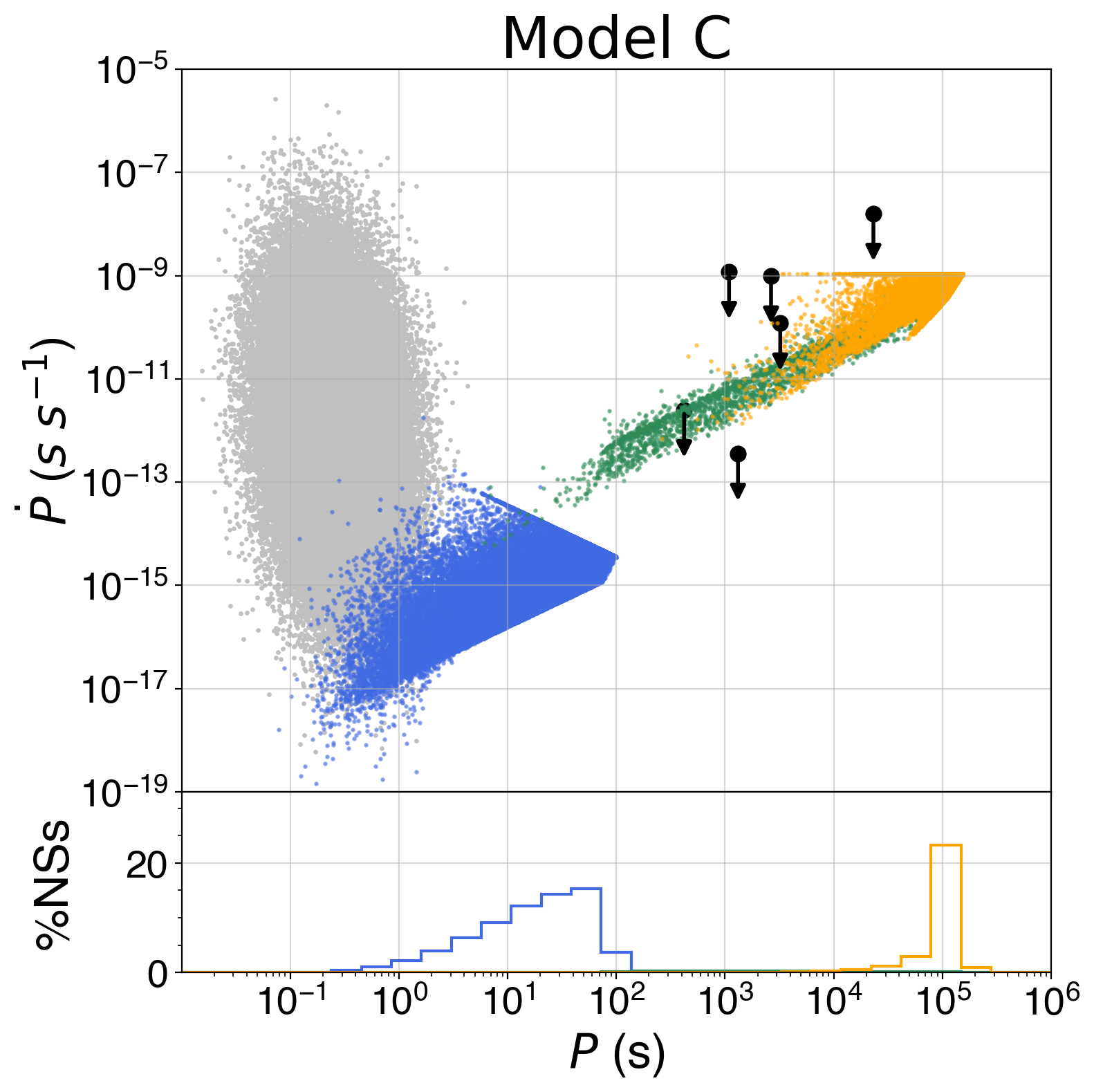}\\
    \includegraphics[width=0.32\textwidth]{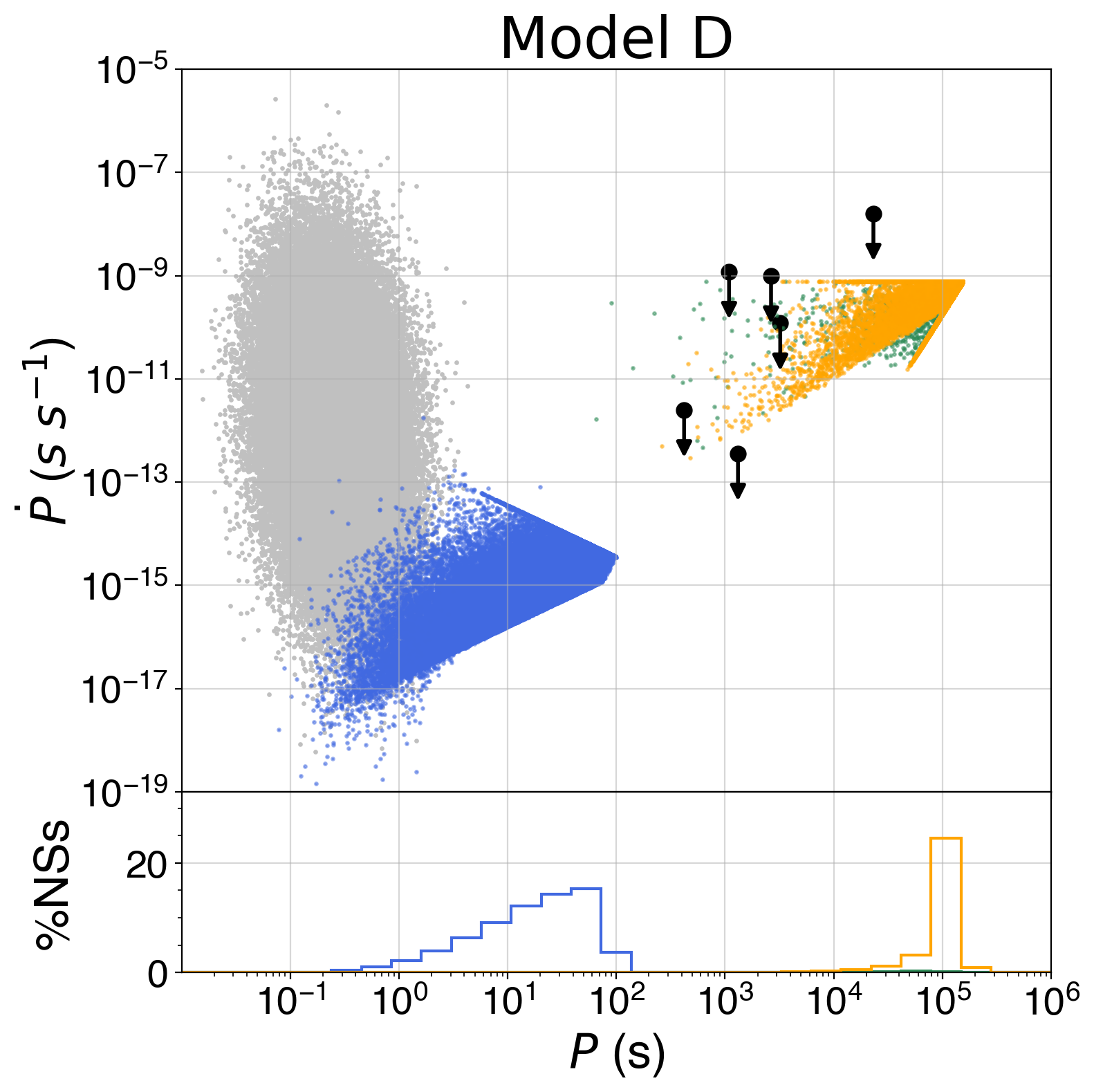}
    \includegraphics[width=0.32\textwidth]{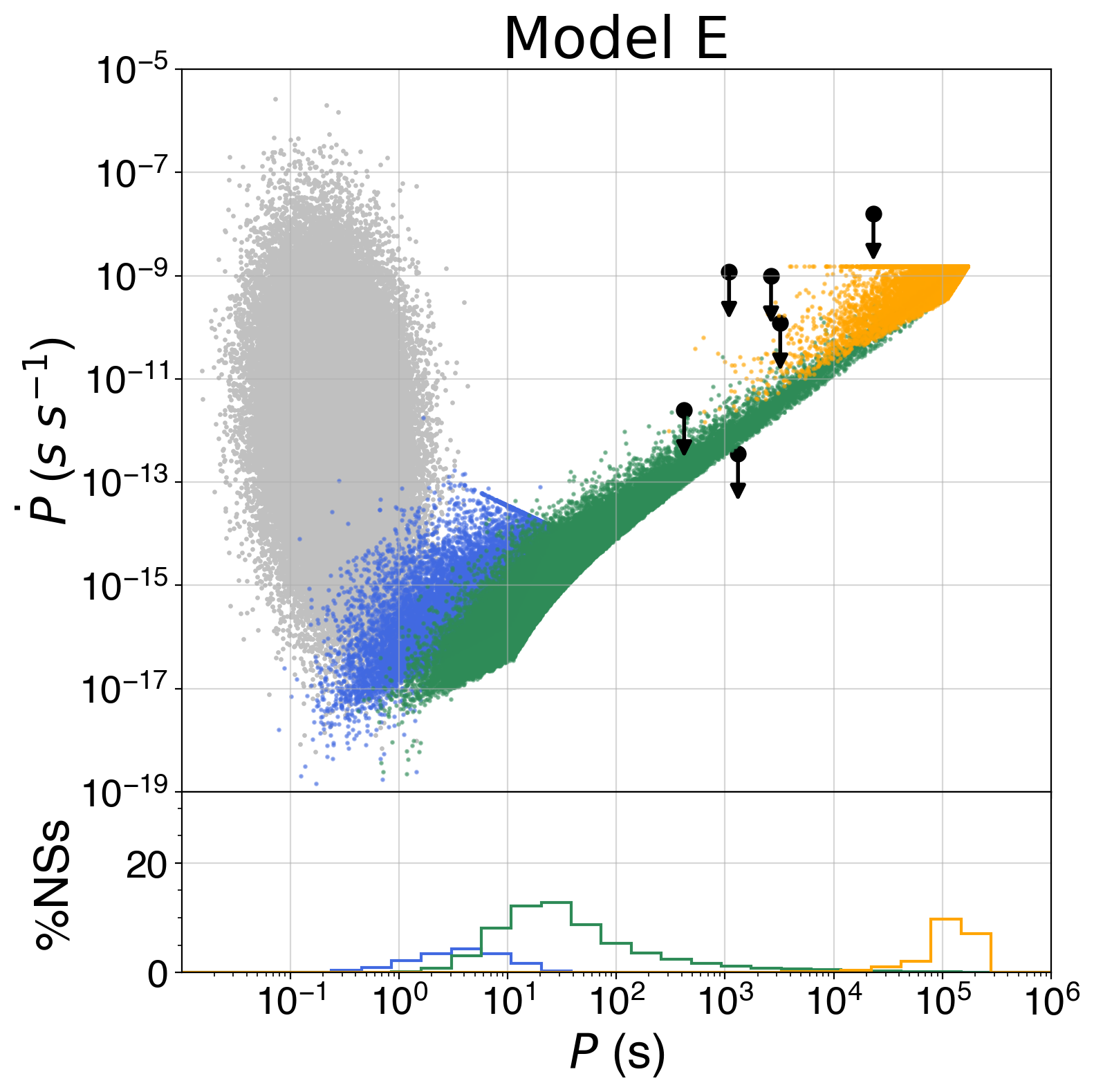}
    \includegraphics[width=0.32\textwidth]{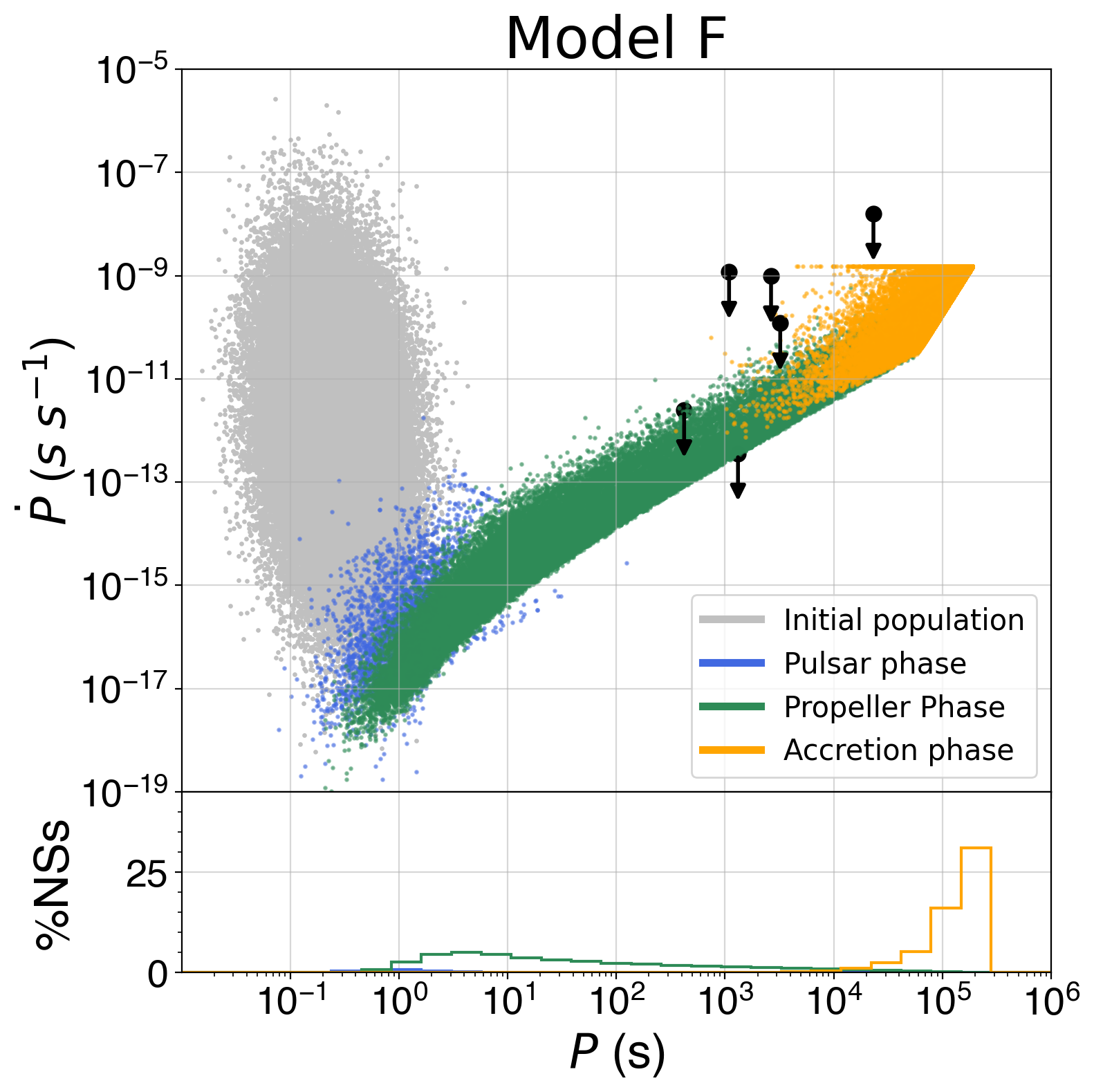}
    \caption{Population synthesis results for the various propeller models plotted in $P$--$\dot{P}$ space. The black dots with downward arrows are the six yet unidentified LPTs with known $\dot{P}$ upper limits (CHIME J0630+25, GLEAM--X J1627--52, GPM J1839--10, ASKAP J1832--09, ASKAP J1935+21, and ASKAP J1839--08).}
    \label{fig:population_synthesis} 
\end{figure}
\begin{deluxetable*}{c|cc|ccc|c}
\tablewidth{0pt} 
\tablecaption{Fraction of NSs found in different phases from the population synthesis study. A `--' indicates that no simulations were found in that phase, corresponding to a $<0.001$\% likelihood of occurring.} \label{tab:population synthesis}
\tablehead{
\colhead{Propeller}  & \multicolumn{2}{c}{pulsar phase} & \multicolumn{3}{c}{Propeller phase} & \colhead{Accretion phase}
\\
\colhead{model} & \colhead{$<10^1$s} & \colhead{$>10^1$s} & \colhead{$<10^3$s} & \colhead{$10^3-10^4$s} & \colhead{$>10^4$s} & \colhead{}}
\startdata 
{A} & 22\% & 47\% & 32\% & -- & -- & --\\
{B} & 22\% & 47\% & 31\% & 0.010\% & 0.002\% & 0.259\%\\
{C} & 22\% & 47\% & 1\% & 1\% & 1\% & 29\% \\
{D} & 22\% & 47\% & 0.02\% & 0.083\% & 1\% & 31\%\\
{E} & 14\% & 2\% & 59\% & 3\% & 1\% & 21\%\\
{F} & 2\% & 0.016\% & 35\% & 4\% & 2\% & 56\%\\
\enddata
\end{deluxetable*}

\subsection{Simulation results}\label{subsec:population_synthesis_results} 

We present the results of the population simulation as scatter plots in $P$--$\dot{P}$ space (Figure \ref{fig:population_synthesis}) where the different colors indicate NSs in the pulsar-phase (blue), propeller-phase (green), and accretion phase (orange). We also present the fraction of simulated NSs that fall within different regimes of period length in Table \ref{tab:population synthesis}.  
These results allow us to draw definitive conclusions about the effectiveness of each model in spinning down NSs to longer periods. In particular, we evaluate whether each model predicts a sizable population of NSs in the $10^3 < P <10^4$ s range, where most LPTs have been found. Models A and B are highly inefficient, producing no NSs with $P>10^3$ s. In contrast, models C and  D appear to be overly efficient as they create a large number of accretion-phase NSs ($\sim 10^5$ s), but only a small population of NSs in the propeller phase. This suggests that, for these models, the timescale of spinning down the NS from $P \sim 10^2$ s to $P \sim 10^4$ s is extremely short relative to the typical NS age of the population ($\lesssim10^7$ yrs $\ll \tau$), making NSs with spin periods in this range ``short-lived" and rare. Models E and F are most successful in that they generate the largest populations of propeller-phase NSs and have $>1\%$ of their initial NS population fall within the $10^3 < P <10^4$ s range.

More generally, a comparison between models B, E, and F suggests that higher $\gamma$ values lead to stronger propeller spin-down torques and enhance the likelihood of producing NSs with longer periods and evolving more NSs into the accretion phase. On the other hand, when comparing models A, B, C and D to models E and F, we see that models A, B, C, and D possess identical pulsar phase fractions, indicating that it is the $\delta$ parameter that controls the timing for the pulsar-to-propeller transition: larger $\delta$ values allow more NSs to enter the propeller phase during their spin evolution. These observations are consistent with how the $\gamma$ and $\delta$ parameters control the NS spin-down evolution in the propeller equations, shown in equation (\ref{eq:R_m}) and (\ref{eq:R_m_grav}), as well as the qualitative arguments made in \S\ref{subsec:model_comparison}. 

We note the gaps appear between the populations of pulsar-phase and propeller-phase NSs in almost all cases, most prominently in models A, C, and D. These gaps are largely caused by the intrinsic $\dot{P}$ discontinuity at the pulsar-propeller transition as seen in Figures \ref{fig:model_comparison} and \ref{fig:qualpar_study}. The smallest $\dot{P}$ discontinuities occur within propeller models with $\gamma=\delta$, such as B, E, and F. We note that Model $D$ in particular also exhibits a large gap in $P$, however, this is not indicative of a physical discontinuity in its period evolution, but rather, another indication of its overly effective propeller mechanism, such that NSs in this model that enter the propeller phase spin-down rapidly to $P>10^4$.

By examining NSs in the $10^3 < P <10^4$ s range for models E and F, we can infer the typical parameters most likely to produce NSs in this range. The average age for this subset is $\tau\sim 4\times10^{8}$ yrs, the average initial B-field is $B_0\sim 4\times10^{13}$ G, and the average velocity is $v_0 \sim 400$ \kms. Additionally, the mean distance traveled within the time scale to reach these spin periods is $\sim 200$ kpc, indicating that NSs would have migrated far from their birth sites by the time they enter the propeller phase.

\begin{deluxetable*}{c|cc|ccc|c}[h!]
\tablewidth{0pt} 
\tablecaption{Fraction of NSs found in different timeslices during the population synthesis study for model E. A `--' indicates that no simulations were found in that phase, corresponding to a $<0.001$\% likelihood of occurring.} \label{tab:timeslice}
\tablehead{
\colhead{Time (yrs)} & \multicolumn{2}{c}{Pulsar phase} & \multicolumn{3}{c}{Propeller phase} & \colhead{Accretion phase}
\\
\colhead{} & \colhead{$<10^1$s} & \colhead{$>10^1$s} & \colhead{$<10^3$s} & \colhead{$10^3-10^4$s} & \colhead{$>10^4$s} & \colhead{}}
\startdata 
{$10^4$}&  97\% & 3\% & 0.004\% & -- & -- & --\\
{$10^5$}&  94\% & 6\% & 0.245\% & -- & -- & --\\
{$10^6$}&  91\% & 7\% & 2\% & -- & -- & --\\
{$10^7$}&  58\% & 33\% & 9\% & 0.002\% & -- & 0.011\%\\
{$10^8$}&  29\% & 6\% & 64\% & 0.218\% & 0.053\% & 1\%\\
{$10^9$}&  5\% & 0.026\% & 45\% & 2\% & 1\% & 47\%\\
\enddata
\end{deluxetable*}

\begin{figure}
    \centering
    \includegraphics[width=0.49\textwidth]{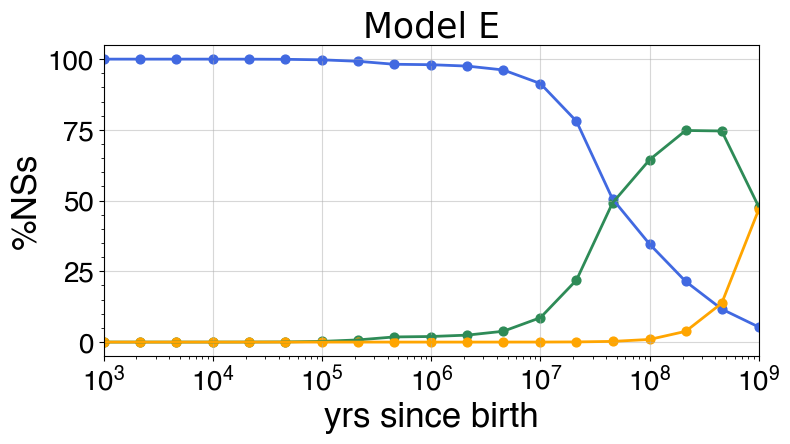}
    \includegraphics[width=0.49\textwidth]{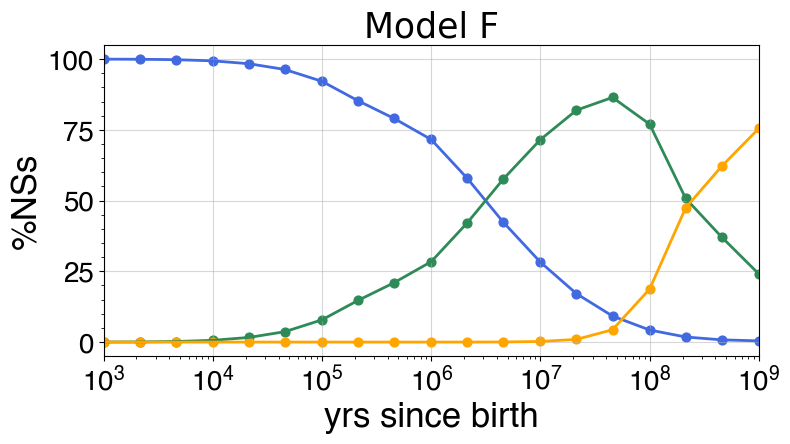}
    \caption{Fraction of simulated NSs in different phases (pulsar phase in blue, propeller phase in green, and accretion phase in orange) at different time-slices for models E and F.}
     \label{fig:timeslice_hist} 
\end{figure}

Aside from analyzing the populations for random NS ages, we also consider the distribution of $P$ and $\dot{P}$ at specific time slices as shown in Appendix \ref{app:timeslice}. Of particular interest is the fraction of NSs within $10^3 < P <10^4$ s for a given time slice. These results are presented for models E and F in Figure \ref{fig:timeslice_hist}. From these results, we see that this fraction peaks at $\sim10^8$ yrs, indicating that NSs reaching LPT-like periods should be quite old. 

\section{Discussion}
\label{sec:discussion}
In this paper, we extensively studied the spin evolution of isolated NSs by considering the propeller effect, which revitalizes significant spin-down torque during the later stages of a NS's lifespan. In each case, we tracked a full spin-down path through the pulsar, propeller, and accretion phases using all known propeller models. In the two subsequent sections, we discuss whether our generic spin-down models can account for the timing properties of the known LPTs and the detectability of old magnetars in the propeller and accretion phases. 

\subsection{Are LPTs magnetars in the propeller phase? } 

Based on the results from our Monte-Carlo simulations in \S\ref{sec:population_synthesis}, the low-efficiency propeller models with $\gamma=\{-1,0\}$ (models A and B) predict no magnetars would spin down to $P>10^3$ s. On the other hand, two models with $\gamma=\{1,2\}$ and $\delta=0$ (models C and D) exhibit large, unphysical jumps in $\dot{P}$ at transitions from the pulsar to propeller phase. In addition, the propeller spin-down is overly effective for model D, and the probability of landing within the observed LPTs' period range of $10^3 \text{ s}<P<10^4 \text{ s}$ is negligible. The most successful models, reproducing the $(P, \dot{P})$ values of LPTs, are models E and F with $\gamma = \delta = \{1,2\}$. These models predict a considerable number of propeller magnetars in the observed LPT range when $\tau \gtrsim 10^8$ yrs and $B \gtrsim 10^{13}$ G, providing a compelling explanation for the origins of these long periods in LPTs.

Overall, our Monte Carlo simulations for models E and F resulted in $\sim$3\% of magnetars within the observed LPT period range. Assuming a birth rate of $\sim1$ NS per century, 
we expect $\sim 3\times10^5$ magnetars in the propeller phase and within the observed LPT period range. This number is significantly larger than the number of known LPTs ($\sim12$). This is not a surprising result, since nearly all LPTs have been found in the Galactic Plane, while magnetars in the propeller phase should have traveled over 200 kpc, and hence, they should be mostly found in the Galactic halo, unless they are gravitationally bound to the Galactic Plane due to smaller velocities. We note that the age and current B-field of a magnetar in the propeller phase cannot be inferred from its known $P$ and $\dot{P}$ values or constraints on $\dot{P}$ using the dipole radiation spin-down formula.

However, these propeller models cannot explain the properties we observe in \textit{all} LPTs. For instance, the radio emission mechanism during the propeller phase is unclear, since persistent pulsar-like radio emission from the polar or outer gaps should be quenched due to the incoming particles during the propeller phase \citep{boldin_popov_2010}. It is still possible that the intermittent radio outbursts observed from LPTs could be produced internally within the NS crust, similar to magnetar flares. In the next subsection, we consider the case of ASKAP J1832--09, as its bright X-ray emission ($L_X \sim 10^{33}$~erg\,s$^{-1}$) during the recent radio outburst favors the NS interpretation \citep{Qu2025}.

\subsubsection{Can the magnetar spin-down model account for the case of ASKAP J1832--09?} \label{sec:ASKAP}

\begin{figure} 
    \centering
    \includegraphics[width=0.92\textwidth]{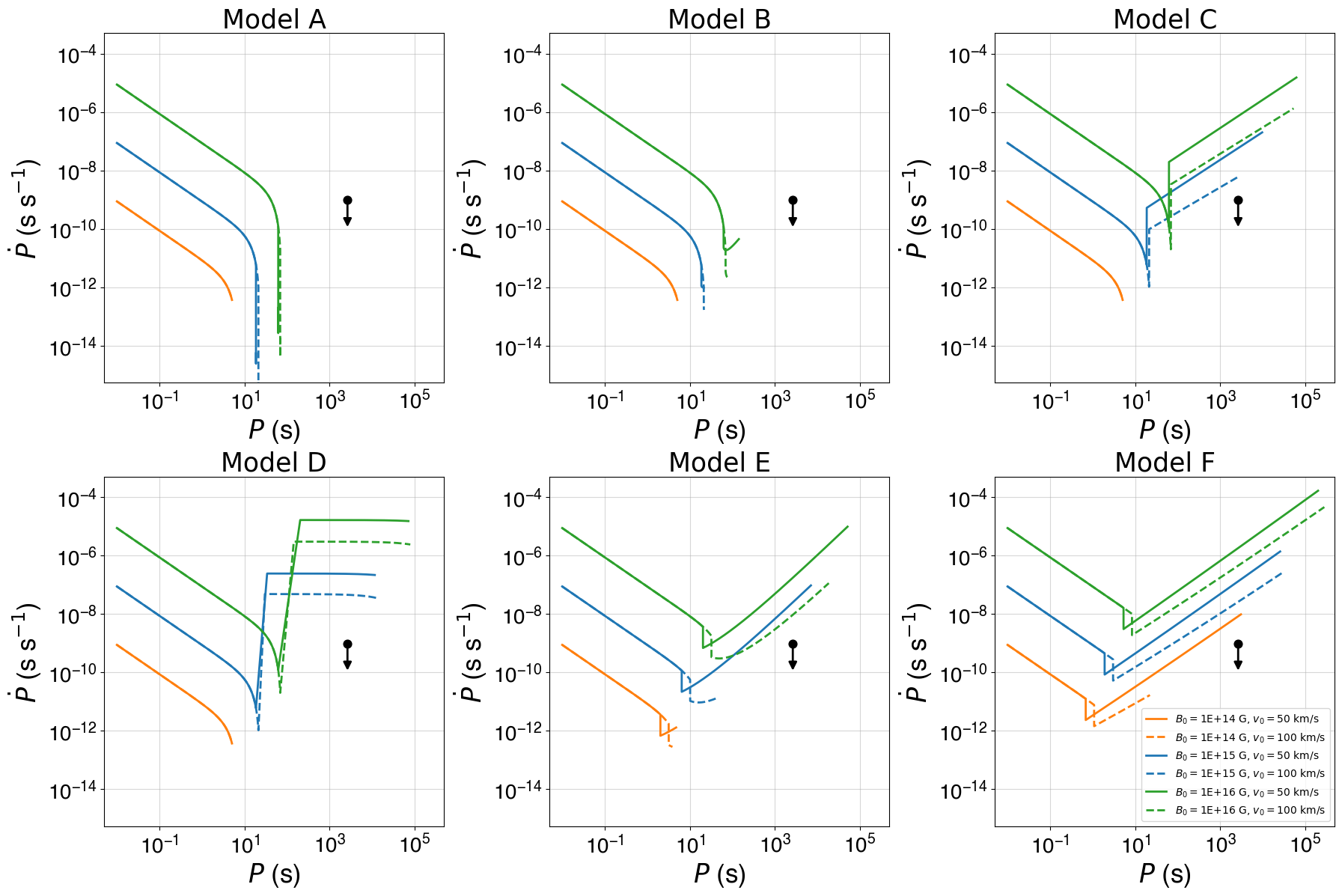}
    \caption{Simulation results for the various propeller models utilizing parameters $\tau = 10^5$ yrs and $n_0 = 680$ cm$^{-3}$ which is representative of a NS interpretation for the ASKAP J1832--09 moving through the molecular cloud GMC 22.6--0.2. The black dot with the downward arrow is the observed $P$ and $\dot{P}$ for ASKAP J1832--09.}
     \label{fig:ASKAP1832}
\end{figure}

Among the known LPTs listed in Table \ref{tab:LPT list}, ASKAP J1832--09 is a peculiar case for its X-ray detection during the last radio outburst in 2024. The short-lived X-ray emission could be akin to those of outbursting magnetars with strong NS magnetic fields ($B \simgt 10^{14}$ G). This implies that ASKAP J1832--09 may be young ($\tau_{\text{NS}}\lesssim 10^5$ yrs), assuming the magnetic field decay is not impeded. The younger age of ASKAP J1832--09 is also supported if ASKAP J1832--09 is associated with a nearby supernova remnant (G22.7$-$0.2) with its estimated age $\sim 10^5$ yrs. 

However, based on the results in Figure \ref{fig:timeslice_hist} and Table \ref{tab:timeslice}, spinning down ASKAP J1832--09 to its current 44-min period within $\sim10^5$ yrs is nearly impossible through a combination of dipole radiation and propeller spin-down mechanisms under the general assumptions of \ref{subsec:initial_population}. Thus, if it is indeed a $\sim10^5$-year-old magnetar, ASKAP J1832--09 should have an extremely high B-field (with a slow magnetic field decay time) and/or spun down in dense molecular clouds. Using the radio position of ASKAP J1832--09 \citep{Wang2024} and assuming that it was born at the center of the SNR, we derived a projection velocity of 50 \kms. Hence, the 3D velocity could reach $\sim 100$ \kms. Furthermore, based on this assumed trajectory of ASKAP J1832--09 with respect to the surrounding molecular clouds, we find that ASKAP J1832--09 is likely to have spent most of its lifetime travelling through the molecular cloud GMC 22.6$-$0.2, whose hydrogen density is 680 cm$^{-3}$ \citep{Su_2014}. Assuming these parameters ($v_0 = 50\rm{-}100$ \kms\ and $n_0 = 680$~cm$^{-3}$), we performed the magnetar spin-down simulations for all the propeller models listed in Table \ref{tab:propmodels}. We presented different spin-down evolution paths of ASKAP J1832--09 on the $P$--$\dot{P}$ diagram in Figure \ref{fig:ASKAP1832}.

From these results, spinning down ASKAP J1832--09 to its current period within $10^5$ yrs is only possible when we assumed (1) high magnetic fields $B_0>10^{15}$ G, (2) slow NS velocities $v_0\sim 50$ \kms, (3) slow magnetic field decay (such as a suppressed Hall effect decay), and/or (4) the most efficient propeller models (D and F) with $\gamma = 2$. However, for all these cases, the predicted $\dot{P}$ values at present are well above the $\dot{P}$ upper limit for ASKAP J1832--09. The only way to suppress the present spin-down rate is to assume a two-phase evolution with ASKAP J1832--09 traveling through the dense molecular cloud GMC 22.6$-$0.2 until recently, when it resides within the ISM. Though the projected position is within GMC 22.6$-$0.2, it is possible that a magnetar moving with $v_0\sim 50$ \kms\ has escaped from the molecular cloud in the line of sight direction. Alternatively, ASKAP J1832--09 may also be a much older magnetar not associated with the SNR.  

\subsection{The detectability of old magnetars in the propeller and accretion phases}

There are seven nearby radio-quiet NSs (RQNSs) or X-ray dim neutron stars (XDINSs) within a few hundred parsecs \citep{Popov2023}. These so-called magnificent seven NSs have been identified by detecting thermal X-ray emission with $kT \simlt 100$ eV. They all exhibit long spin periods of $P_{\rm spin} \sim 3\rm{-}10$ sec, and the $\dot{P}$ measurements suggest dipole B-fields of $B \sim 10^{13}\rm{-}10^{14}$~G and estimated ages of $\tau \sim 10^{5-6}$~yrs. The closest RQNS (RX J1856.5$-$3754) was likely formed in a nearby OB association as inferred from its observed proper motion. These RQNSs are believed to represent an evolved stage of magnetars after their magnetic fields decayed \citep{Pons2007}. By taking into account misaligned radio beams and faint thermal X-ray emission, the number of RQNS in the galaxy should be significantly larger than the seven currently known objects. 
The two plausible spin-down models (E and F) predict that magnetars should still be in the pulsar phase or potentially close to transitioning to the propeller phase at $\tau\sim10^{5-6}$ yrs. Thus, our models indicate that RQNSs will eventually enter the propeller phase within $\tau \simlt 10^7$ yrs. By that stage, while older magnetars would have traveled further over $d\simgt 1$~kpc, their surface temperature may be kept hot at $kT\sim 0.1$~keV due to on-going magnetic field decay \citep{Pons2007}. 

As predicted by \citet{rutledge_2001} and our spin-down model, propelling magnetars will eventually accrete from the ISM and emit soft thermal X-rays. As seen in figure \ref{fig:population_synthesis}, these accretion-phase NSs must be spinning extremely slowly ($\Omega\sim 10^{-4}$ s$^{-1}$) such that their behavior should asymptotically approach models of non-spinning accreting NSs. Thus, we follow \citet{rutledge_2001} to estimate the X-ray luminosity and surface temperature from the poles during the accretion phase. For the NSs we are considering, which move through the ISM at some velocity, the accretion rate is essentially the density of the surrounding material multiplied by the velocity of the NS multiplied by the cross-sectional area of the magnetosphere that captures the material for accretion:
\begin{equation}
    \dot{M} = \pi R_m^2 v_m \rho_m
\end{equation}
It is important to note that, if we allow $v_m$ and $\rho_m$ to take on their gravitational corrected definitions from \S\ref{subsec:grav_correction}, in the regime where $v_0$ is low compared to the Keplerian velocity at $R_m$, the above accretion rate will be equivalent to that for Bondi accretion -- the gravitational corrections allow us to naturally capture both accretion modes. We can derive the X-ray luminosity $L_X$ due to this accretion by calculating the energy release of infalling gas to the NS surface with an accretion rate of $\dot{M}$, which is given by 
\begin{equation} \label{eq:L_X}
    L_X = (1.8\times10^{28})\Biggl(\frac{\mu}{10^{33}\text{G cm$^3$}}\Biggr)^{2/3} \Biggl(\frac{v_m}{100 \text{\kms}}\Biggr)^{1/3} \Biggl(\frac{n_m}{0.1\text{ cm}^{-3}}\Biggr)^{2/3} \Biggl(\frac{M}{1.4 M_{\odot}}\Biggr) \Biggl(\frac{R}{10 \rm{km}}\Biggr)^{-1} \text{ ergs/s}. 
\end{equation}
Finally, we can also obtain the effective temperature of the X-ray emissions by utilizing the Stefan-Boltzmann law and assuming that the emission area is a small portion of the polar cap
\begin{equation}
        kT_{\text{eff}} = 0.39\Biggl(\frac{\mu}{10^{33}\text{G cm$^3$}}\Biggr)^{1/4} \Biggl(\frac{n_m}{0.1\text{ cm}^{-3}}\Biggr)^{1/8}\Biggl(\frac{M}{1.4 M_{\odot}}\Biggr)^{1/4} \Biggl(\frac{R}{10 \rm{km}}\Biggr)^{-1} \text{ keV}.
\end{equation}
By performing calculations for model E, we find that the mean X-ray luminosity from the simulations is $1.1\times 10^{28}$ ergs/s and the effective temperature is 0.2 keV, suggesting that accreting magnetars could be detected in the soft X-ray band. However, one should note that equation (\ref{eq:L_X}) assumes a 100\% efficiency of converting gravitational potential energy of accreting gas into X-ray emissions. In reality, this is not the case, and we expect lower X-ray luminosities and temperatures than our estimates.

Based on the above estimates, the upcoming UVEX observatory and the proposed AXIS probe mission may be able to detect thermal emission from the surface of old magnetars either in the propeller or accretion phase ($\tau\sim10^7$ yrs), provided that these evolved magnetars are nearby and neutral hydrogen absorption is not significant ($N_{\rm H} \simlt 10^{20}$~cm$^{-2}$). To robustly estimate the population size and UV detectability of these old magnetars, a detailed Monte Carlo simulation study, as in \citet{boldin_popov_2010}, is necessary, accounting for the distribution of their birth sites in our galaxy.

\section{Conclusion} 

LPTs are a new class of Galactic radio sources characterized by long periods ($P > 10^3$ s) and highly variable radio emission. The number of LPTs is expected to increase rapidly over the next few years due to ongoing extensive searches in the radio band. We explored one of the two proposed scenarios for LPTs involving old magnetars that spin down via propeller interactions with the ISM or molecular clouds. Our fully parameterized magnetar spin-down model shows that, in general, all magnetars are expected to enter the propeller phase, with higher initial B-fields and higher ambient densities favoring an earlier transition. Monte-Carlo simulations demonstrated that two particular propeller models (models E and F) are consistent with most of the $P$ and $\dot{P}$ distribution of LPTs. Furthermore, these models exhibit small discontinuities in $\dot{P}$ during the propeller transition, making them more physically plausible \citep{mori_2003}, and model F is supported by MHD simulations of the propeller effect \citep{modelF_romanova_2003}. These simulations also suggest that, if LPTs are indeed propeller magnetars, many of them should be as old as $\tau \sim 10^8$ yrs and may be found in the Galactic halo, thus it is possible that subsequent searches will reveal new LPTs located away from the Galactic Plane. However, the spin-down process is sped up if magnetars move through denser regions, so we may find more LPTs, like ASKAP J1832--09, which are located near or within molecular clouds. In addition, we suspect that some of the fast radio bursts (FRBs) could be old magnetars in the propeller phase originating from our galaxy. Further MHD simulations, such as those in \cite{modelF_romanova_2003}, will be important for pinpointing the most physically accurate propeller model, thereby allowing us to better calculate the propeller spin-down torque and refine our predictions. 

Furthermore, we can improve the simulations from section \ref{sec:population_synthesis} by incorporating more realistic galactic birth and kick velocity distributions of NSs, the galactic gravitational potential, and the galactic ISM density distribution. A more sophisticated study of the galactic magnetar population, similar to \cite{Sautron_2025}, would allow us to better understand the dynamics of propeller-phase magnetars and yield concrete predictions on their number density within our detectable neighborhood. Additional observational constraints, such as the non-detection of LPTs in $10 \text{ s}<P<50 \text{ s}$ \citep{Sherman_2025}, could help us to constrain these propeller models. In the future, we may be able to identify magnetars in the propeller phase by detecting irregular spin-down rates or timing noise from known or new LPTs. The search for new LPTs, the continued monitoring of known LPTs, and the capability for multi-wavelength follow-ups of LPT outbursts are all crucial efforts to understand their emission mechanisms, spin evolution, and populations. The recent discovery of an X-ray counterpart to ASKAP J1832--09 \citep{44min_wang_rea_2024} and multi-wavelength follow-up observations of CHIME J1634+44 \citep{Lundy2025} mark the beginning of our explorations of LPTs.

\section{Acknowledgments}
This work was partially supported by NASA NuSTAR AO-11 grant NNH24ZDA001N-NUSTAR. 
This paper is dedicated to Prof. Malvin (Mal) Ruderman, who first proposed the idea of spinning down magnetars through propeller effects. The foundations of our understanding of numerous neutron star phenomena owe much to his extensive contributions and pioneering theoretical work since the 1960s. We acknowledge discussions with the CHIME team at McGill University and Kyung Duk Yoon for his early contribution to this project. We thank Nanda Rea, Yuri Levin, and Ali Alpar for useful discussions and suggestions.

\bibliography{Magnetar}{}
\bibliographystyle{aasjournal}

\appendix
\section{Qualitative Parameter Space Study for all Models}\label{app:qual_par_study}
\begin{figure}[h!]
    \centering
    \includegraphics[width=0.95\textwidth]{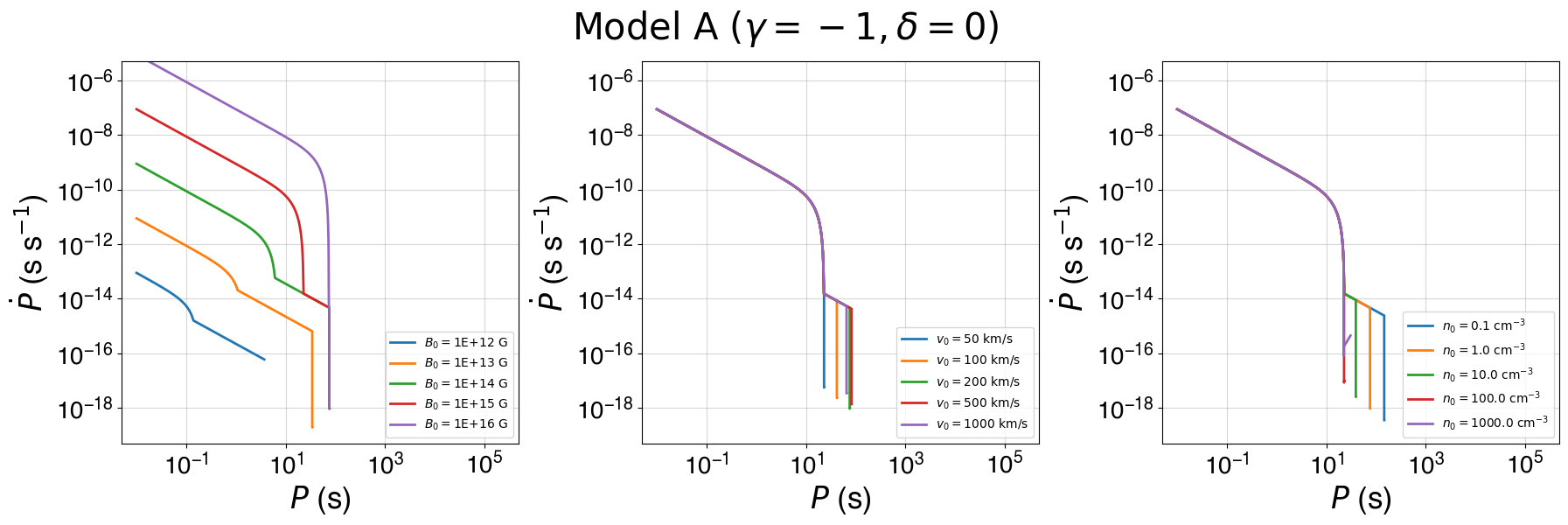}\\
    \includegraphics[width=0.95\textwidth]{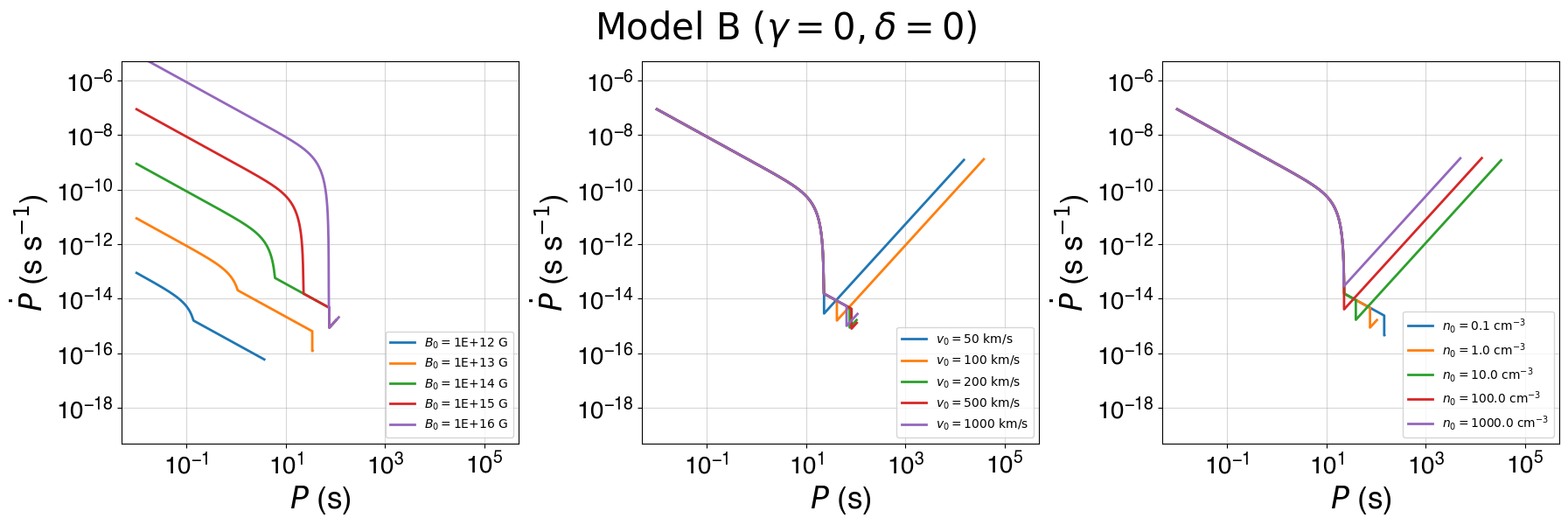}\\
    \includegraphics[width=0.95\textwidth]{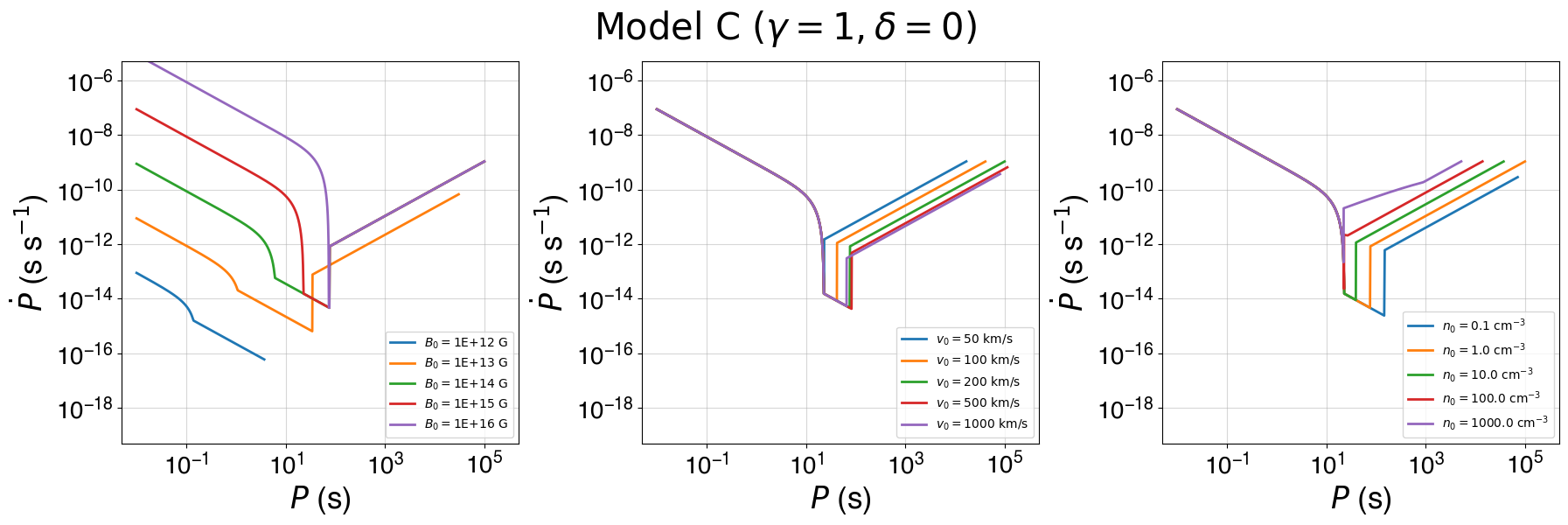}
    
\end{figure}
\begin{figure}[h!]
    \centering
    \includegraphics[width=0.95\textwidth]{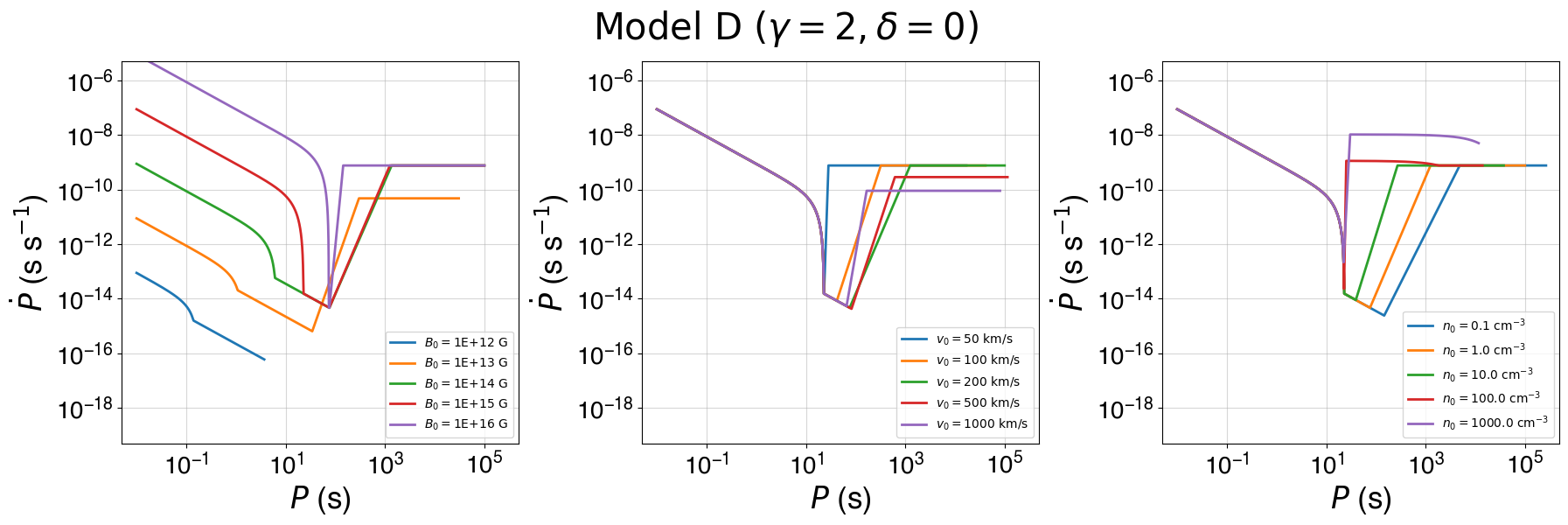}\\
    \includegraphics[width=0.95\textwidth]{model_E_parameter_study.png}\\
    \includegraphics[width=0.95\textwidth]{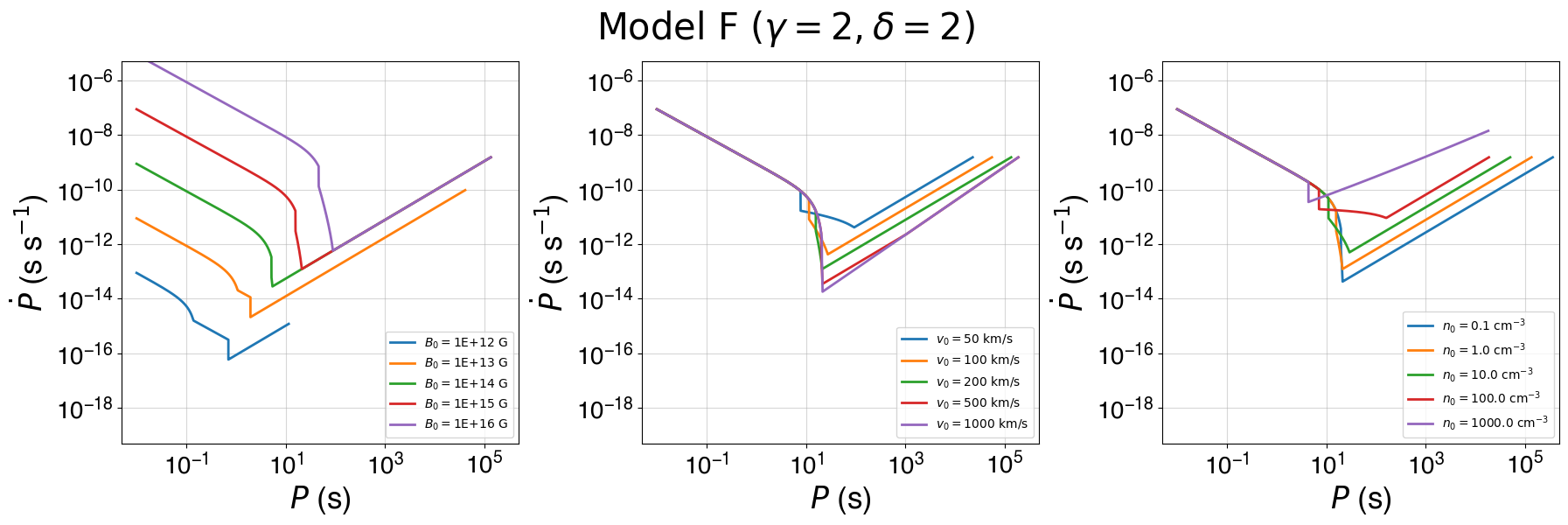}  
    \caption{A qualitative study of the parameter space for all the propeller model E, which traces the evolution of an NS in the $P-\dot{P}$ space. The leftmost plot varies the initial B-field $B_0$, the center plot varies the translational velocity of the NS $v_0$, and the right column varies the density of the surrounding material $\rho_0$. The remaining parameters, which are not being varied, are fixed to $B_0=10^{15}$ G, $v_m = 200$ \kms and $n_0$ = 1~cm$^{-3}$.}
\end{figure} 
\clearpage
s
\section{Time-slice Plots}\label{app:timeslice}
\subsection{Model A}
\begin{figure}[h!]
    \centering
    \includegraphics[width=0.32\textwidth]{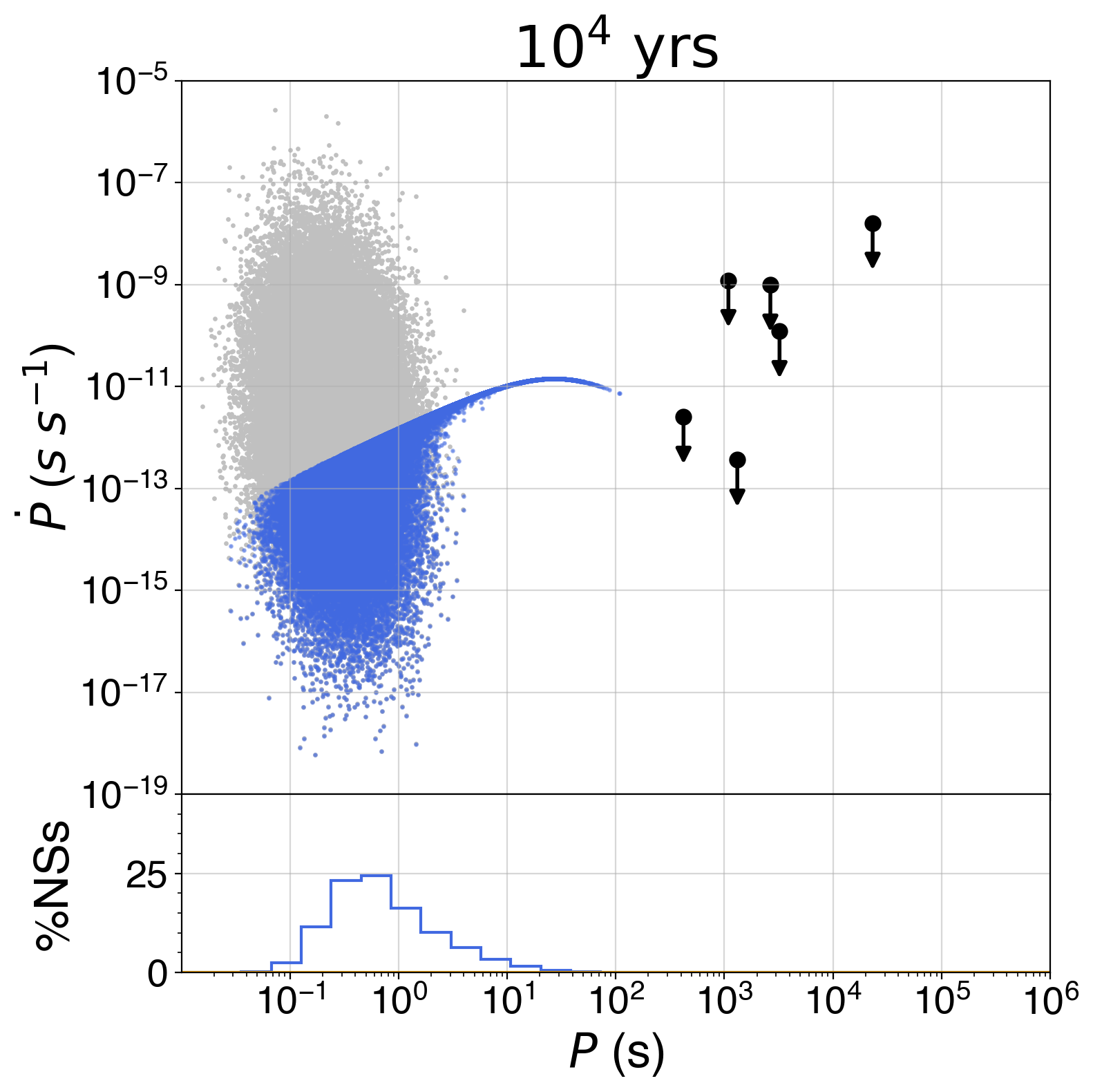}
    \includegraphics[width=0.32\textwidth]{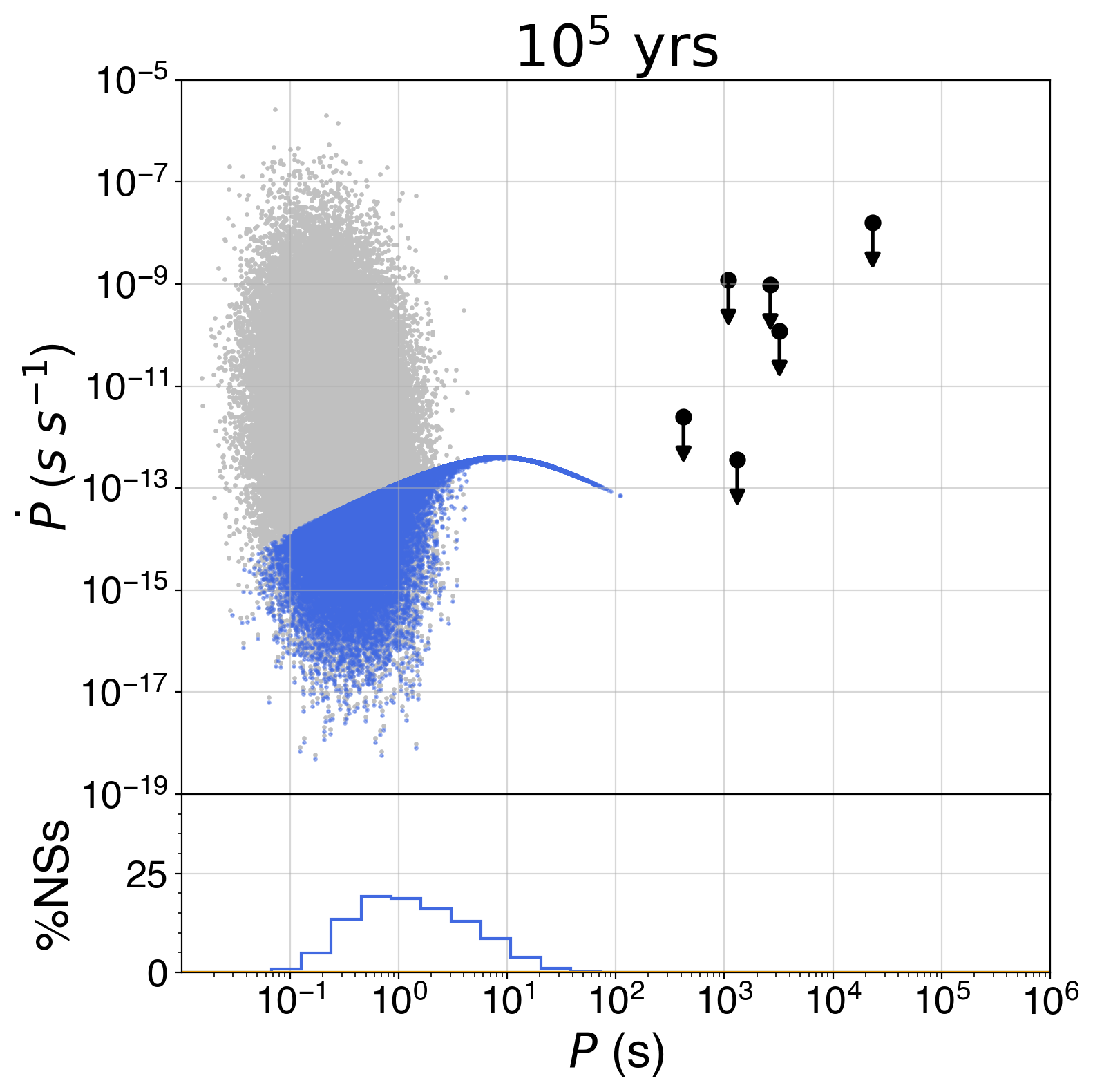}
    \includegraphics[width=0.32\textwidth]{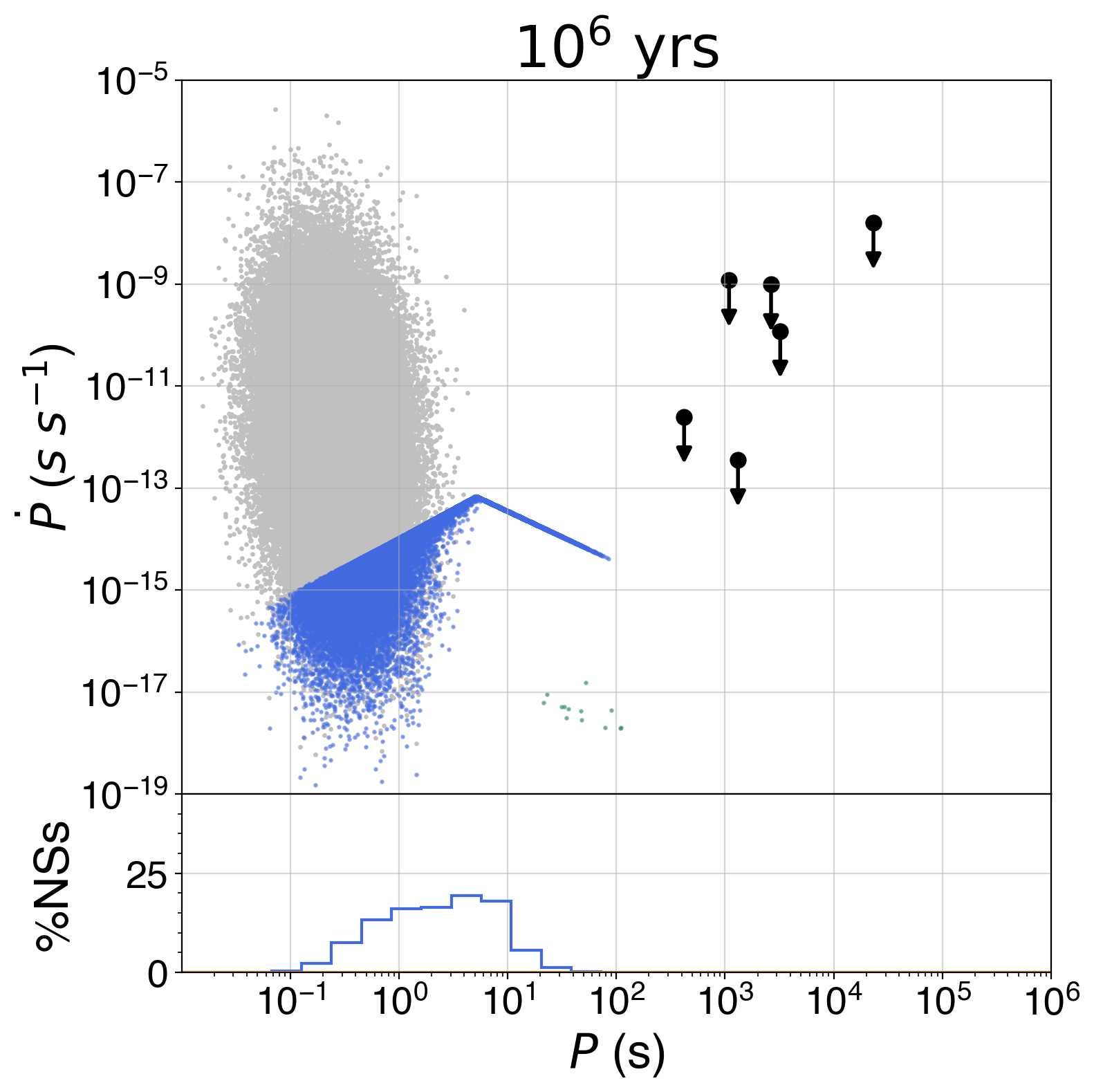}\\
    \includegraphics[width=0.32\textwidth]{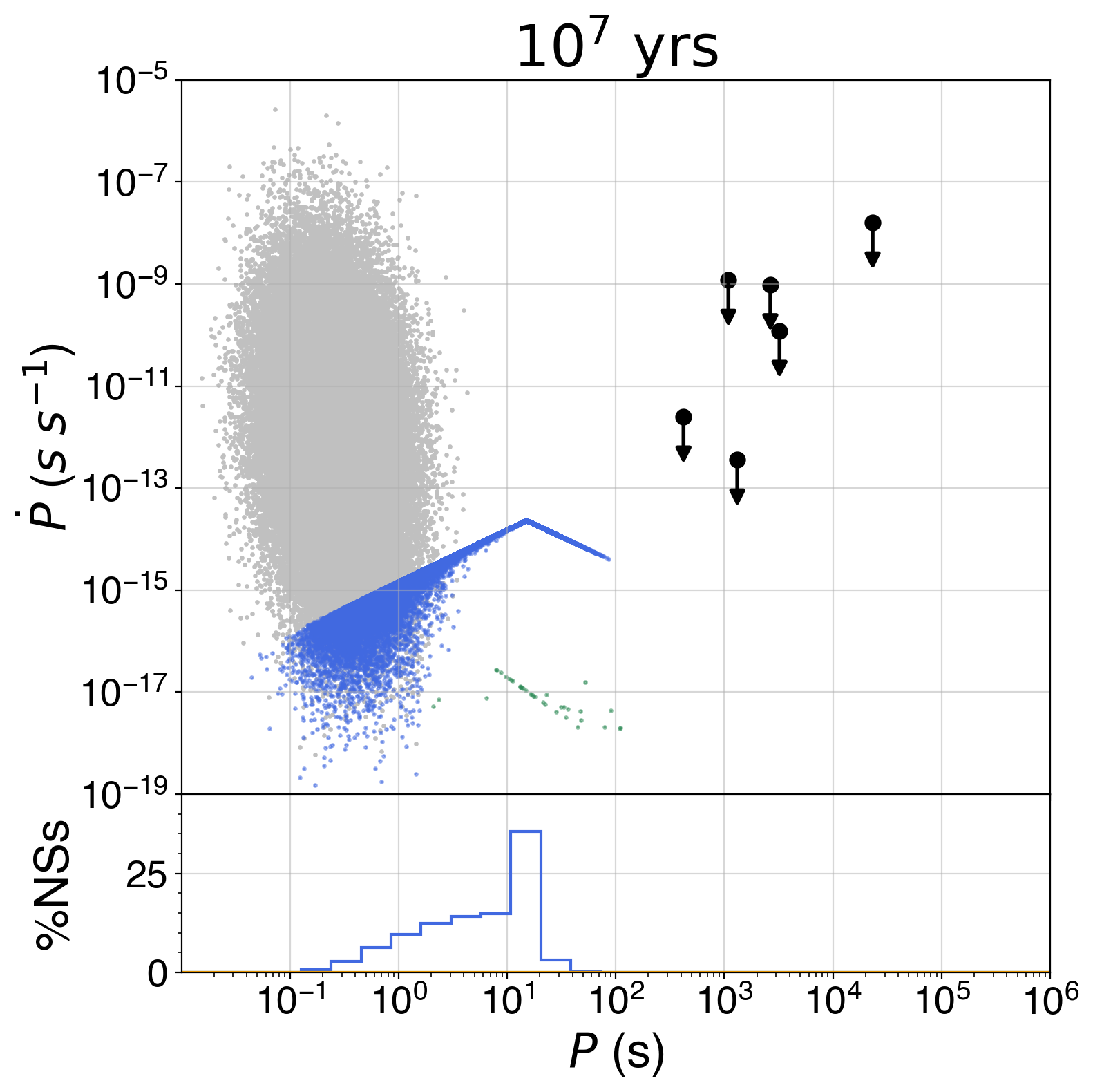}
    \includegraphics[width=0.32\textwidth]{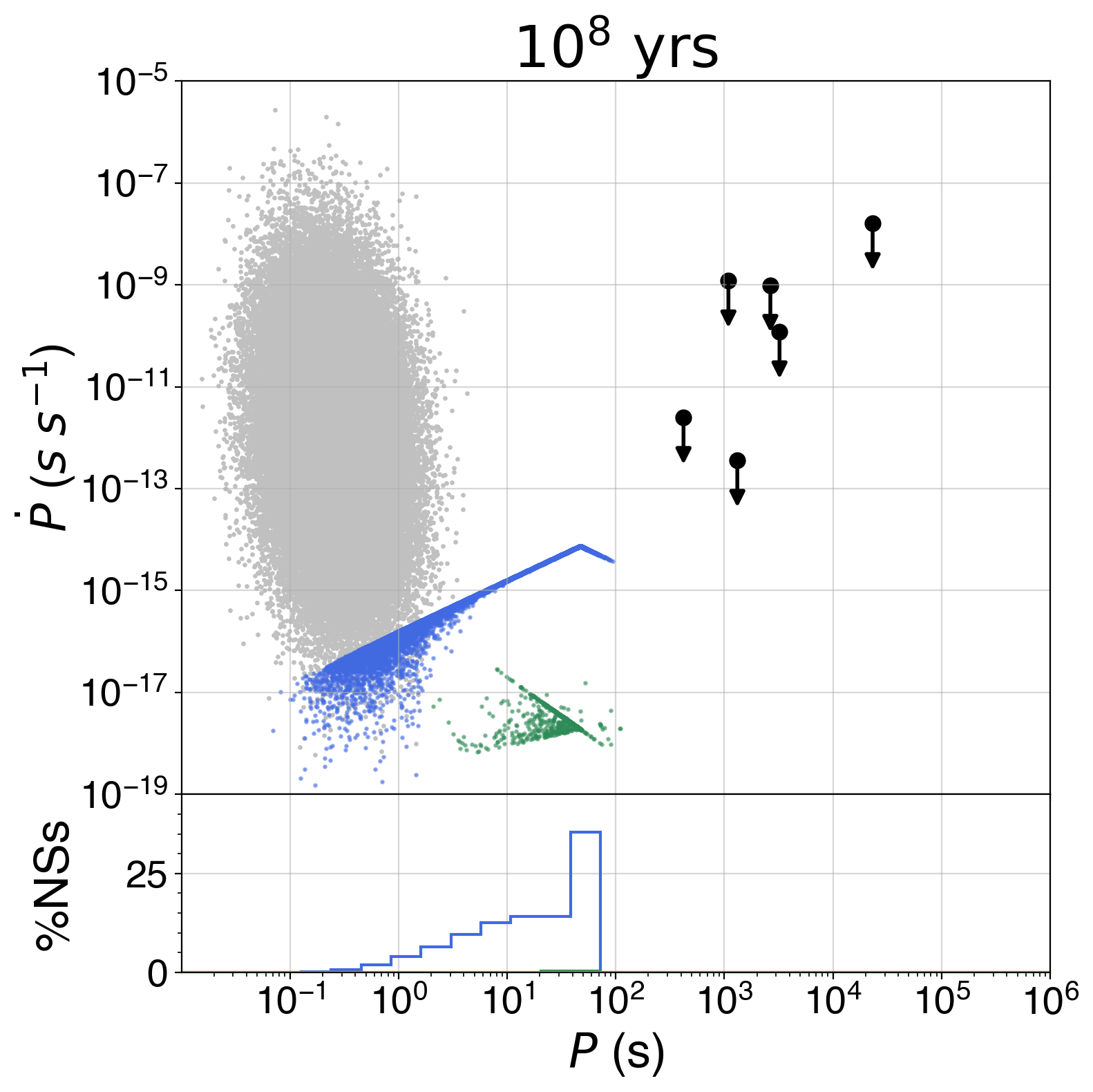}
    \includegraphics[width=0.32\textwidth]{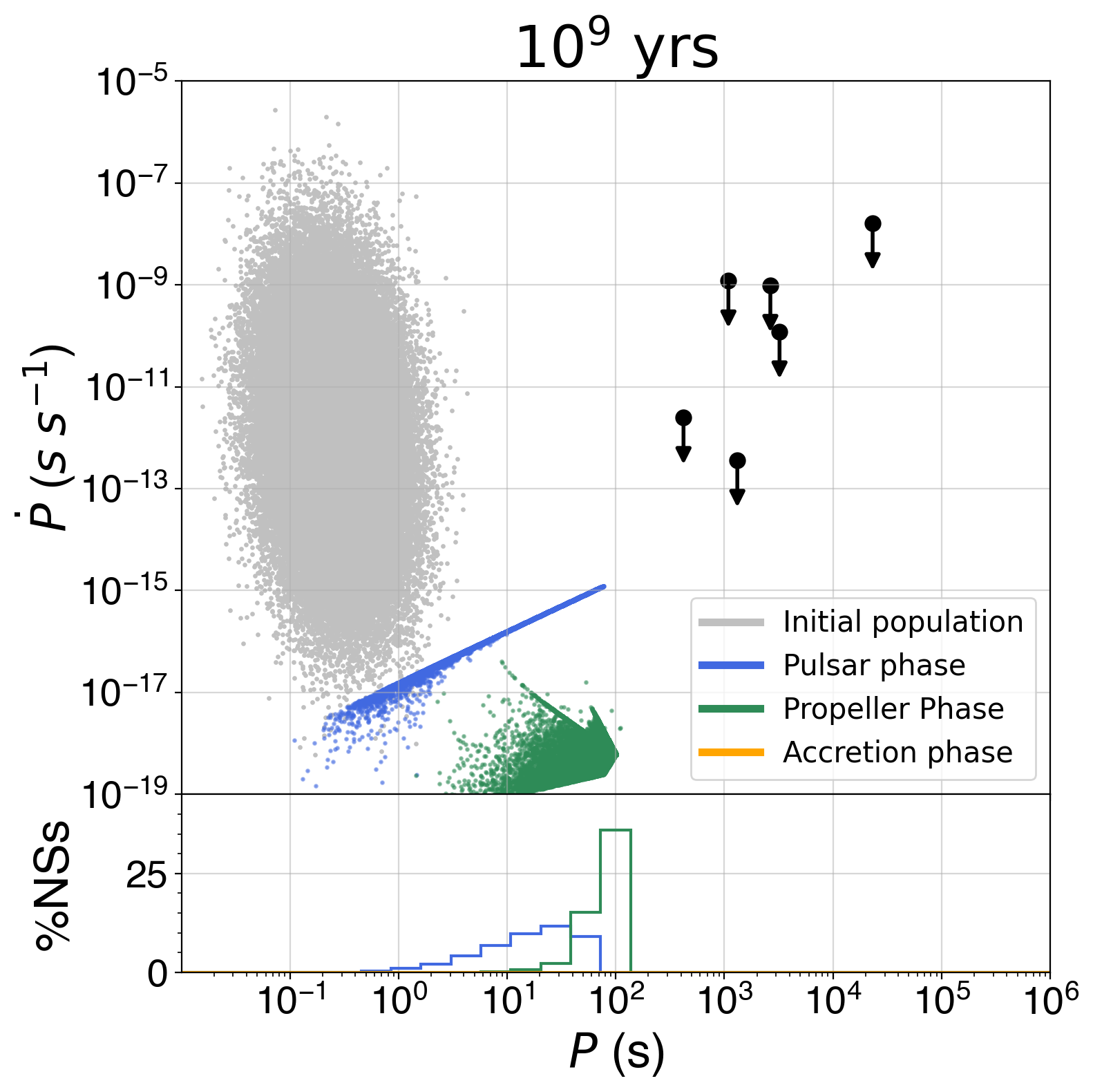}
    \caption{Population snapshots at various timeslices for for model A ($\gamma=-1$, $\delta=0$) plotted in $P$--$\dot{P}$ space.}
\end{figure}
\begin{deluxetable*}{c|cc|ccc|c}[h!]
\tablewidth{0pt} 
\tablecaption{Fraction of NSs found in different timeslices during the population synthesis study for model A. A `--' indicates that no simulations were found in that phase, corresponding to a $<0.001$\% likelihood of occuring.} \label{tab:time-slice_A}
\tablehead{
\colhead{Time (yrs)} & \multicolumn{2}{c}{Pulsar phase} & \multicolumn{3}{c}{Propeller phase} & \colhead{Accretion phase}
\\
\colhead{} & \colhead{$<10^1$s} & \colhead{$>10^1$s} & \colhead{$<10^3$s} & \colhead{$10^3-10^4$s} & \colhead{$>10^4$s} & \colhead{}}
\startdata 
{$10^4$}&  97\% & 3\% & -- & -- & -- & --\\
{$10^5$}&  94\% & 6\% & -- & -- & -- & --\\
{$10^6$}&  92\% & 8\% & 0.013\% & -- & -- & --\\
{$10^7$}&  59\% & 41\% & 0.026\% & -- & -- & --\\
{$10^8$}&  34\% & 65\% & 1\% & -- & -- & --\\
{$10^3$}&  14\% & 32\% & 55\% & -- & -- & --\\
\enddata
\end{deluxetable*}

\subsection{Model B}
\begin{figure} [h!]
    \centering
    \includegraphics[width=0.32\textwidth]{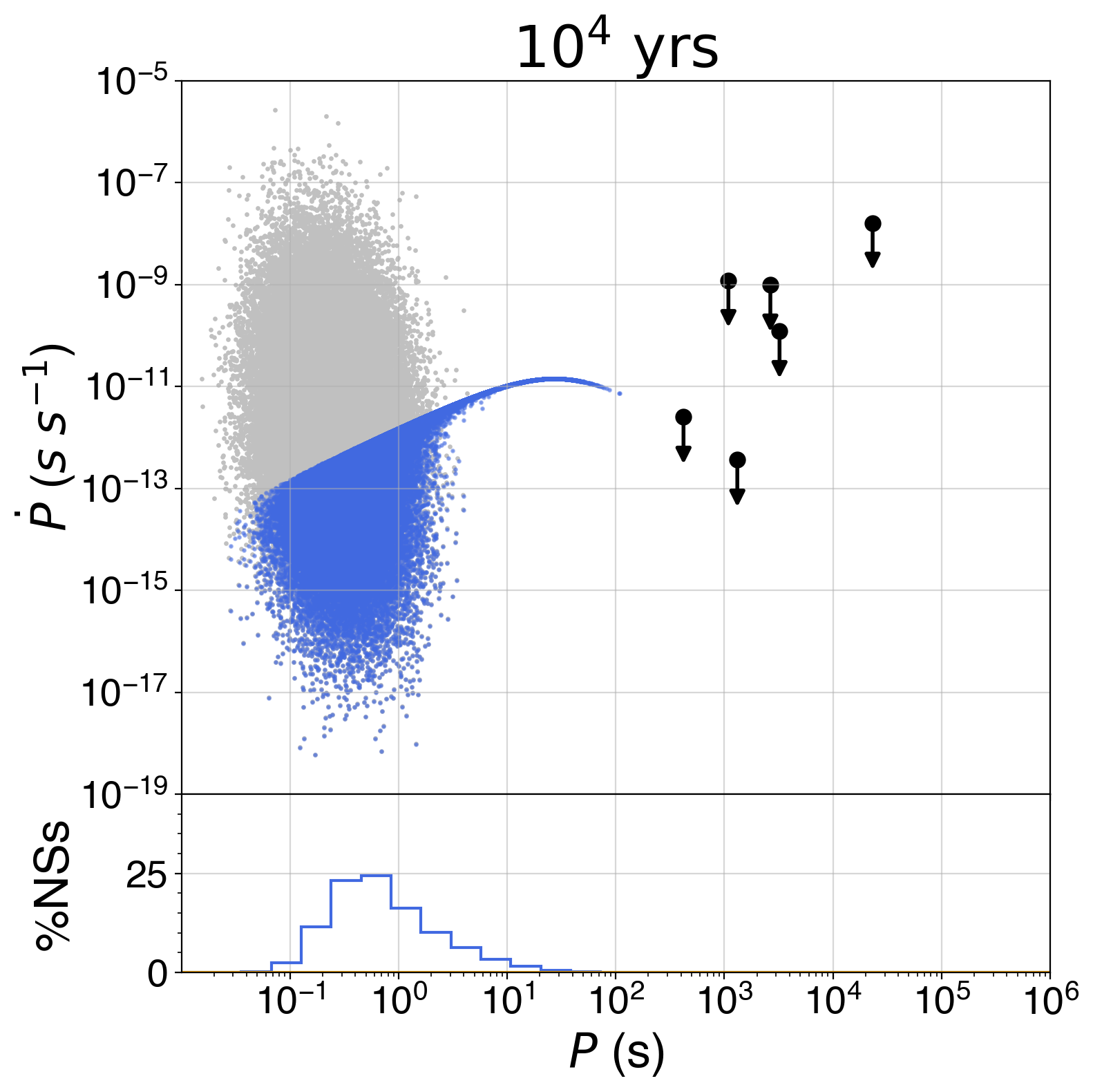}
    \includegraphics[width=0.32\textwidth]{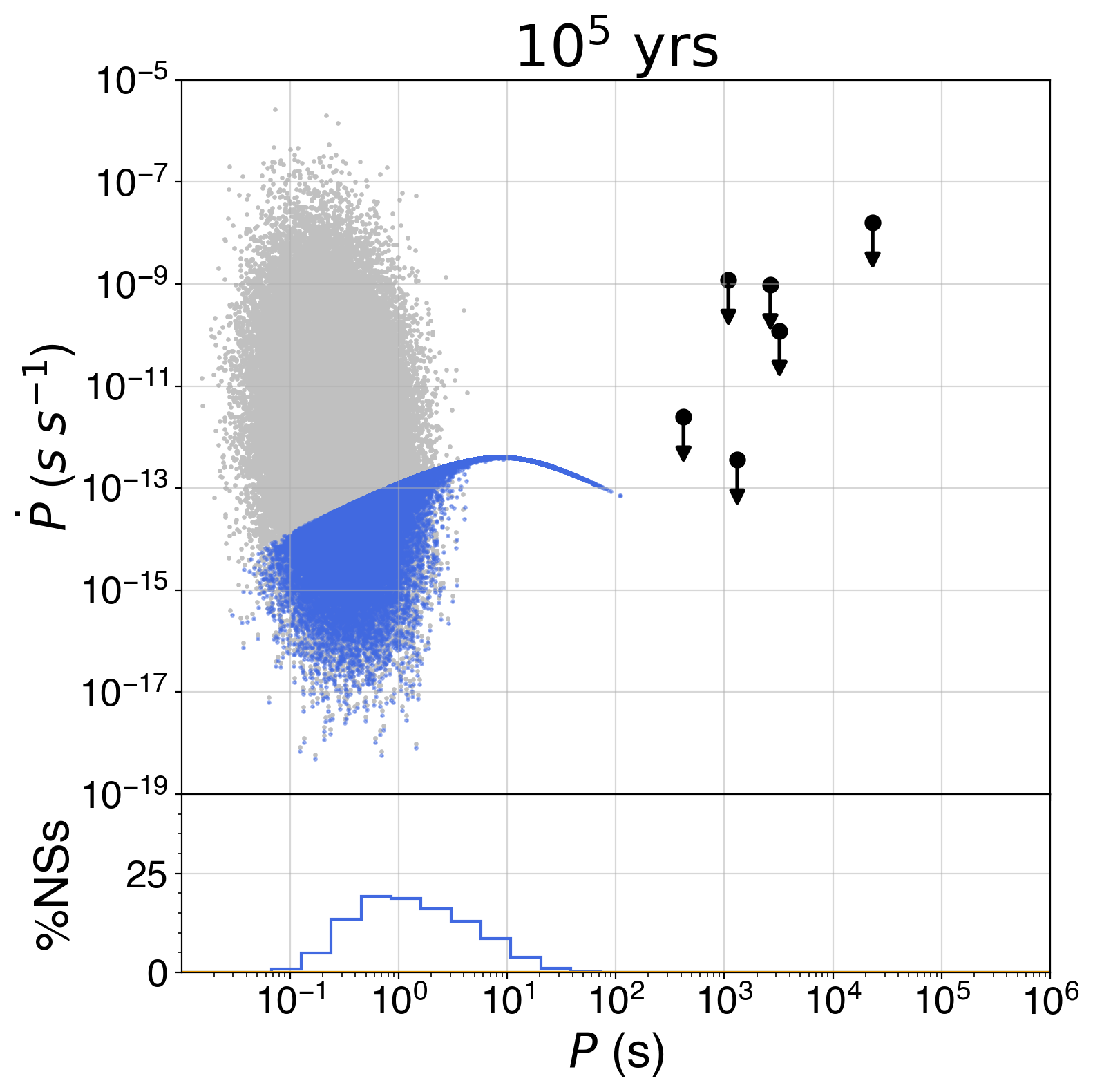}
    \includegraphics[width=0.32\textwidth]{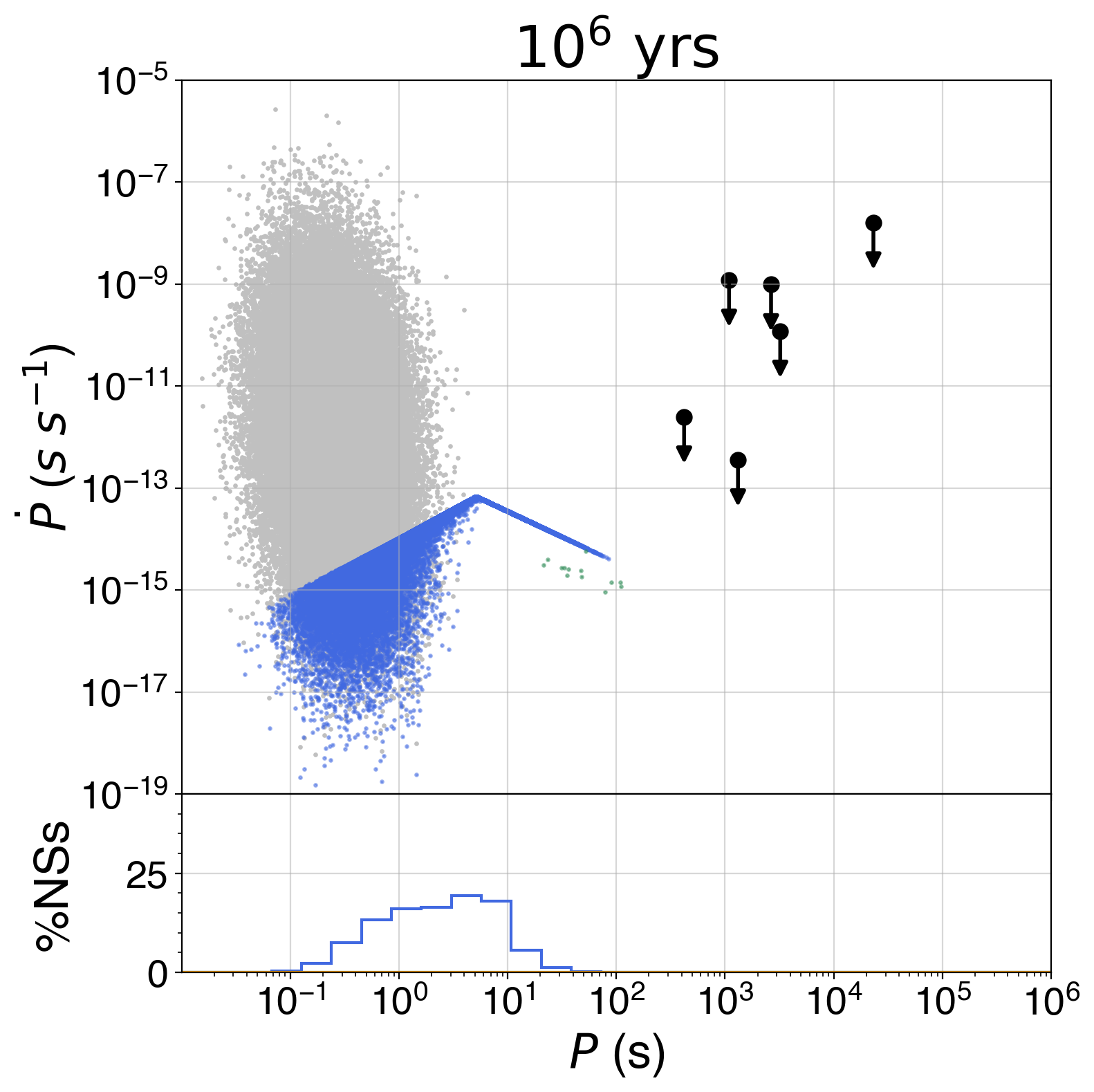}\\
    \includegraphics[width=0.32\textwidth]{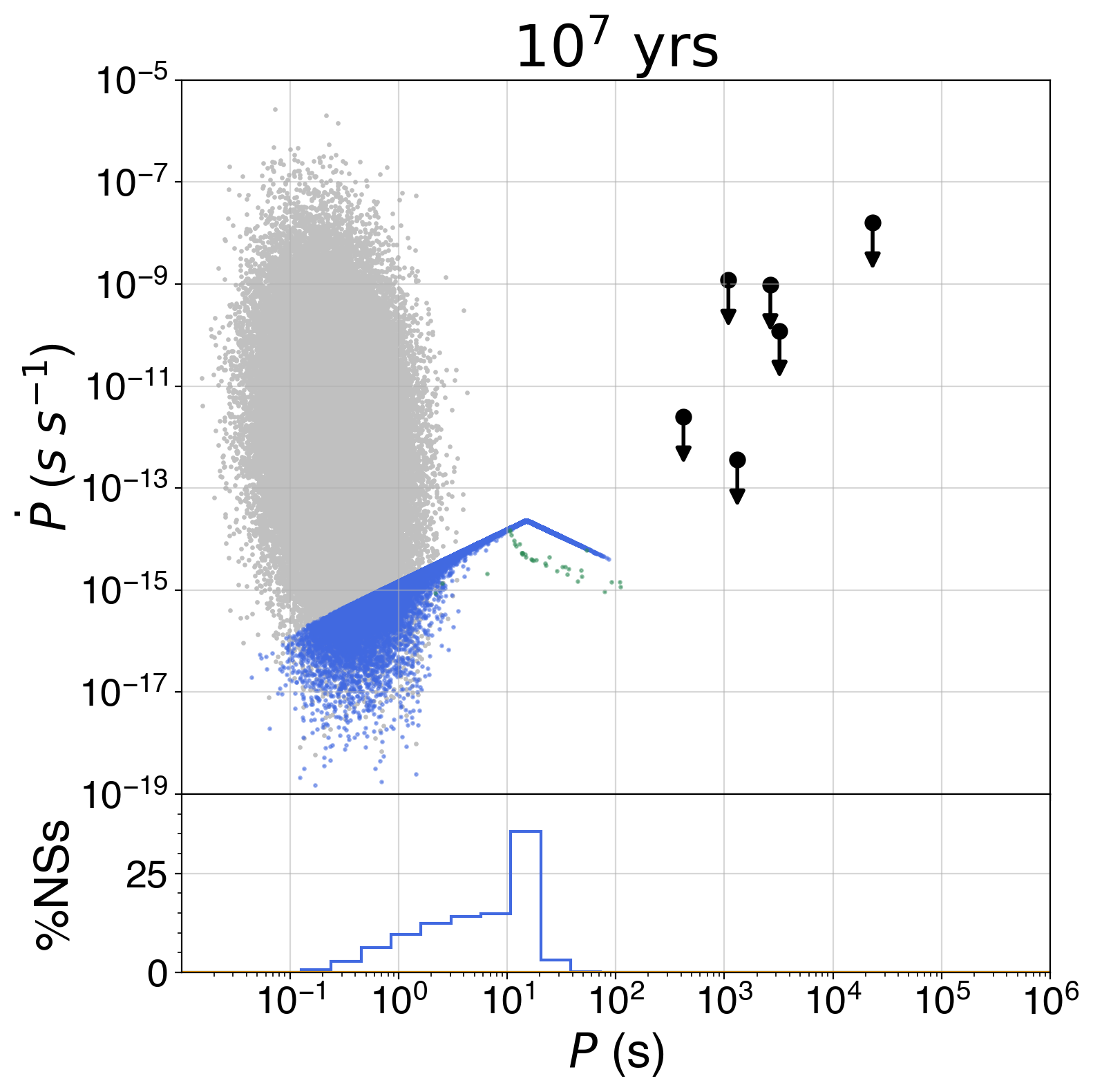}
    \includegraphics[width=0.32\textwidth]{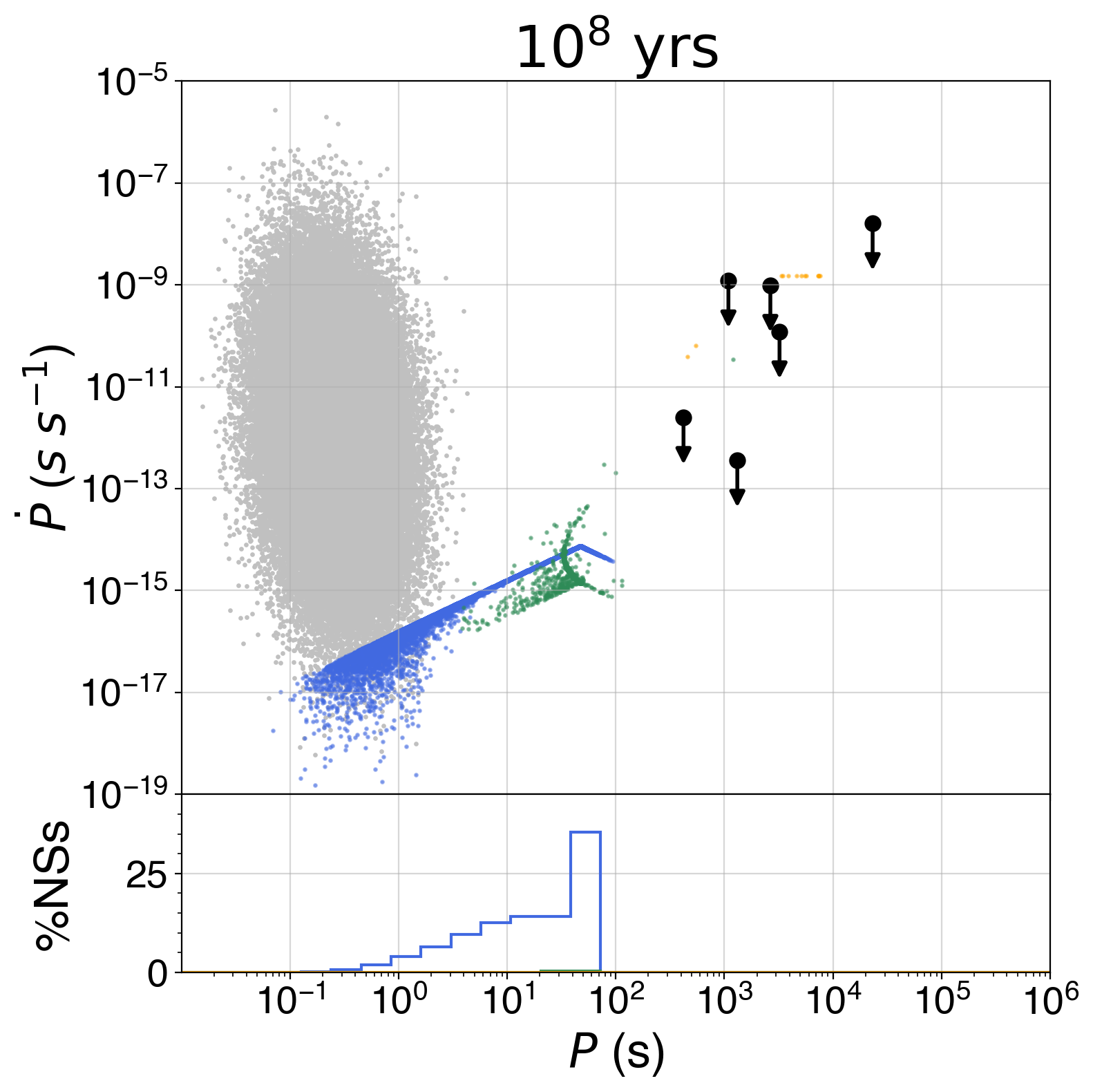}
    \includegraphics[width=0.32\textwidth]{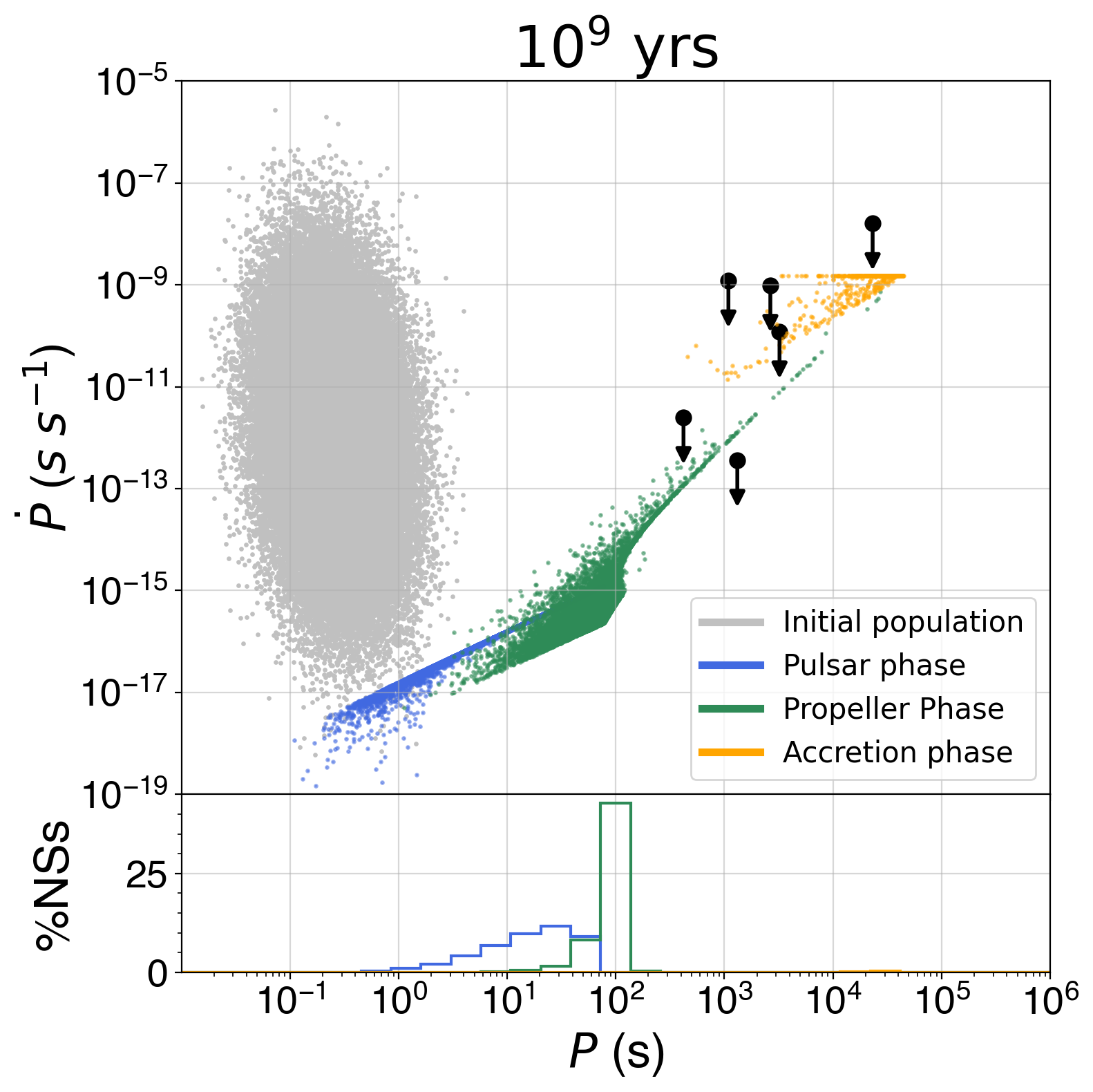}
    \caption{Population snapshots at various timeslices for for model B ($\gamma=0$, $\delta=0$) plotted in $P$--$\dot{P}$ space.}
\end{figure}
\begin{deluxetable*}{c|cc|ccc|c}[h!]
\tablewidth{0pt} 
\tablecaption{Fraction of NSs found in different timeslices during the population synthesis study for model B. A `--' indicates that no simulations were found in that phase, corresponding to a $<0.001$\% likelihood of occuring.} \label{tab:time-slice_B}
\tablehead{
\colhead{Time (yrs)} & \multicolumn{2}{c}{Pulsar phase} & \multicolumn{3}{c}{Propeller phase} & \colhead{Accretion phase}
\\
\colhead{} & \colhead{$<10^1$s} & \colhead{$>10^1$s} & \colhead{$<10^3$s} & \colhead{$10^3-10^4$s} & \colhead{$>10^4$s} & \colhead{}}
\startdata 
{$10^4$}&  97\% & 3\% & -- & -- & -- & --\\
{$10^5$}&  94\% & 6\% & -- & -- & -- & --\\
{$10^6$}&  92\% & 8\% & 0.013\% & -- & -- & --\\
{$10^7$}&  59\% & 41\% & 0.036\% & -- & -- & --\\
{$10^8$}&  34\% & 65\% & 1\% & 0.001\% & -- & 0.013\%\\
{$10^3$}&  14\% & 32\% & 54\% & 0.04\% & 0.004\% & 1\%\\
\enddata
\end{deluxetable*}
\clearpage
\pagebreak
\subsection{Model C}
\begin{figure} [h!]
    \centering
    \includegraphics[width=0.32\textwidth]{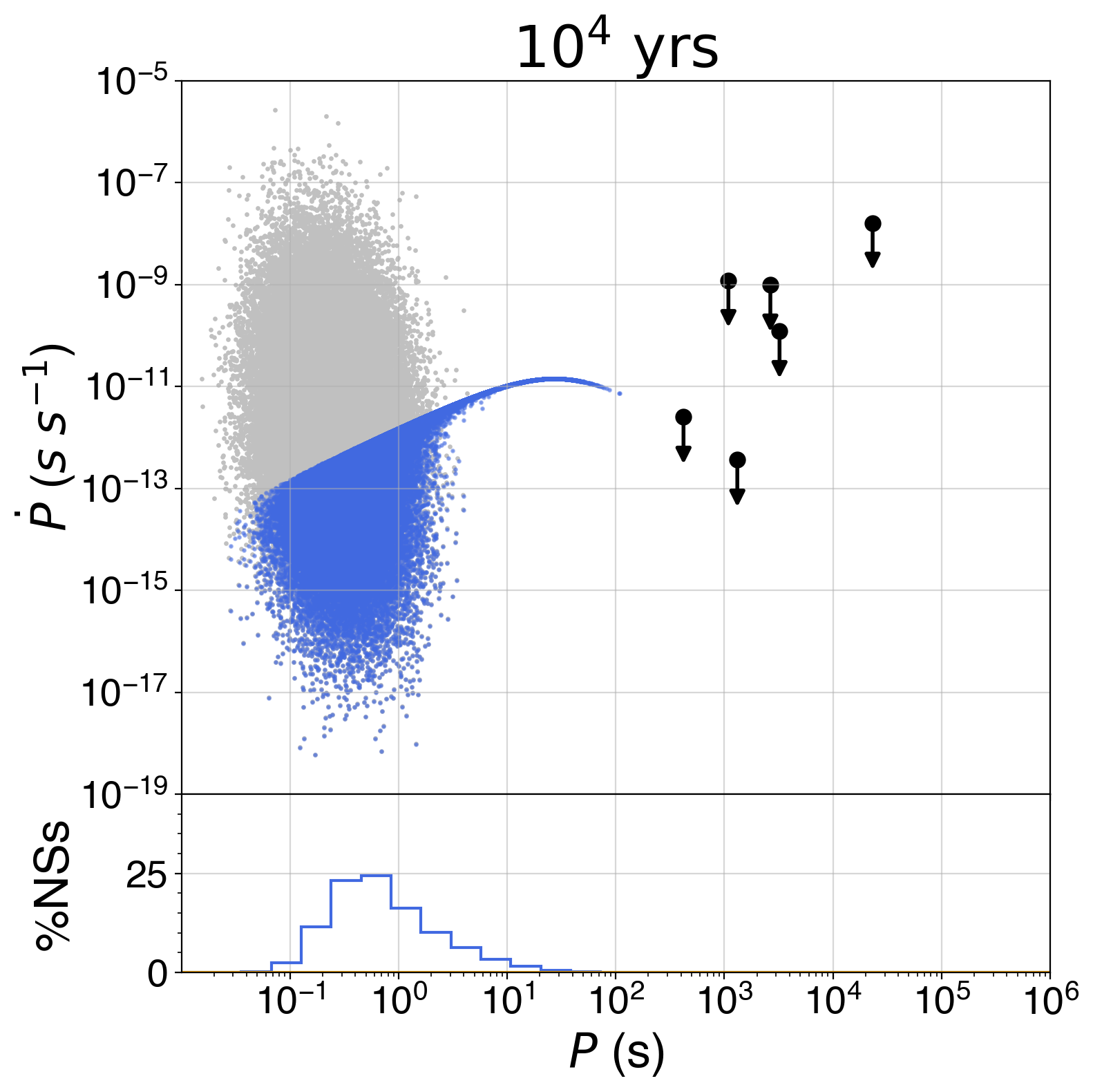}
    \includegraphics[width=0.32\textwidth]{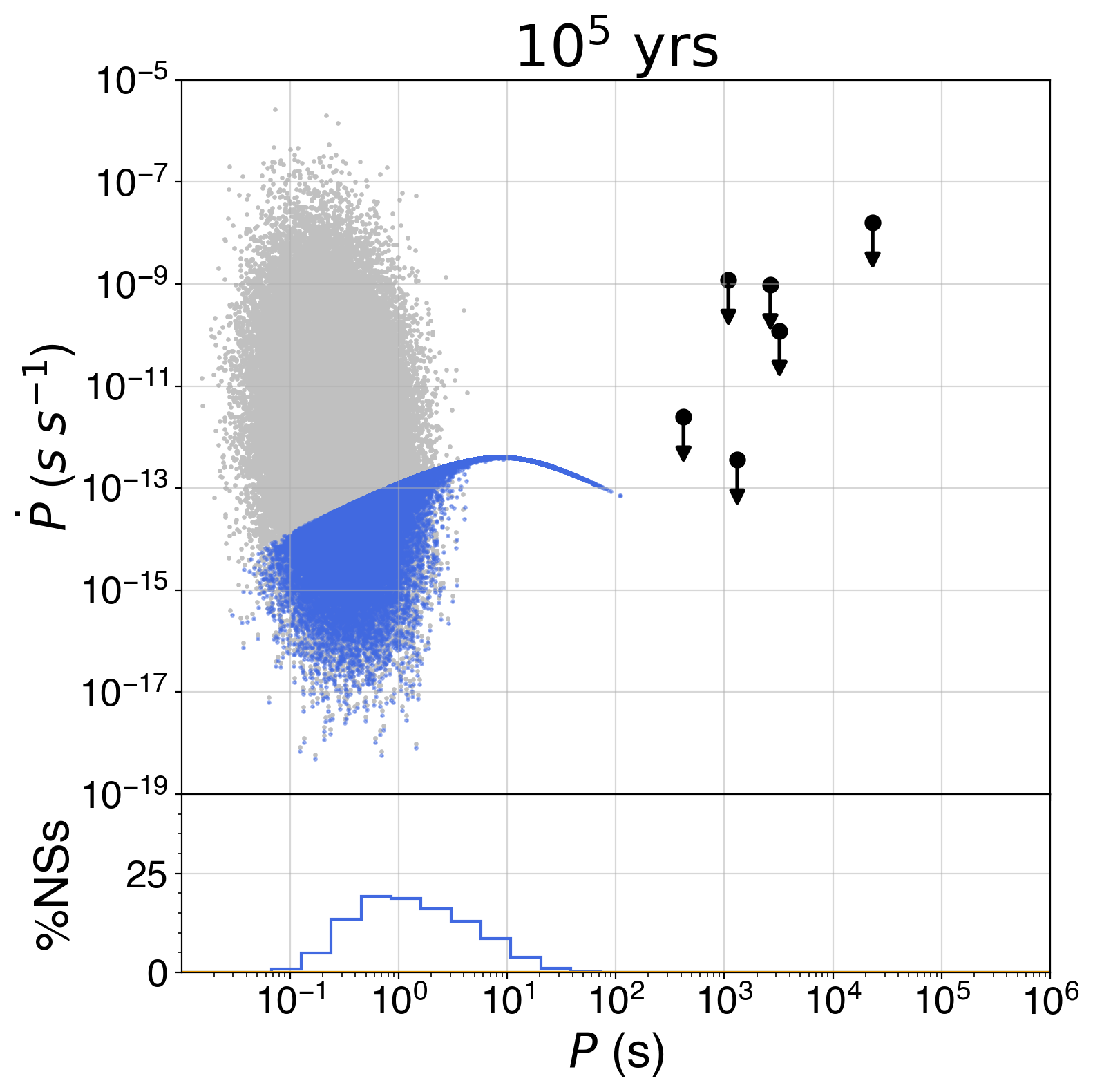}
    \includegraphics[width=0.32\textwidth]{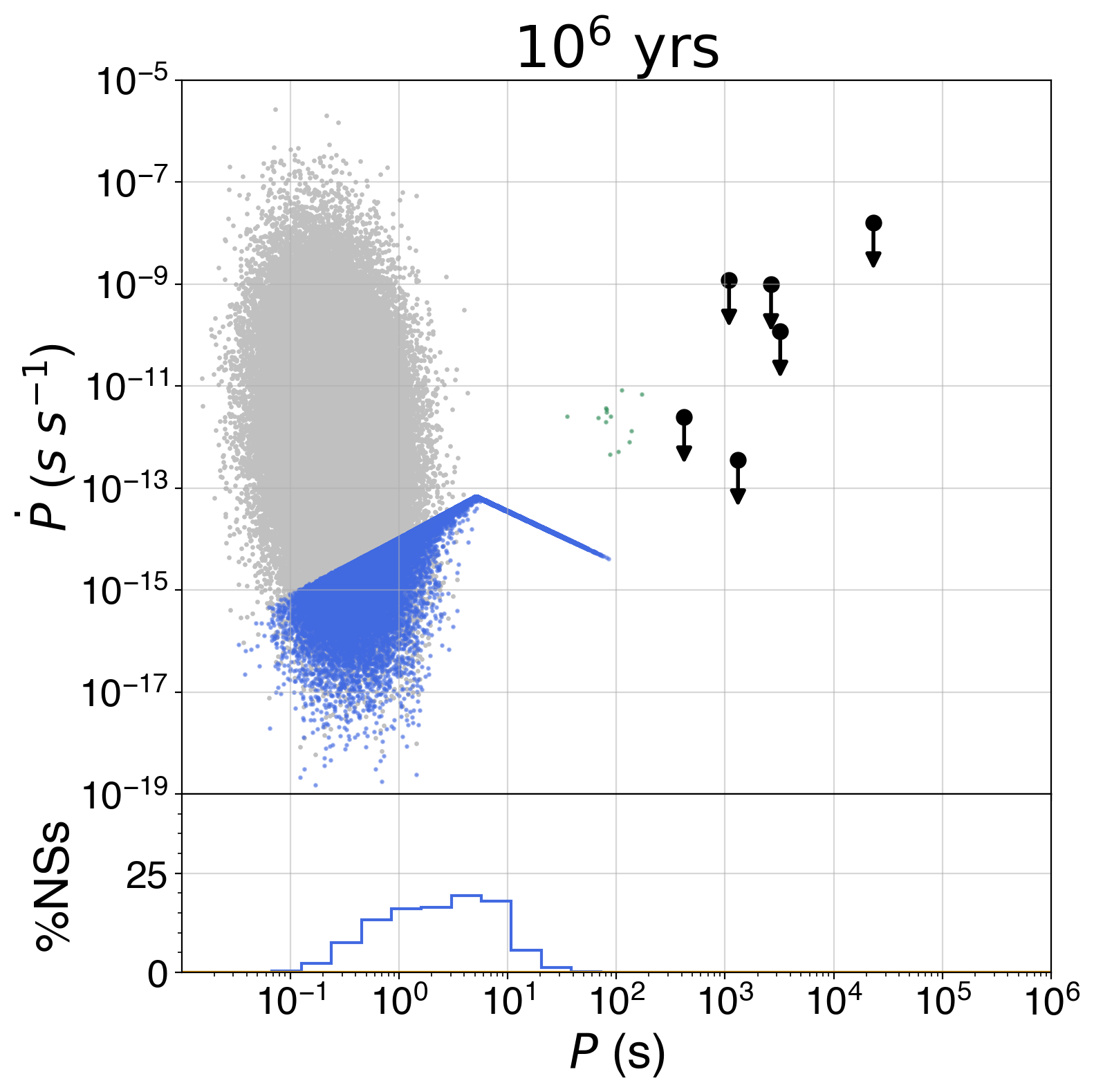}\\
    \includegraphics[width=0.32\textwidth]{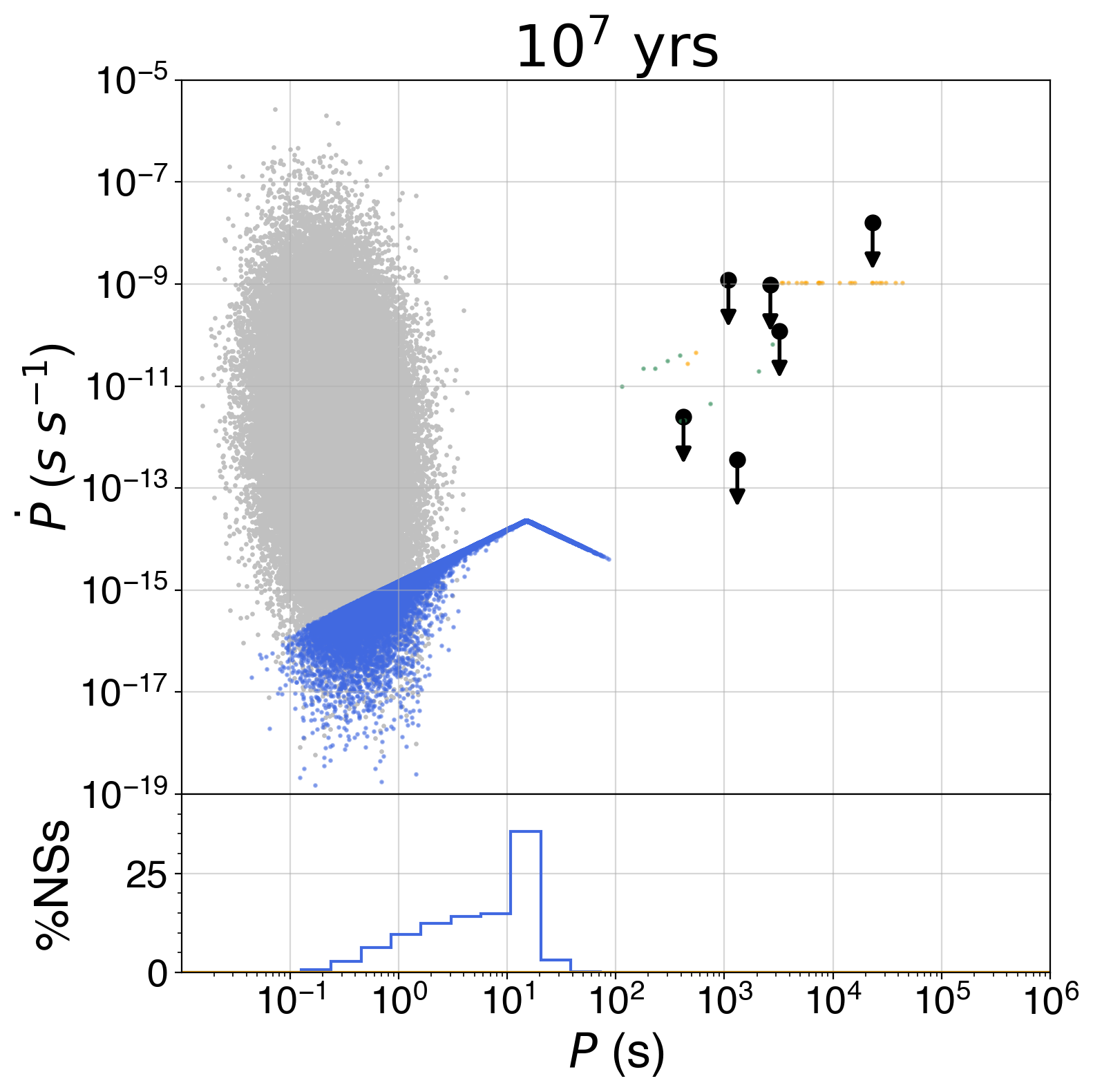}
    \includegraphics[width=0.32\textwidth]{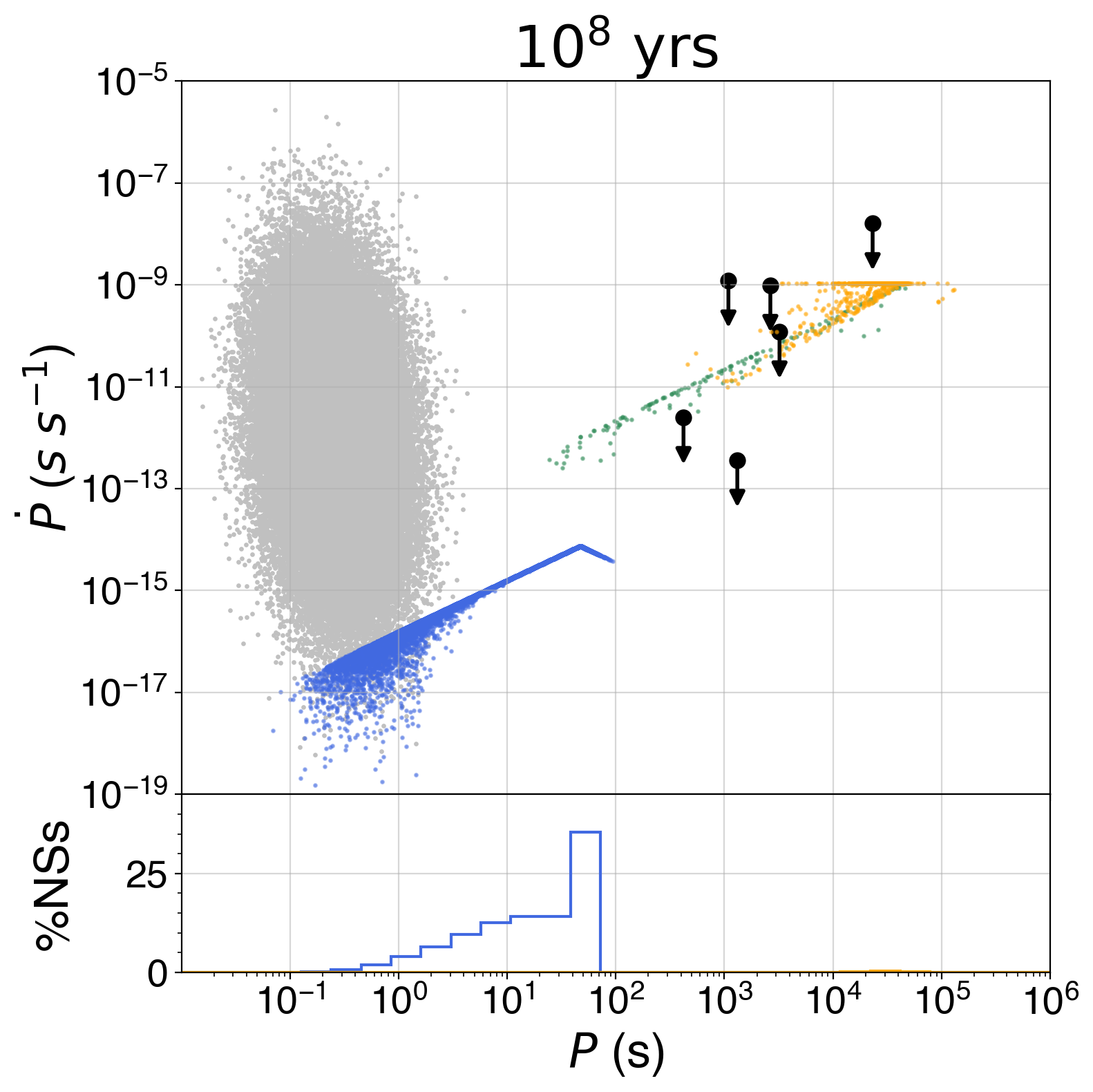}
    \includegraphics[width=0.32\textwidth]{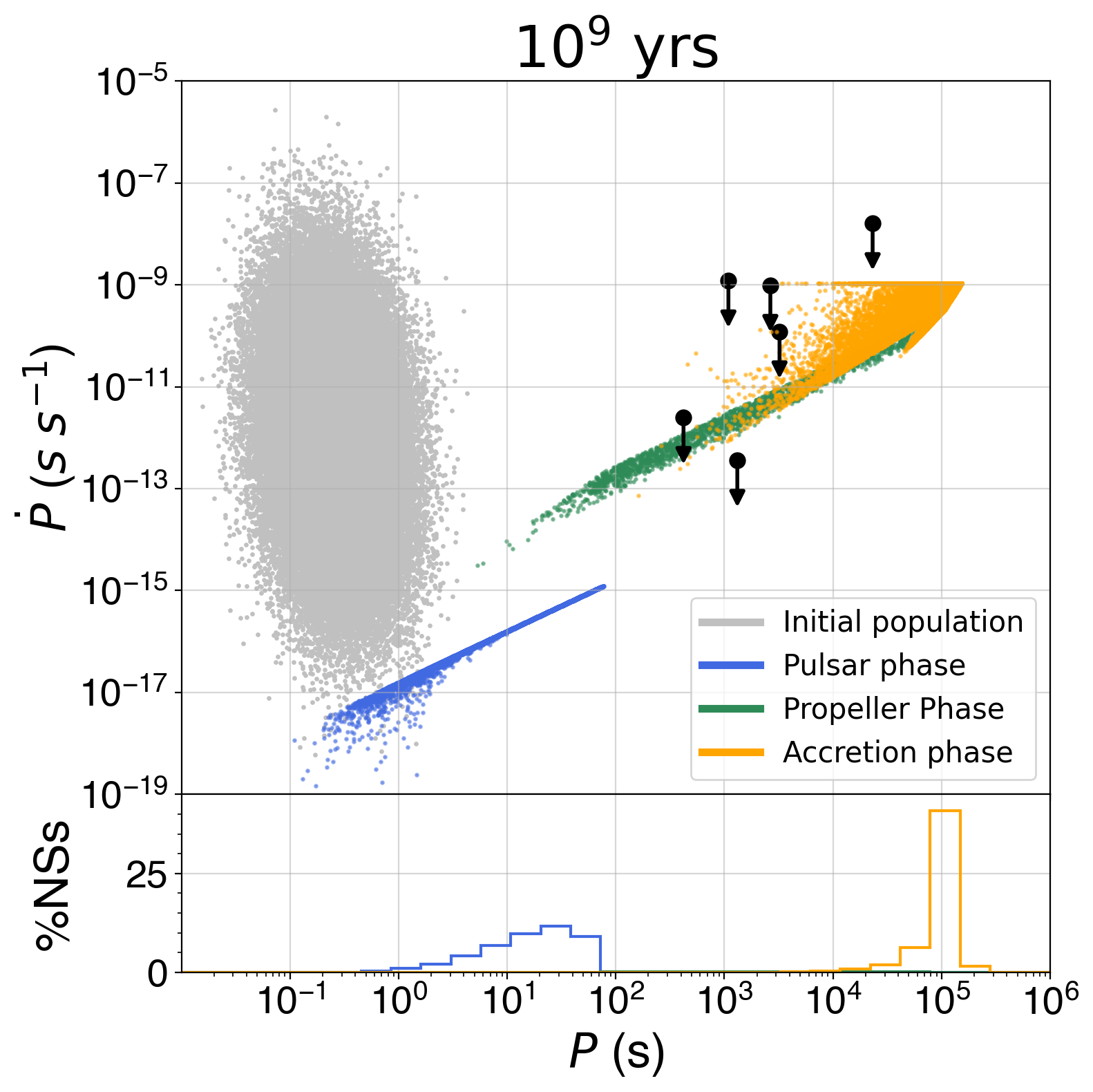}
    \caption{Population snapshots at various timeslices for for model B ($\gamma=1$, $\delta=0$) plotted in $P$--$\dot{P}$ space.}
\end{figure}
\begin{deluxetable*}{c|cc|ccc|c}[h!]
\tablewidth{0pt} 
\tablecaption{Fraction of NSs found in different timeslices during the population synthesis study for model C. A `--' indicates that no simulations were found in that phase, corresponding to a $<0.001$\% likelihood of occuring.} \label{tab:time-slice_C}
\tablehead{
\colhead{Time (yrs)} & \multicolumn{2}{c}{Pulsar phase} & \multicolumn{3}{c}{Propeller phase} & \colhead{Accretion phase}
\\
\colhead{} & \colhead{$<10^1$s} & \colhead{$>10^1$s} & \colhead{$<10^3$s} & \colhead{$10^3-10^4$s} & \colhead{$>10^4$s} & \colhead{}}
\startdata 
{$10^4$}&  97\% & 3\% & -- & -- & -- & --\\
{$10^5$}&  94\% & 6\% & -- & -- & -- & --\\
{$10^6$}&  92\% & 8\% & 0.013\% & -- & -- & --\\
{$10^7$}&  59\% & 41\% & 0.008\% & 0.002\% & -- & 0.026\%\\
{$10^8$}&  34\% & 65\% & 0.079\% & 0.041\% & 0.028\% & 1\%\\
{$10^9$}&  14\% & 32\% & 1\% & 1\% & 1\% & 53\%\\
\enddata
\end{deluxetable*}
\clearpage
\pagebreak
\subsection{Model D}
\begin{figure}[h!]
    \centering
    \includegraphics[width=0.32\textwidth]{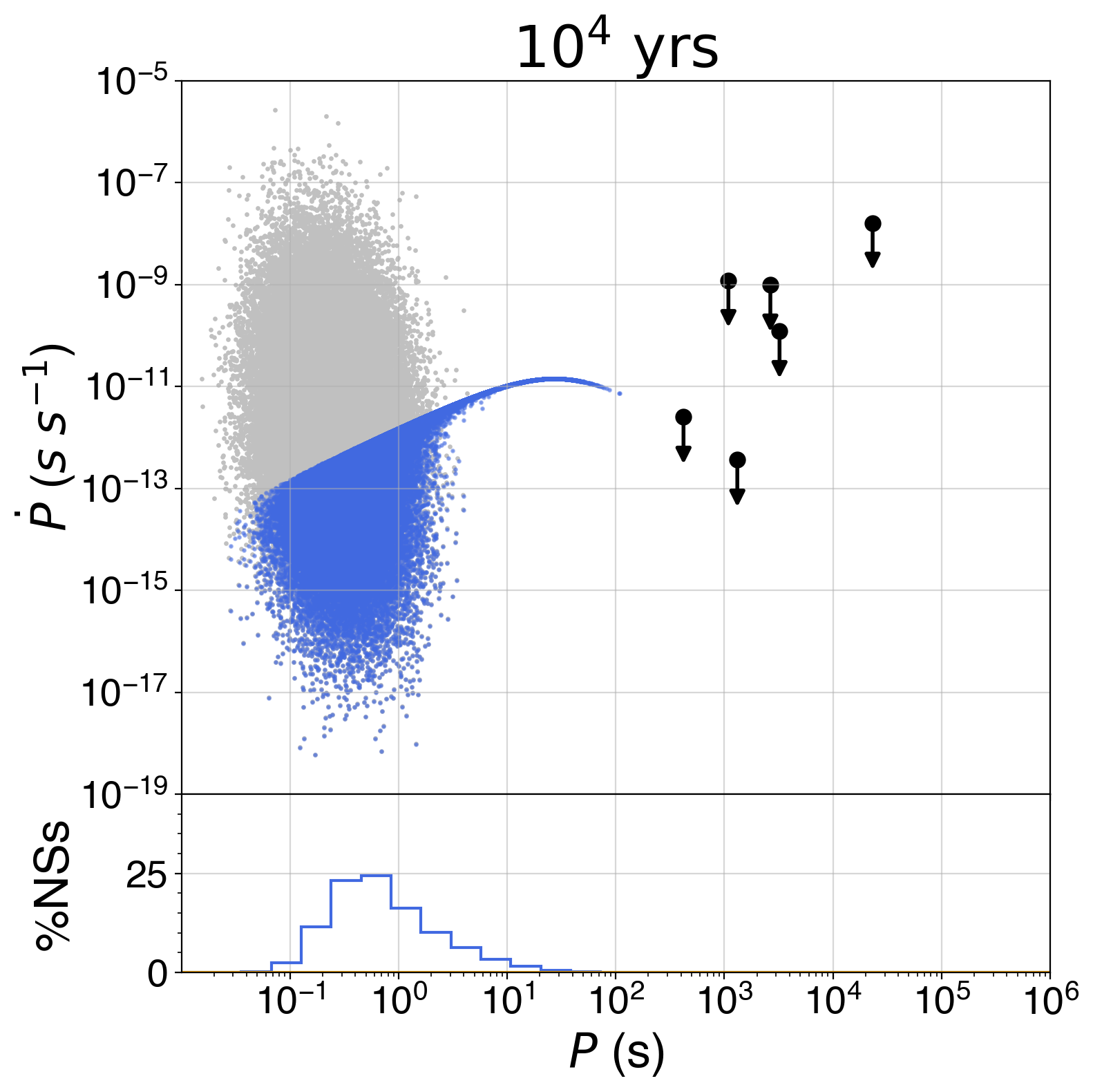}
    \includegraphics[width=0.32\textwidth]{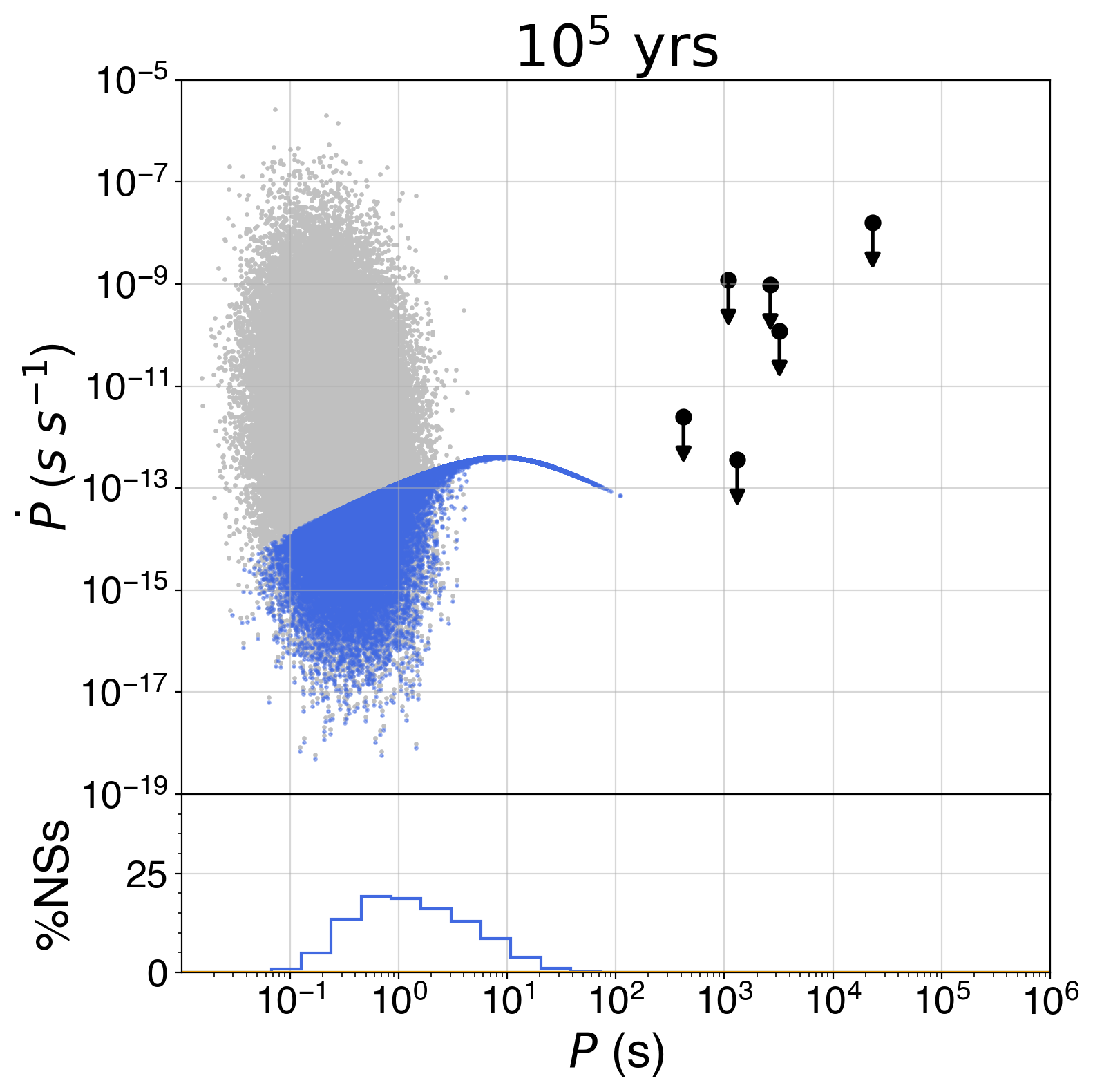}
    \includegraphics[width=0.32\textwidth]{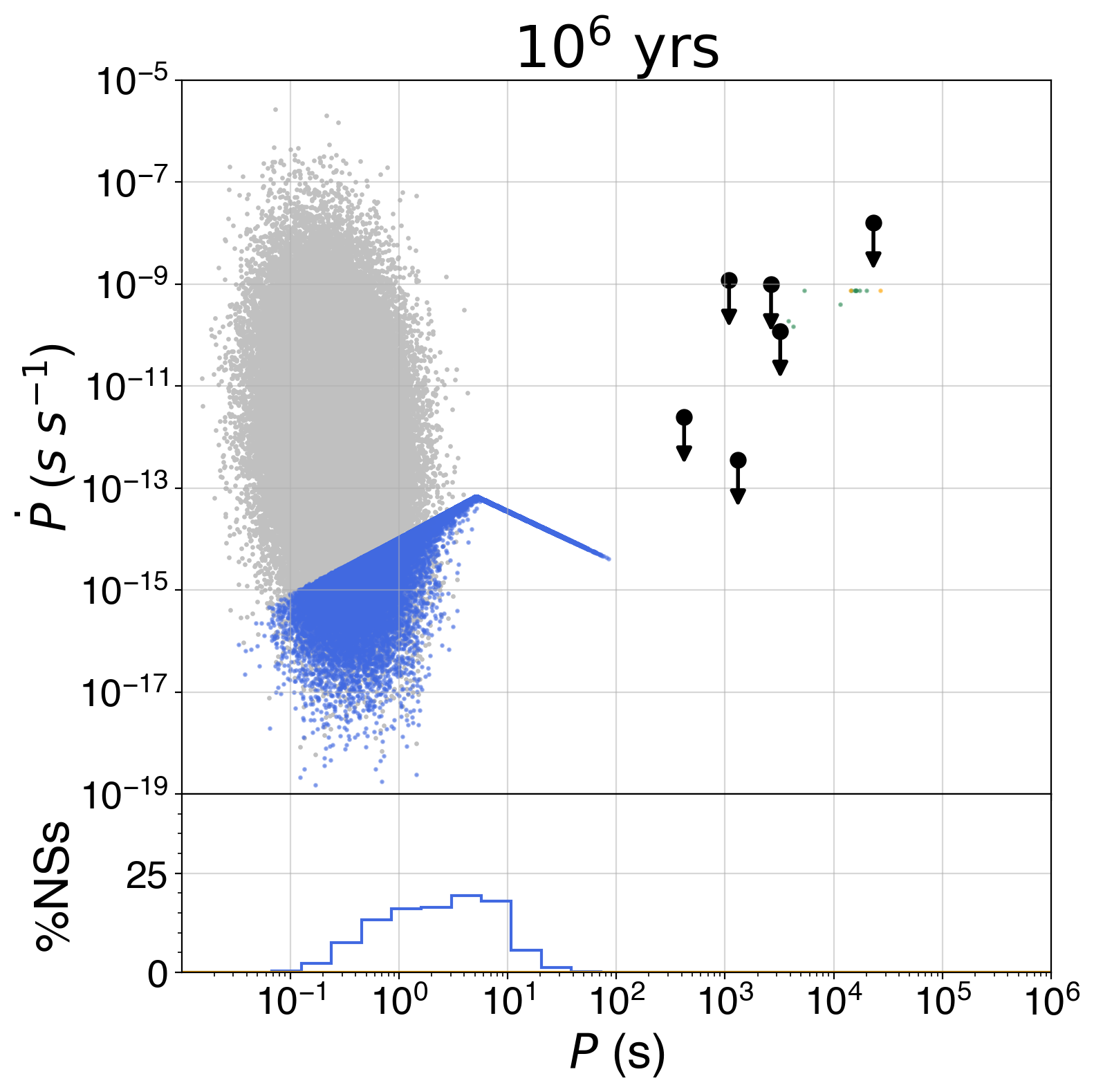}\\
    \includegraphics[width=0.32\textwidth]{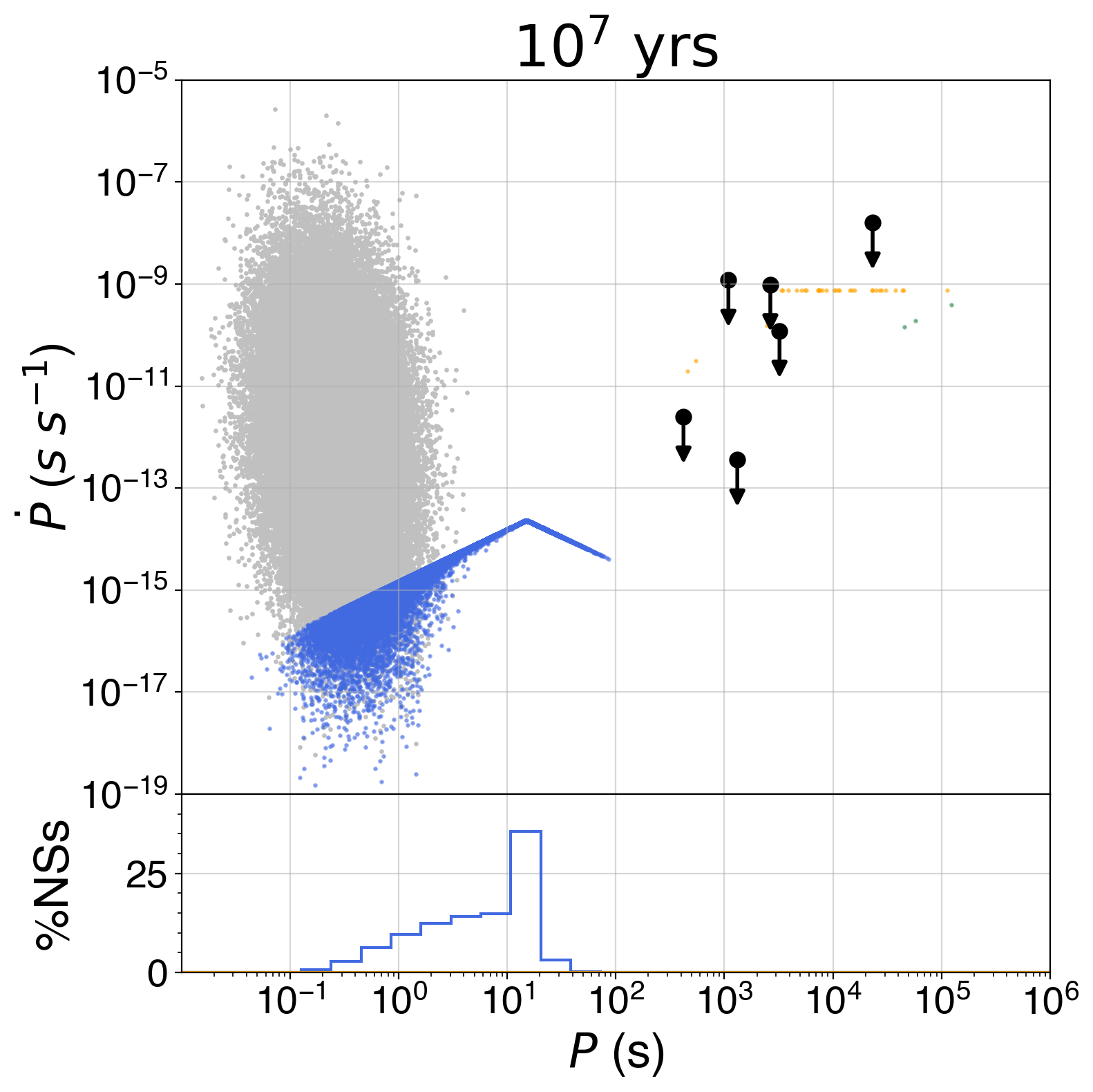}
    \includegraphics[width=0.32\textwidth]{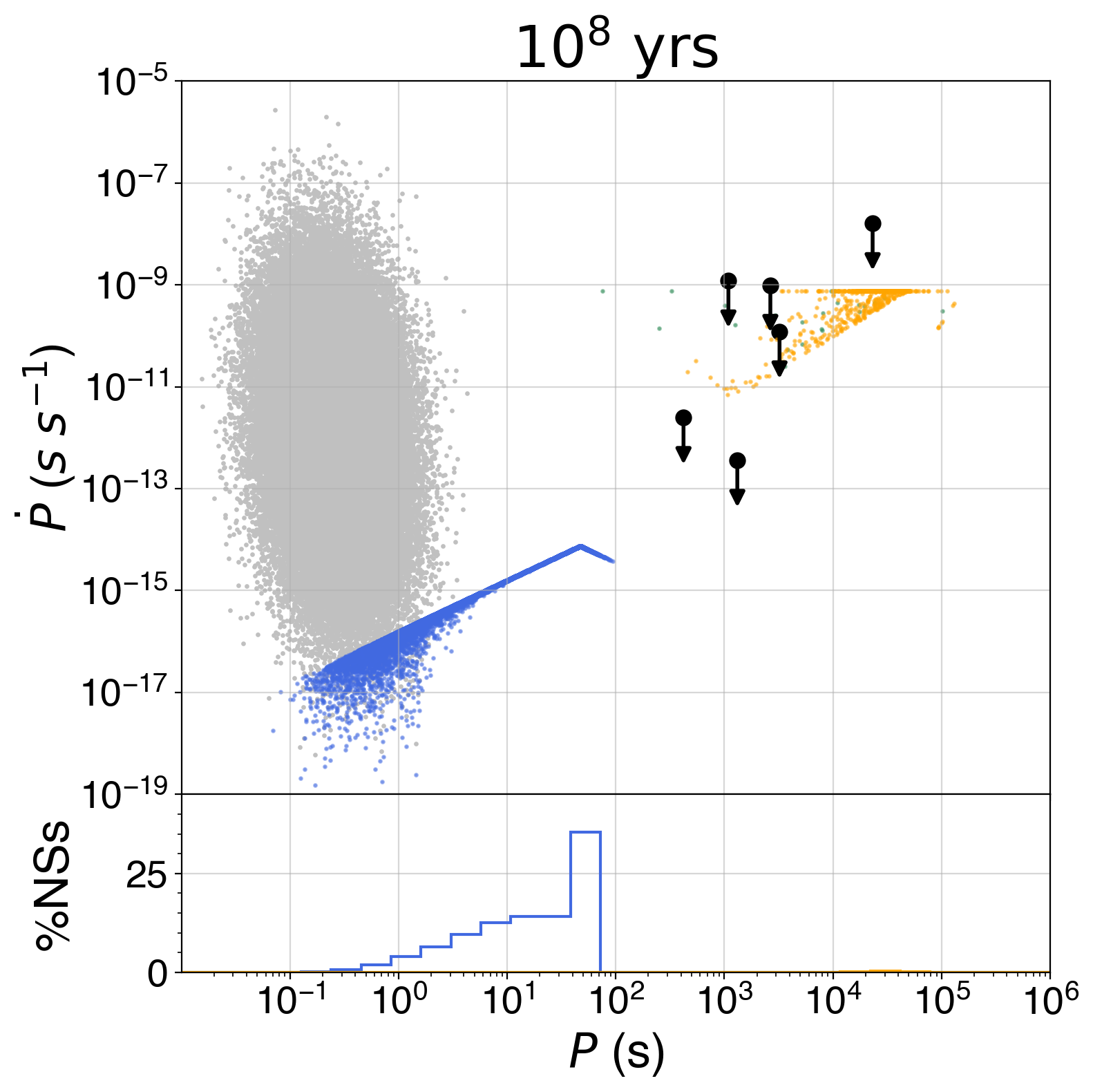}
    \includegraphics[width=0.32\textwidth]{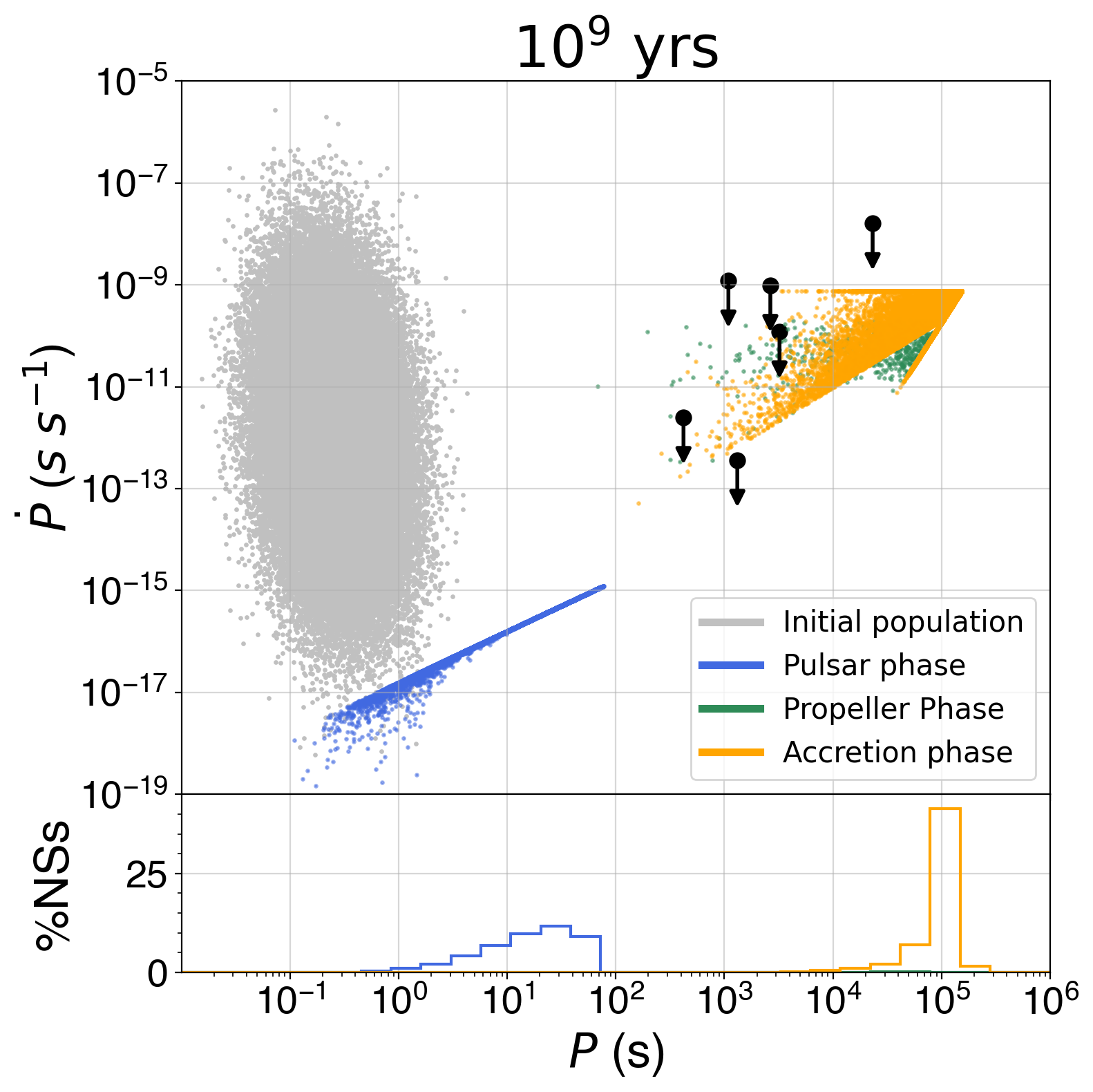}
    \caption{Population snapshots at various timeslices for for model D ($\gamma=2$, $\delta=0$) plotted in $P$--$\dot{P}$ space.}
\end{figure}
\begin{deluxetable*}{c|cc|ccc|c}[h!]
\tablewidth{0pt} 
\tablecaption{Fraction of NSs found in different timeslices during the population synthesis study for model D. A `--' indicates that no simulations were found in that phase, corresponding to a $<0.001$\% likelihood of occuring.} \label{tab:time-slice_D}
\tablehead{
\colhead{Time (yrs)} & \multicolumn{2}{c}{Pulsar phase} & \multicolumn{3}{c}{Propeller phase} & \colhead{Accretion phase}
\\
\colhead{} & \colhead{$<10^1$s} & \colhead{$>10^1$s} & \colhead{$<10^3$s} & \colhead{$10^3-10^4$s} & \colhead{$>10^4$s} & \colhead{}}
\startdata 
{$10^4$}&  97\% & 3\% & -- & -- & -- & --\\
{$10^5$}&  94\% & 6\% & -- & -- & -- & --\\
{$10^6$}&  92\% & 8\% & -- & 0.003\% & 0.008\% & 0.002\%\\
{$10^7$}&  59\% & 41\% & -- & -- & 0.003\% & 0.033\%\\
{$10^8$}&  34\% & 65\% & 0.003\% & 0.011\% & 0.016\% & 1\%\\
{$10^9$}&  14\% & 32\% & 0.018\% & 0.124\% & 0.473\% & 54\%\\
\enddata
\end{deluxetable*}
\clearpage
\pagebreak
\subsection{Model E}
\begin{figure}[h!]
    \centering
    \includegraphics[width=0.32\textwidth]{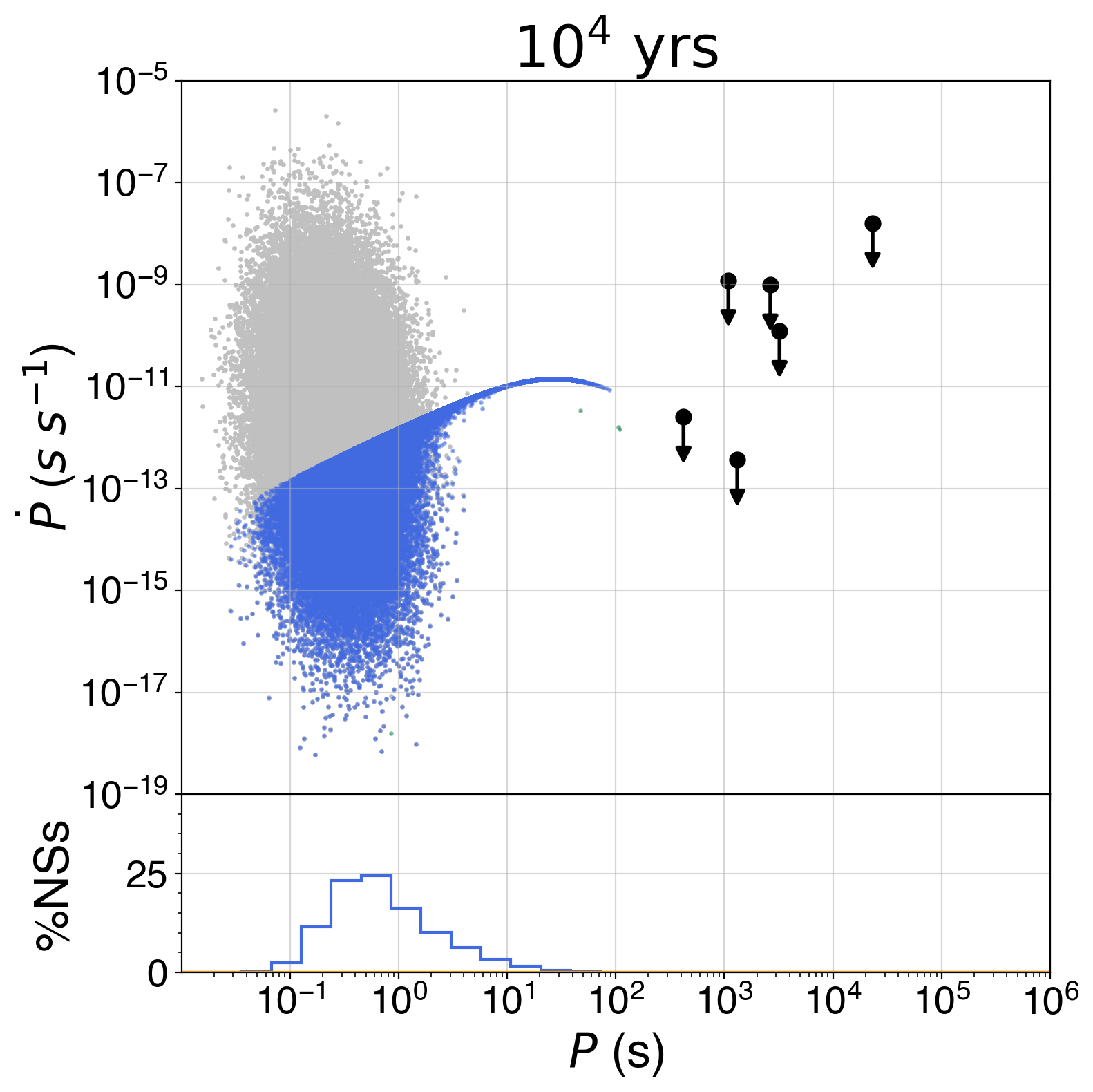}
    \includegraphics[width=0.32\textwidth]{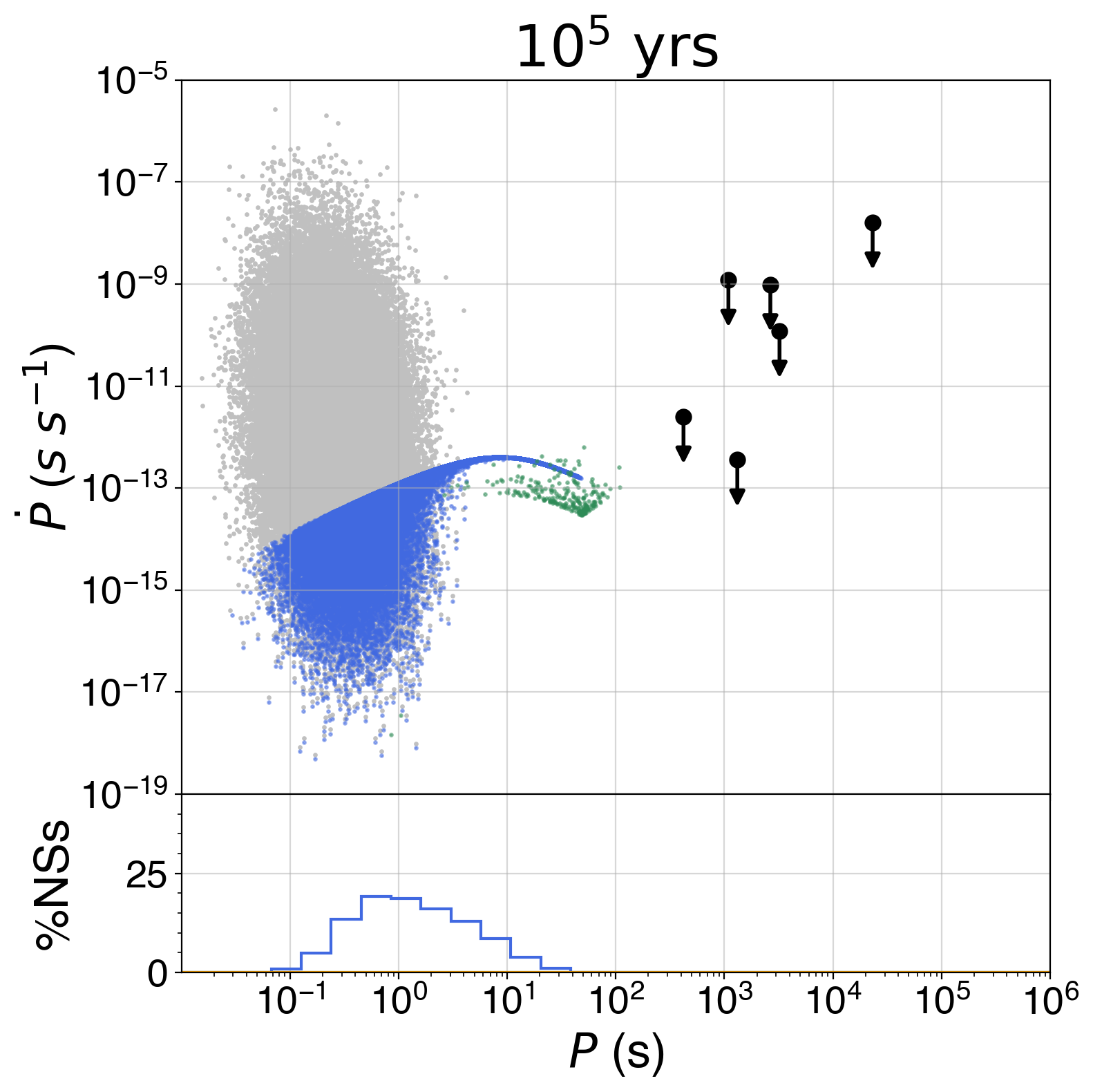}
    \includegraphics[width=0.32\textwidth]{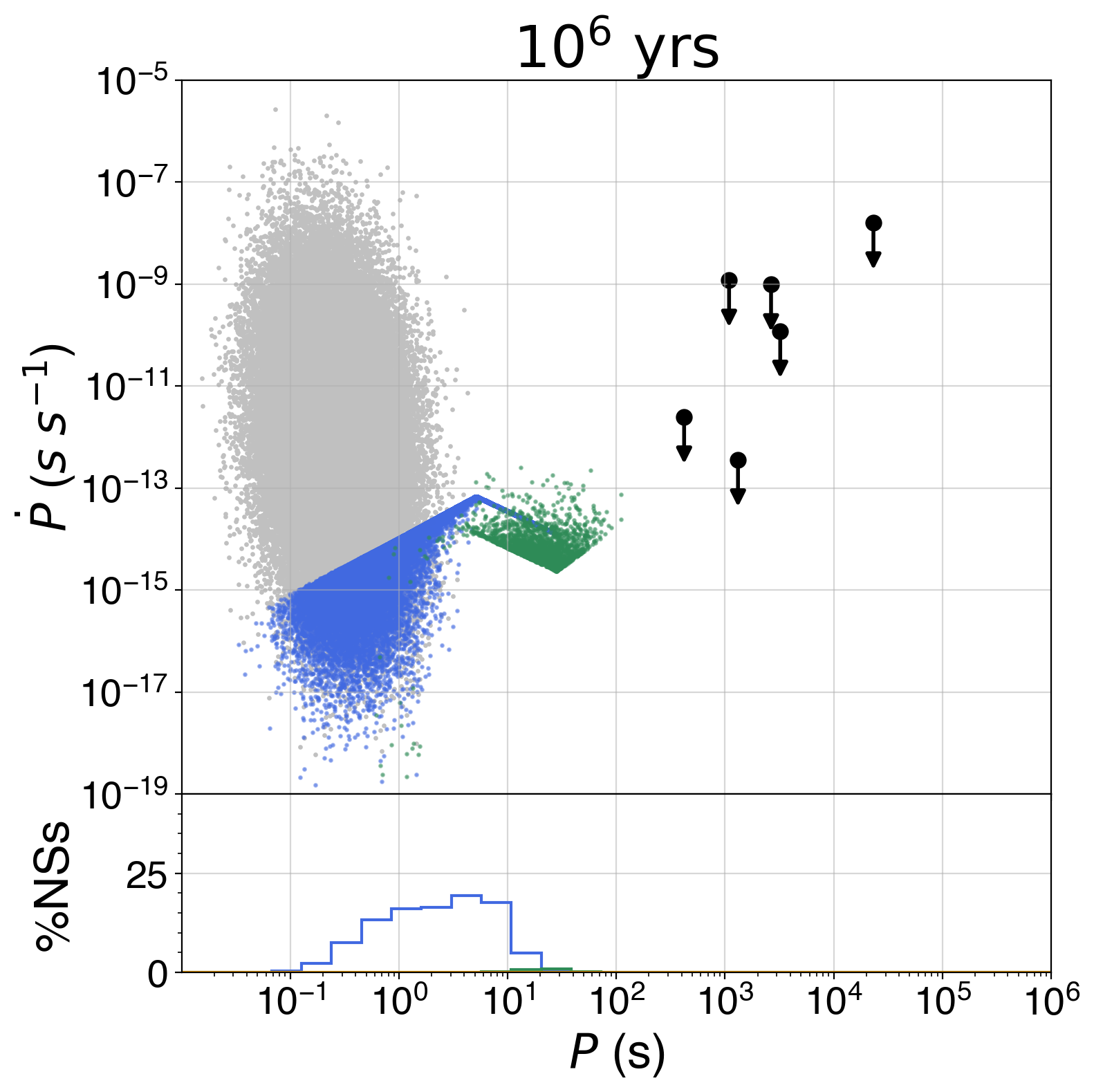}\\
    \includegraphics[width=0.32\textwidth]{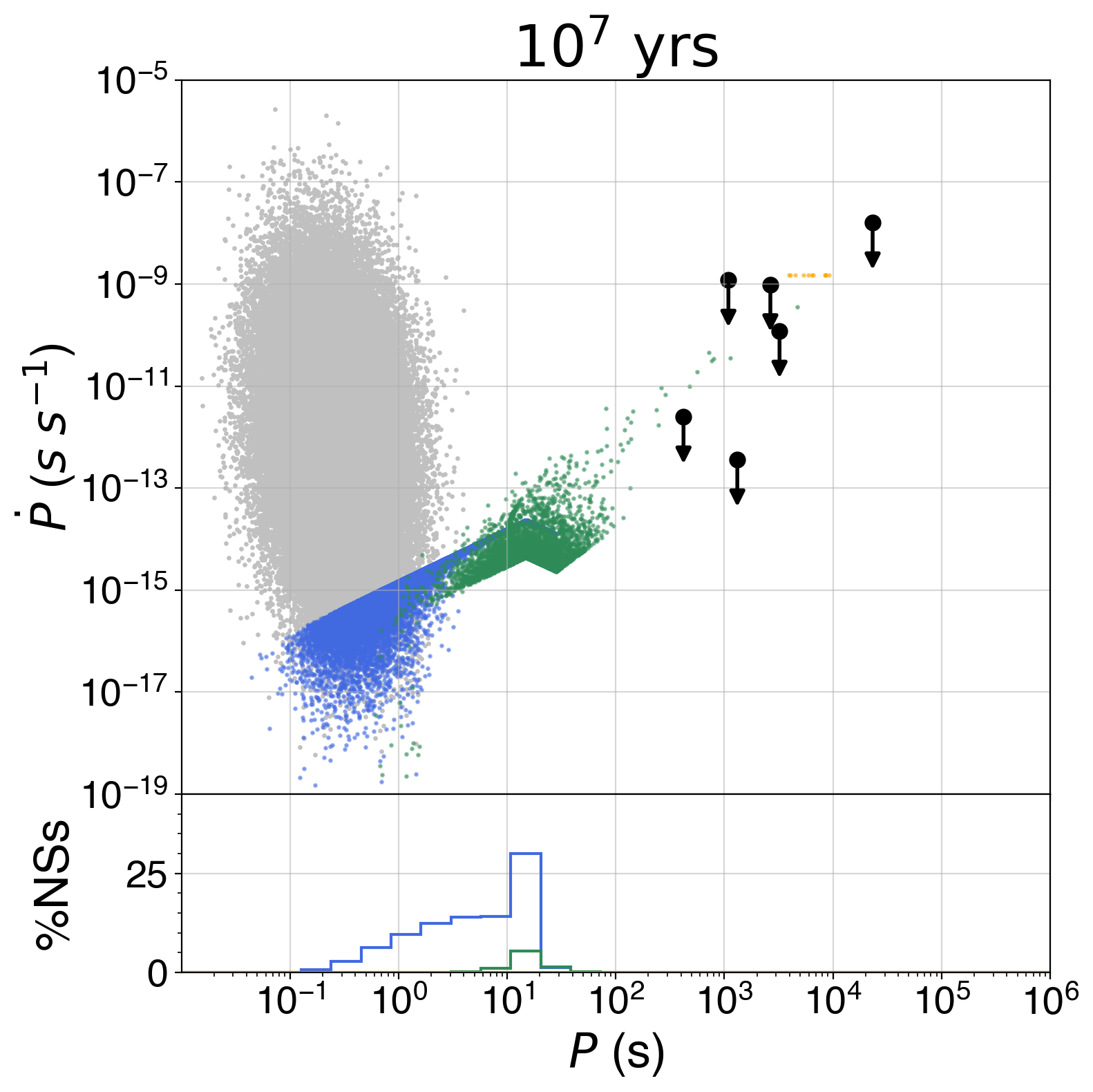}
    \includegraphics[width=0.32\textwidth]{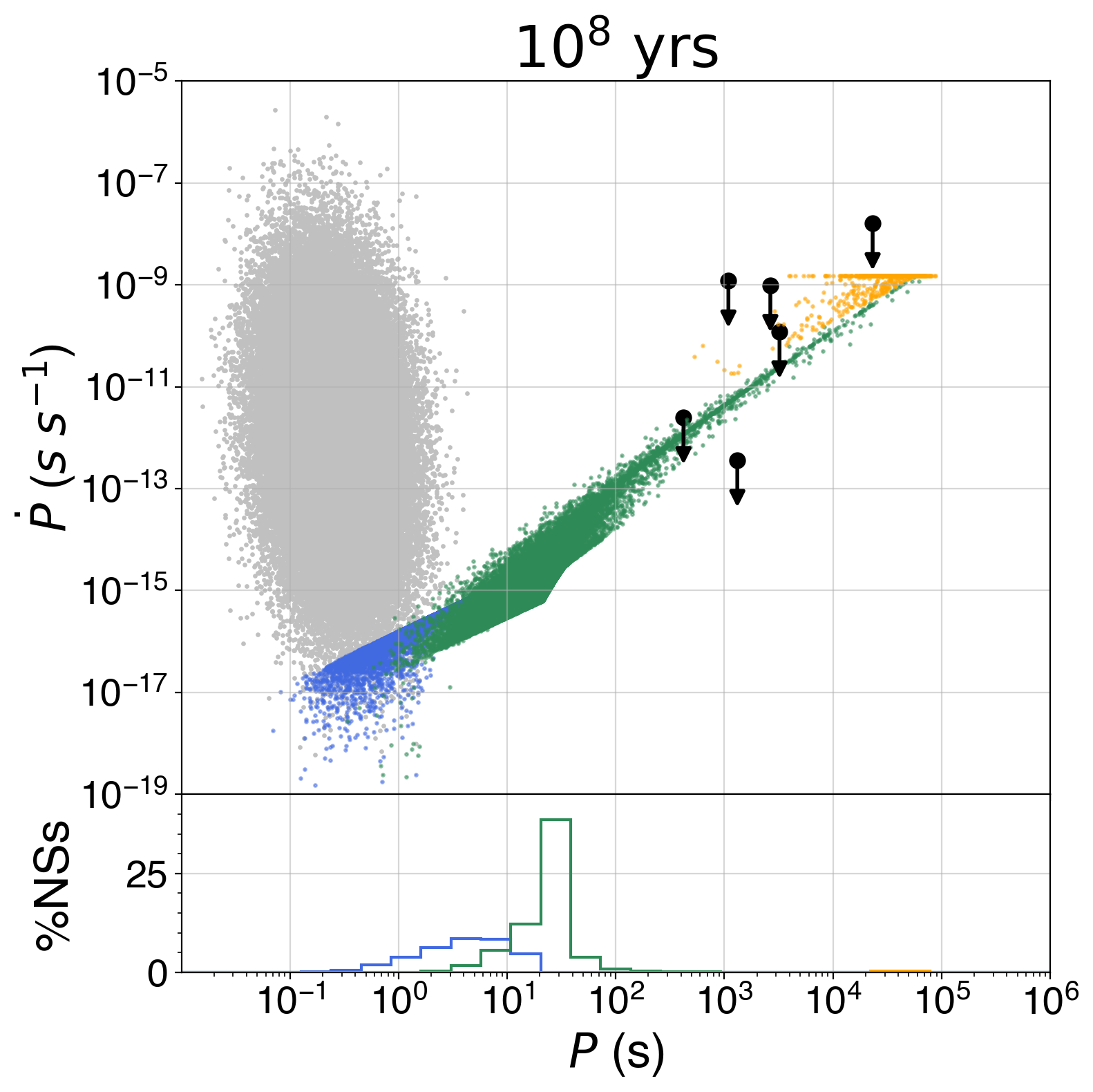}
    \includegraphics[width=0.32\textwidth]{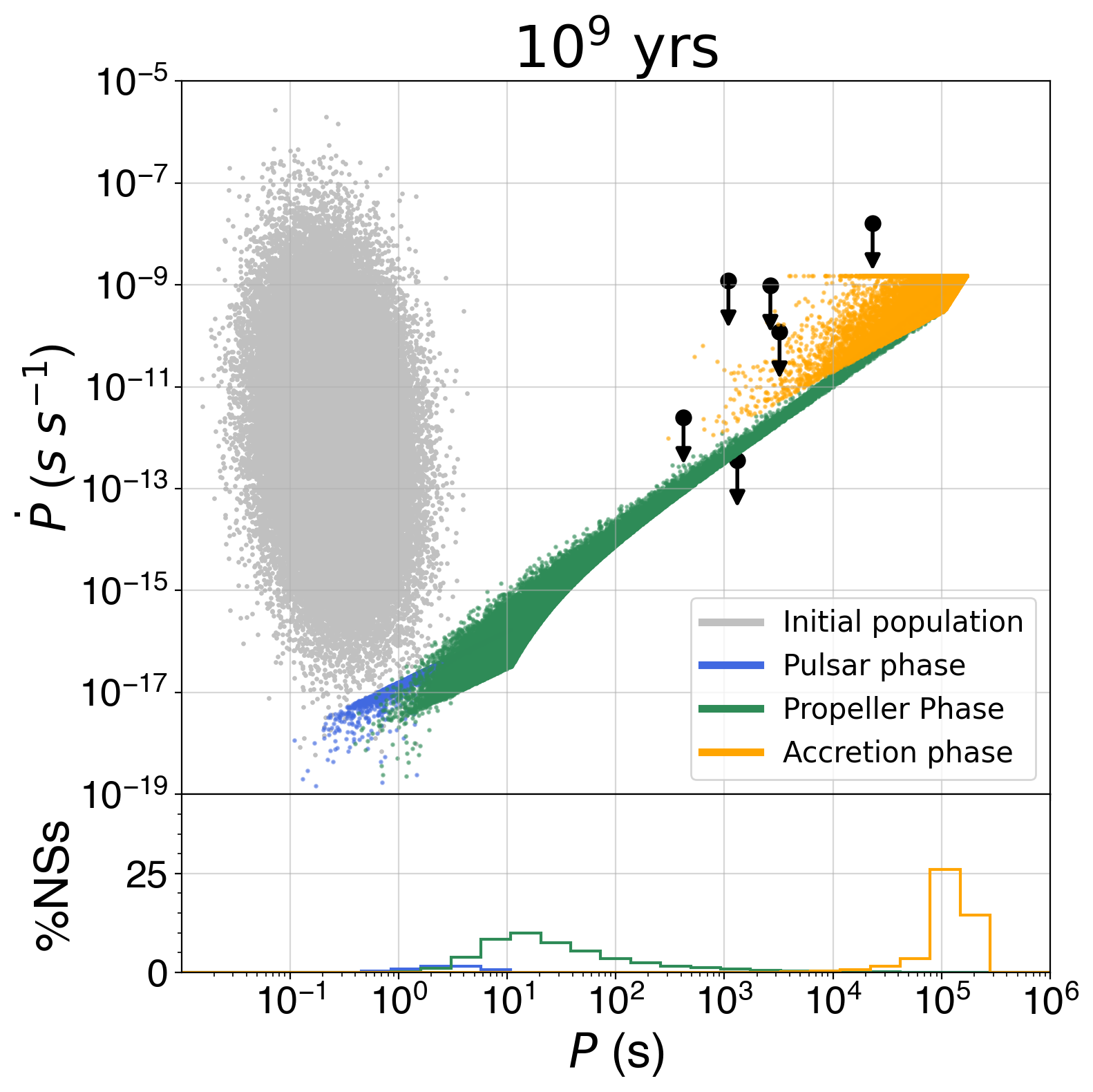}
    \caption{Population snapshots at various timeslices for for model E ($\gamma=1$, $\delta=1$) plotted in $P$--$\dot{P}$ space.}
\end{figure}
\begin{deluxetable*}{c|cc|ccc|c}[h!]
\tablewidth{0pt} 
\tablecaption{Fraction of NSs found in different timeslices during the population synthesis study for model E. A `--' indicates that no simulations were found in that phase, corresponding to a $<0.001$\% likelihood of occuring.} 
\tablehead{
\colhead{Time (yrs)} & \multicolumn{2}{c}{Pulsar phase} & \multicolumn{3}{c}{Propeller phase} & \colhead{Accretion phase}
\\
\colhead{} & \colhead{$<10^1$s} & \colhead{$>10^1$s} & \colhead{$<10^3$s} & \colhead{$10^3-10^4$s} & \colhead{$>10^4$s} & \colhead{}}
\startdata 
{$10^4$}&  97\% & 3\% & 0.004\% & -- & -- & --\\
{$10^5$}&  94\% & 6\% & 0.245\% & -- & -- & --\\
{$10^6$}&  91\% & 7\% & 2\% & -- & -- & --\\
{$10^7$}&  58\% & 33\% & 9\% & 0.002\% & -- & 0.011\%\\
{$10^8$}&  29\% & 6\% & 64\% & 0.218\% & 0.053\% & 1\%\\
{$10^9$}&  5\% & 0.026\% & 45\% & 2\% & 1\% & 47\%\\
\enddata
\end{deluxetable*}
\clearpage
\pagebreak
\subsection{Model F}
\begin{figure}[h!]
    \centering
    \includegraphics[width=0.32\textwidth]{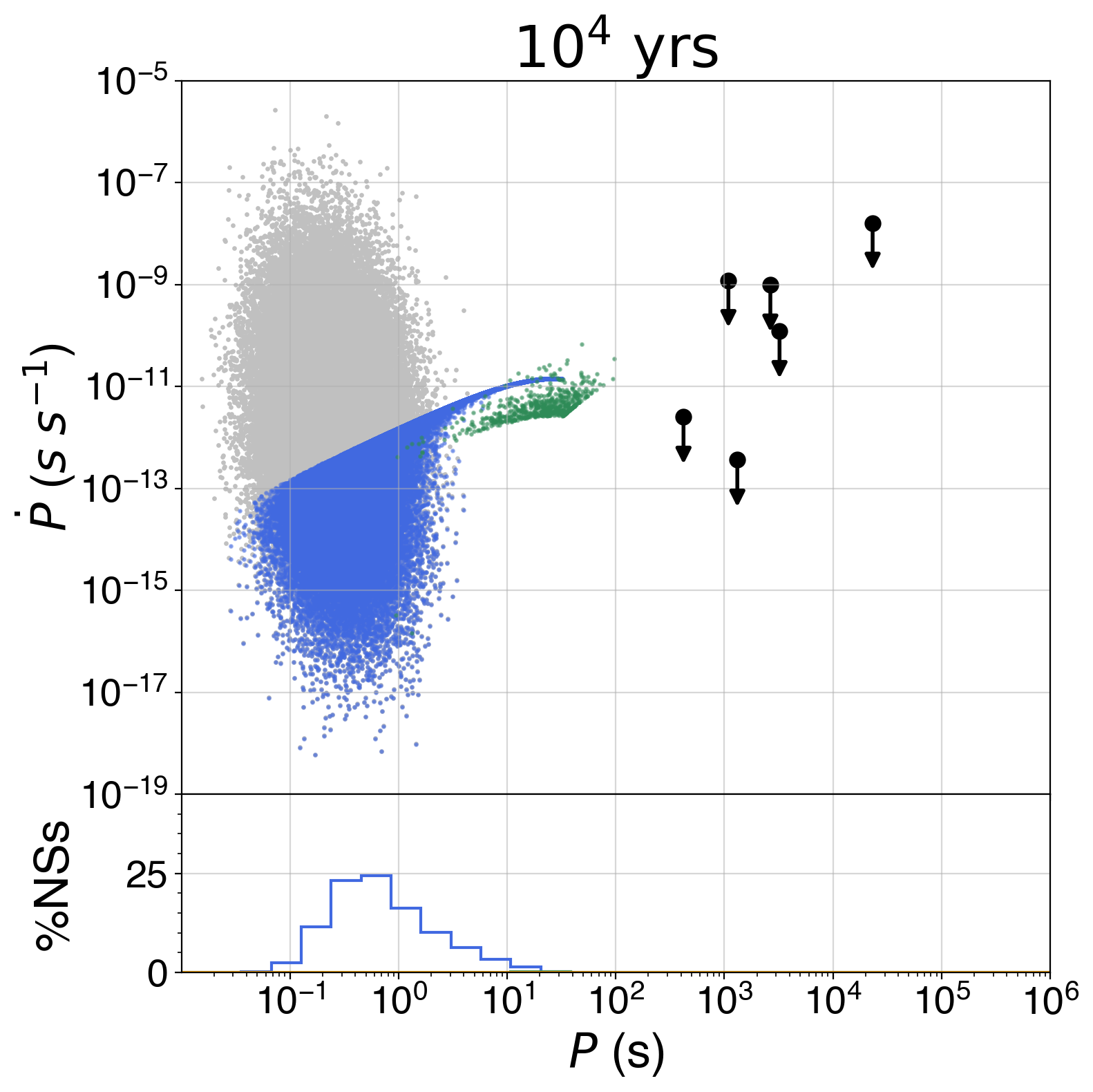}
    \includegraphics[width=0.32\textwidth]{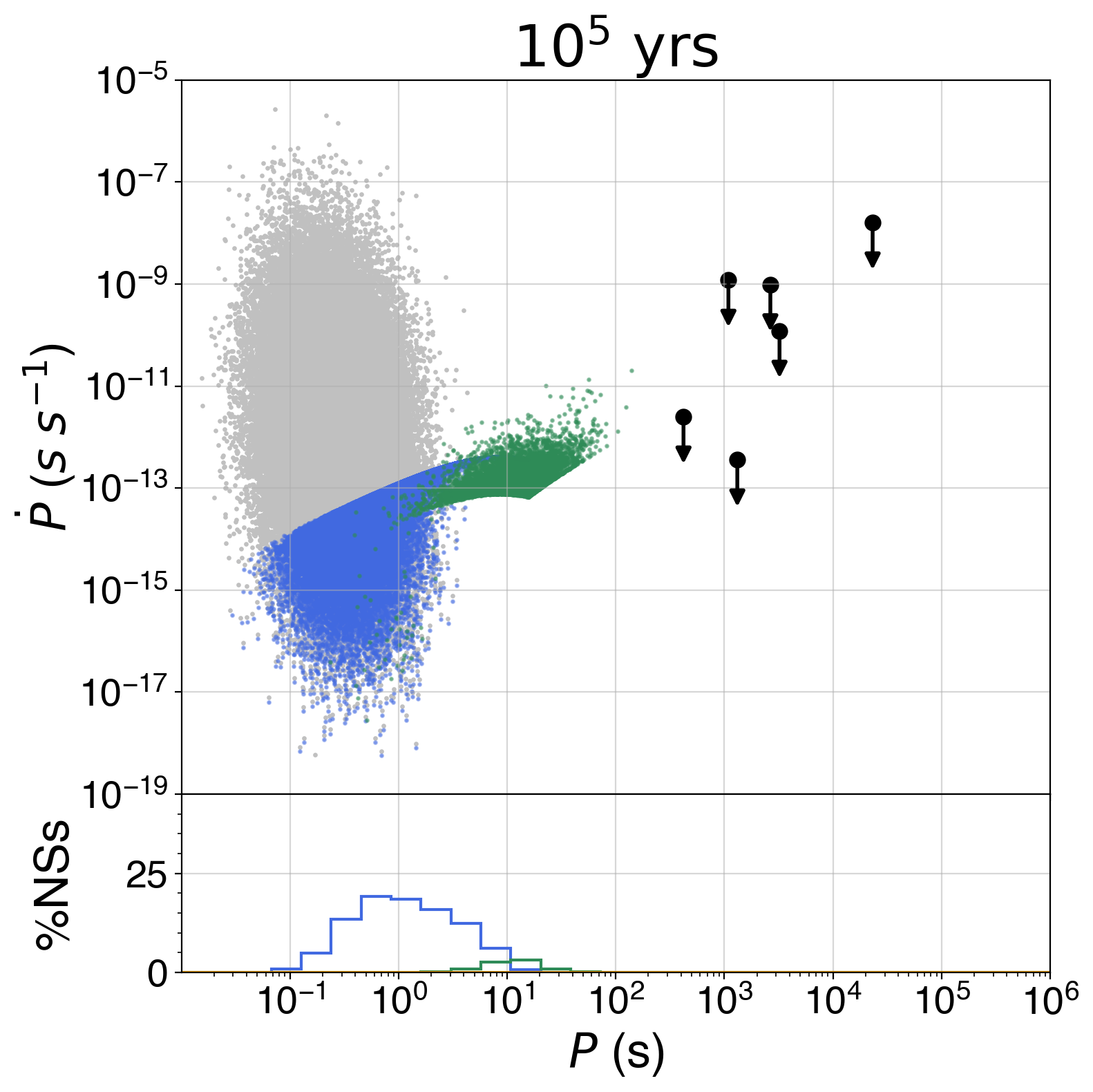}
    \includegraphics[width=0.32\textwidth]{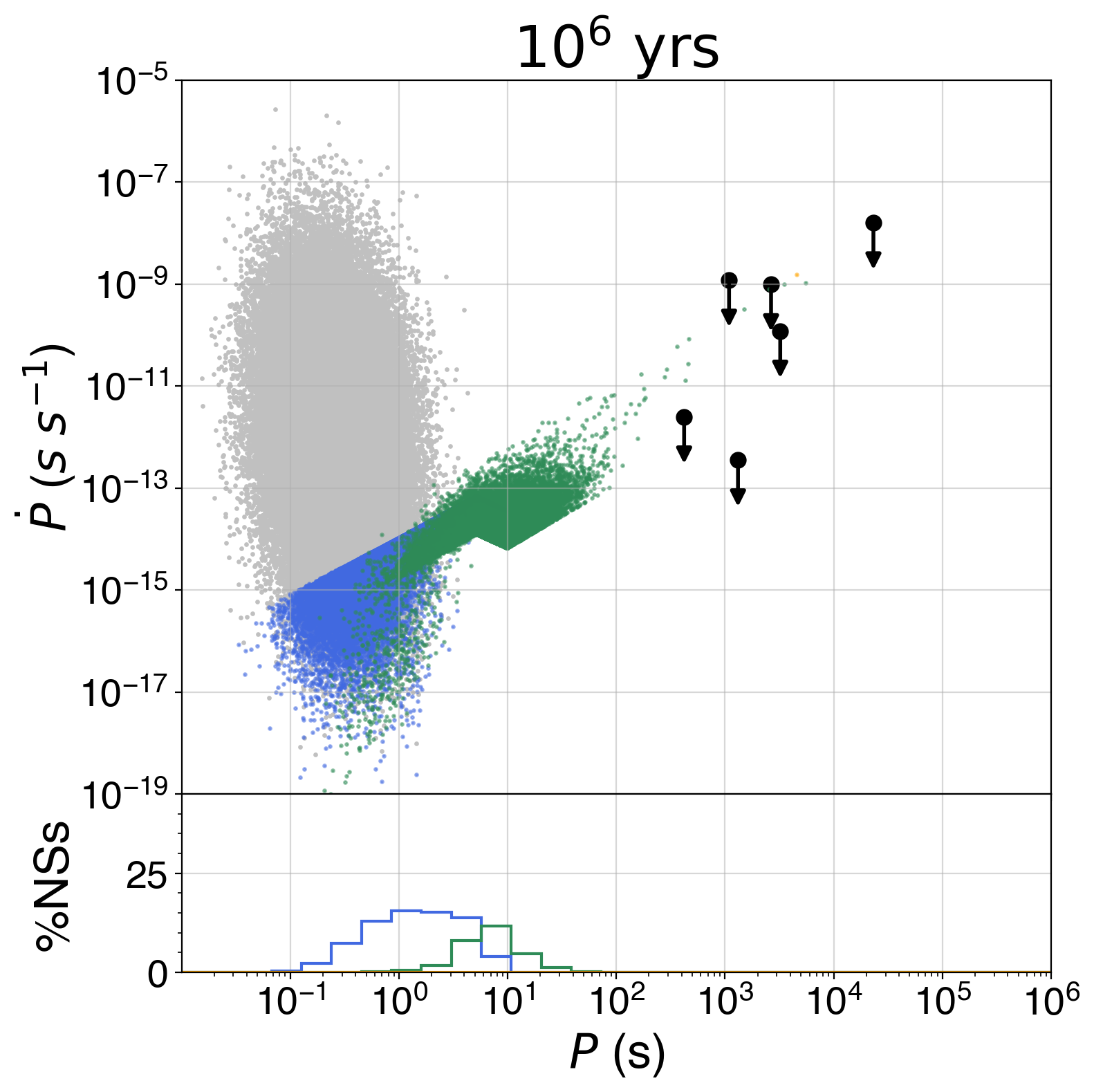}\\
    \includegraphics[width=0.32\textwidth]{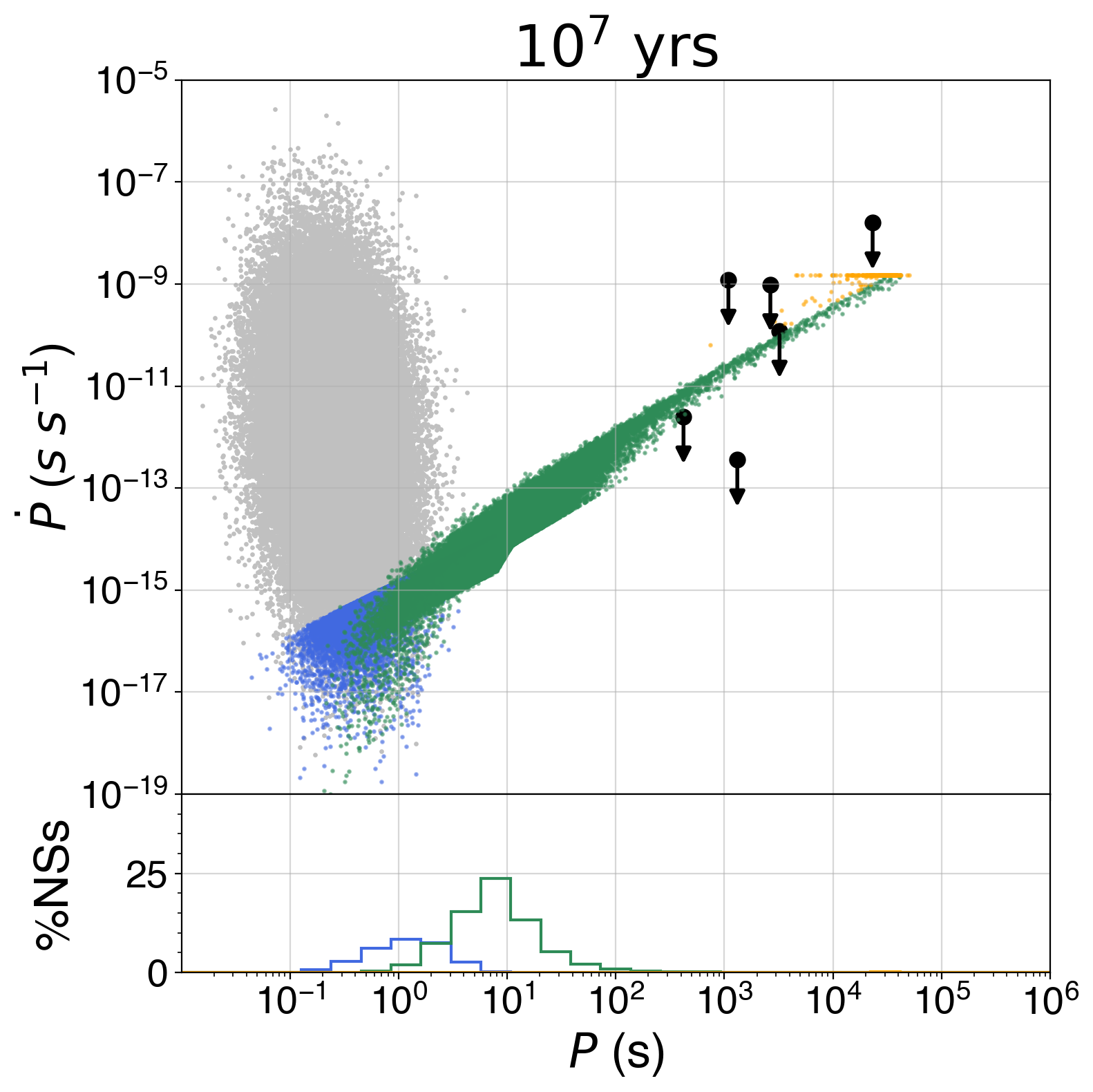}
    \includegraphics[width=0.32\textwidth]{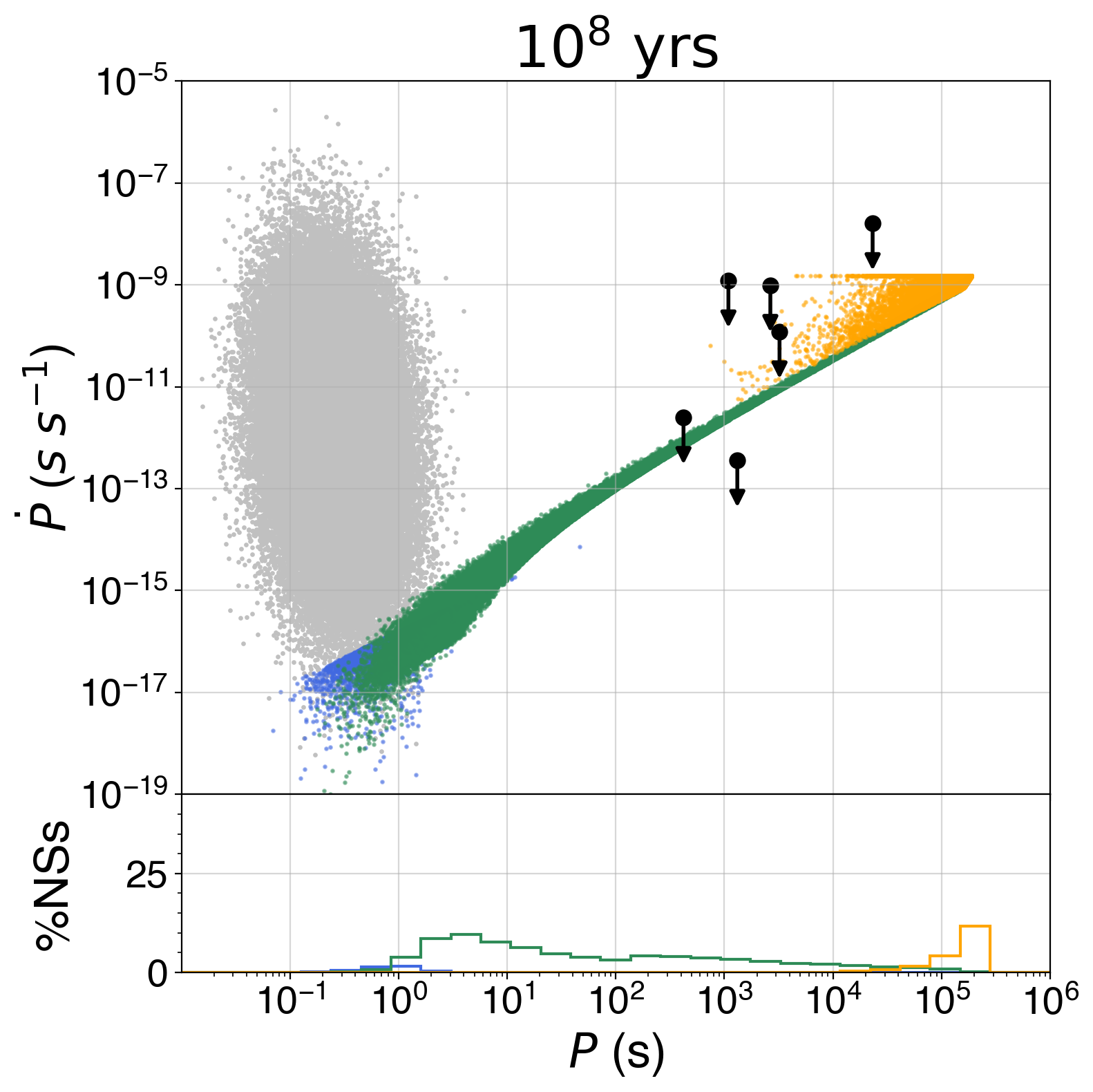}
    \includegraphics[width=0.32\textwidth]{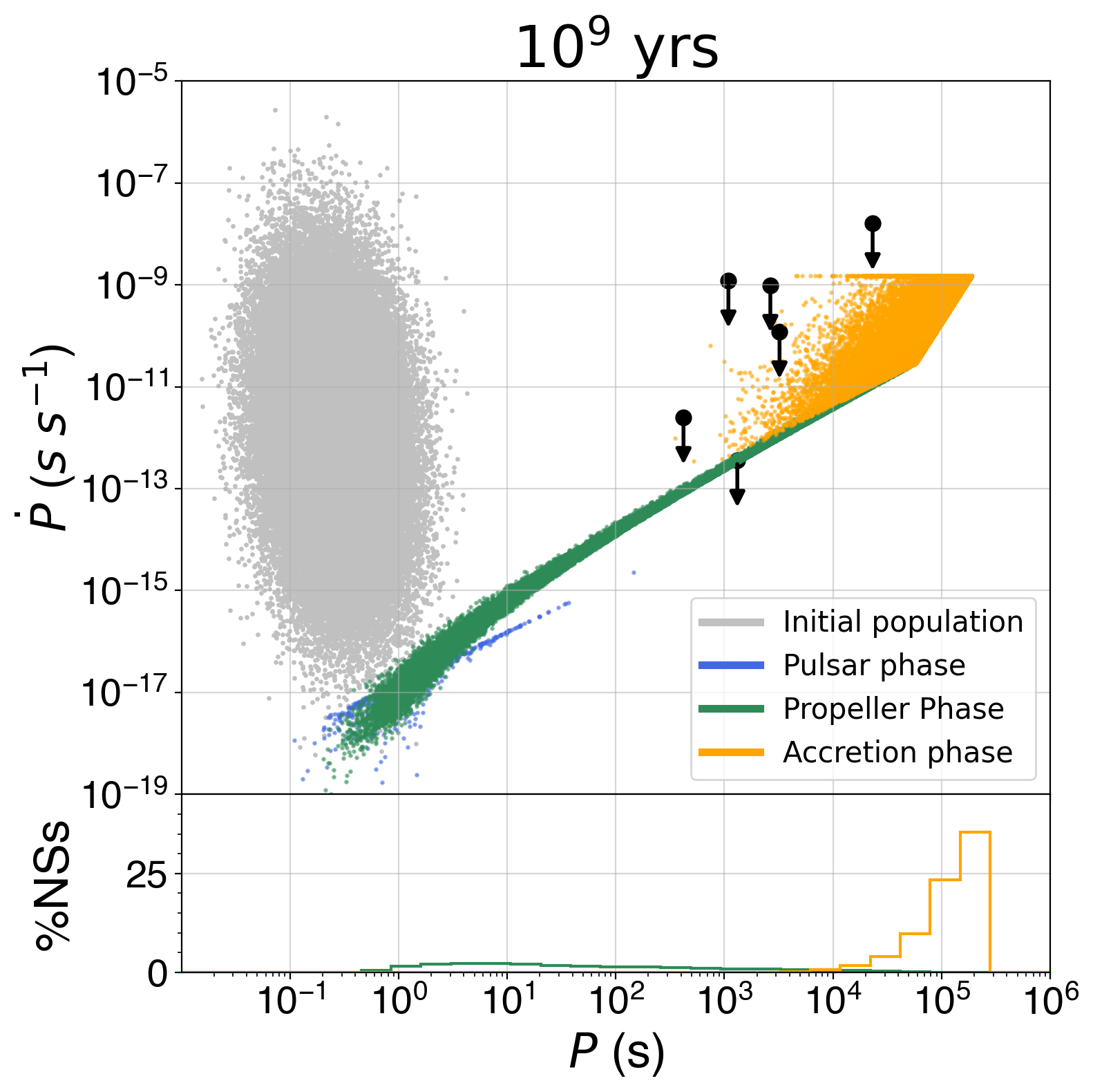}
    \caption{Population snapshots at various timeslices for for model F ($\gamma=2$, $\delta=2$) plotted in $P$--$\dot{P}$ space.}
\end{figure}
\begin{deluxetable*}{c|cc|ccc|c}[h!]
\tablewidth{0pt} 
\tablecaption{Fraction of NSs found in different timeslices during the population synthesis study for model F. A `--' indicates that no simulations were found in that phase, corresponding to a $<0.001$\% likelihood of occuring.} \label{tab:time-slice_F}
\tablehead{
\colhead{Time (yrs)} & \multicolumn{2}{c}{Pulsar phase} & \multicolumn{3}{c}{Propeller phase} & \colhead{Accretion phase}
\\
\colhead{} & \colhead{$<10^1$s} & \colhead{$>10^1$s} & \colhead{$<10^3$s} & \colhead{$10^3-10^4$s} & \colhead{$>10^4$s} & \colhead{}}
\startdata 
{$10^4$}&  97\% & 2\% & 1\% & -- & -- & --\\
{$10^5$}&  91\% & 1\% & 8\% & -- & -- & --\\
{$10^6$}&  72\% & -- & 28\% & 0.004\% & -- & 0.001\%\\
{$10^7$}&  28\% & 0.001\% & 71\% & 0.236\% & 0.065\% & 0.213\%\\
{$10^8$}&  4\% & 0.004\% & 61\% & 10\% & 6\% & 19\%\\
{$10^9$}&  0.397\% & 0.027\% & 20\% & 3\% & 1\% & 76\%\\
\enddata
\end{deluxetable*}

\end{document}